\documentclass[english,preprint,aip,jcp]{revtex4-1}
\usepackage[T1]{fontenc}
\usepackage[latin9]{inputenc}
\usepackage{amsmath}
\usepackage{amssymb}
\usepackage{graphicx}
\PassOptionsToPackage{normalem}{ulem}
\usepackage{ulem}

\makeatletter

\newcommand{\lyxdot}{.}

\usepackage{enumitem}
\usepackage{amsmath}
\usepackage{accents}
\newlength{\dhatheight}
\newcommand{\doublehat}[1]{%
    \settoheight{\dhatheight}{\ensuremath{\hat{#1}}}%
    \addtolength{\dhatheight}{-0.15ex}%
    \widehat{\vphantom{\rule{1pt}{\dhatheight}}%
    \smash{\widehat{#1}}}}

\makeatother

\usepackage{babel}
\begin{document}

\title{Ehrenfest+R Dynamics I: A Mixed Quantum-Classical Electrodynamics
Simulation of Spontaneous Emission}

\author{Hsing-Ta Chen}
\email{hsingc@sas.upenn.edu}

\selectlanguage{english}%

\affiliation{Department of Chemistry, University of Pennsylvania, Philadelphia,
Pennsylvania 19104, U.S.A.}

\author{Tao E. Li}

\affiliation{Department of Chemistry, University of Pennsylvania, Philadelphia,
Pennsylvania 19104, U.S.A.}

\author{Maxim Sukharev}

\affiliation{Department of Physics, Arizona State University, Tempe, Arizona 85287,
USA}

\affiliation{College of Integrative Sciences and Arts, Arizona State University,
Mesa, AZ 85212, USA}

\author{Abraham Nitzan}

\affiliation{Department of Chemistry, University of Pennsylvania, Philadelphia,
Pennsylvania 19104, U.S.A.}

\author{Joseph E. Subotnik}

\affiliation{Department of Chemistry, University of Pennsylvania, Philadelphia,
Pennsylvania 19104, U.S.A.}
\begin{abstract}
The dynamics of an electronic system interacting with an electromagnetic
field is investigated within mixed quantum-classical theory. Beyond
the classical path approximation (where we ignore all feedback from
the electronic system on the photon field), we consider all electron\textendash photon
interactions explicitly according to Ehrenfest (i.e. mean\textendash field)
dynamics and a set of coupled Maxwell\textendash Liouville equations.
Because Ehrenfest dynamics cannot capture certain quantum features
of the photon field correctly, we propose a new \emph{Ehrenfest+R}
method that can recover (by construction) spontaneous emission while
also distinguishing between electromagnetic fluctuations and coherent
emission. 

\end{abstract}
\maketitle

\section{Introduction\label{sec:Introduction}}

Light\textendash matter interactions are of pivotal importance to
the development of physics and chemistry. The optical response of
matter provides a useful tool for probing the structural and dynamical
properties of materials, with one possible long term goal being the
manipulation of light to control microscopic degrees of freedom. Now,
we usually describe light\textendash matter interactions through linear
response theory; the electromagnetic (EM) field is considered a perturbation
to the matter system and the optical response is predicted by extrapolating
the behavior of the system without illumination. Obviously, this
scheme does not account for the feedback of the matter system on the
EM field, and many recent experiments cannot be modeled through this
lens. For instance, in situations involving strong light\textendash matter
coupling, such as molecules in an optical cavity, spectroscopic observations
of nonlinearity have been reported as characteristic of quantum effects.\citep{thompson_nonlinear_1998,solano_strong-driving-assisted_2003,fink_climbing_2008,gibbs_excitonpolariton_2011,lodahl_interfacing_2015}
As another example, for systems composed of many quantum emitters,
collective effects from light\textendash matter interactions lead
to phenomena incompatible with linear response theory, such as coupled
exciton\textendash plasma optics\citep{torma_strong_2015,puthumpally-joseph_dipole-induced_2014,puthumpally-joseph_theoretical_2015,sukharev_optics_2017,vasa_strong_2018}
and superradiance lasers.\citep{dicke_coherence_1954,andreev_collective_1980,oppel_directional_2014}

The phenomena above raise an exciting challenge to existing theories;
one needs to treat the matter and EM fields within a consistent framework.
Despite great progress heretofore using simplified quantum models,\citep{dirac_quantum_1927,dirac_quantum_1927-1}
semiclassical simulations provide an important means for studying
subtle light\textendash matter interactions in realistic systems.\citep{milonni_semiclassical_1976}
Most semiclassical simulations are based on a mixed quantum\textendash classical
separation treating the electronic/molecular system with quantum mechanics
and the bath degrees of freedom with classical mechanics. While there
are many semiclassical approaches for coupled electronic\textendash nuclear
systems offering intuitive interpretations and meaningful predictions,\citep{kapral_mixed_1999,tully_mixed_1998,tully_molecular_1990,wang_semiclassical_1998,wang_semiclassical_1999}
the feasibility of analogous semiclassical techniques for coupled
electron\textendash radiation dynamics remains an open question. With
that in mind, recent semiclassical advances, including numerical implementations
of the Maxwell\textendash Liouville equations,\citep{ziolkowski_ultrafast_1995,slavcheva_coupled_2002,fratalocchi_three-dimensional_2008,sukharev_numerical_2011}
symmetrical quantum-classical dynamics,\citep{miller_classical/semiclassical_1978,li_mixed_2018,provazza_communication:_2018}
and mean-field Ehrenfest dynamics,\citep{li_mixed_2018} have now
begun exploring exciting collective effects, even when spontaneous
emission is included.

For electron\textendash radiation dynamics, the most natural approach
is the Ehrenfest method, combining the quantum Liouville equation
with classical electrodynamics in a mean-field manner; this approach
should be reliable given the lack of a time-scale separation between
electronic and EM dynamics. Nevertheless, Ehrenfest dynamics are known
to suffer from several drawbacks. First, it is well-known that, for
electronic\textendash nuclear dynamics, Ehrenfest dynamics do not
satisfy detailed balance.\citep{parandekar_detailed_2006} This drawback
will usually lead to incorrect electronic populations at long times.
The failure to maintain detailed balance results in anomalous energy
flow (that can even sometimes violate the second law of thermodynamics
at equilibrium.\citep{jain_vibrational_2018}) For scattering of light
from electronic materials, this problem may not be fatal since the
absorption and emission of a radiation field may be considered relatively
fast compared to electronic\textendash nuclear dynamics and other
relaxation processes.

Apart from any concerns about detail balance, Ehrenfest dynamics
has a second deficiency related to spontaneous and stimulated emission.\citep{li_mixed_2018}
Consider a situation where the electronic system has zero average
current initially and exists within a vacuum environment without external
fields; if the electronic state is excited, one expects spontaneous
emission to occur. However, according to Ehrenfest dynamics, the electron\textendash radiation
coupling will remain zero always, so that Ehrenfest dynamics will
not predict any spontaneous emission. In this paper, our goal is
to investigate the origins of this Ehrenfest failure by analyzing
the underlying mixed quantum\textendash classical theory; even more
importantly we will propose a new \emph{ad~hoc} algorithm for adding
spontaneous emission into an Ehrenfest framework.

This paper is organized as follows. In Sec.~\ref{sec:Review}, we
review the quantum electrodynamics (QED) theory of spontaneous emission.
In Sec.~\ref{sec:Ehrenfest-Dynamics}, we review Ehrenfest dynamics
as an ansatz for semiclassical QED and quantify the failure of the
Ehrenfest method to recover spontaneous emission. In Sec.~\ref{sec:Ehrenfest+R-method},
we propose a new Ehrenfest+R approach to correct some of the deficiencies
of the standard Ehrenfest approach. In Sec.~\ref{sec:Results}, we
present Ehrenfest+R results for spontaneous emission emanating from
a two-level system in 1D and 3D space. In Sec.~\ref{sec:Conclusions},
we discuss extensions of the proposed Ehrenfest+R approach, including
applications to energy transfer and Raman spectroscopy. 

Regarding notation, we use a bold symbol to denote a space vector
$\mathbf{r}=x\hat{\mathbf{x}}+y\hat{\mathbf{y}}+z\hat{\mathbf{z}}$
in Cartesian coordinate. Vector functions are denoted as $\mathbf{A}\left(\mathbf{r}\right)=A_{x}\left(\mathbf{r}\right)\hat{\mathbf{x}}+A_{y}\left(\mathbf{r}\right)\hat{\mathbf{y}}+A_{z}\left(\mathbf{r}\right)\hat{\mathbf{z}}$
and $\widehat{\mathbf{A}}$ denotes the corresponding quantum operator.
We use $\int dv=\int dxdydz$ for integration over 3D space.  We
work in SI units.

\section{Review of Quantum theory for spontaneous emission\label{sec:Review}}

Spontaneous emission is an irreversible process whereby a quantum
system makes a transition from an excited state to the ground state,
while simultaneously emitting a photon into the vacuum. The general
consensus is that spontaneous emission cannot fully be described by
any classical electromagnetic theory; almost by definition, a complete
description of spontaneous emission requires quantization of the photon
field. In this section, we review the Weisskopf\textendash Wigner
theory\citep{weisskopf_berechnung_1930,scully_quantum_1997} of spontaneous
emission, evaluating both the expectation value of the electric field
and the emission intensity.

\subsection{Power-Zienau-Woolley Hamiltonian\label{subsec:PZW_Hamiltonian}}

Before studying spontaneous emission in detail, one must choose a
Hamiltonian and a gauge for QED calculations. We will work with the
Power-Zienau-Woolley (PZW) Hamiltonian\citep{power_coulomb_1959,atkins_interaction_1970,cohen-tannoudji_photons_1997}
in the Coulomb gauge (so that $\mathbf{A}_{\parallel}=0$ and $\mathbf{A}=\mathbf{A}_{\perp}$)
because we believe this combination naturally offers a semiclassical
interpretation.\citep{cohen-tannoudji_photons_1997} Here, the total
Hamiltonian is:
\begin{equation}
\widehat{H}_{\text{PZW}}=\widehat{H}_{P}+\widehat{H}_{R}+\widehat{H}_{I},\label{eq:Power-Zienau-Woolley}
\end{equation}
where the particle Hamiltonian is 
\begin{equation}
\widehat{H}_{P}=\widehat{H}_{s}+\frac{1}{2\epsilon_{0}}\int\mathrm{d}v\left|\widehat{\mathbf{P}}_{\perp}\left(\mathbf{r}\right)\right|^{2},
\end{equation}
the transverse radiation field Hamiltonian is
\begin{equation}
\widehat{H}_{R}=\int\mathrm{d}v\left\{ \frac{1}{2\epsilon_{0}}\widehat{\mathbf{D}}_{\perp}\left(\mathbf{r}\right)^{2}+\frac{1}{2\mu_{0}}\left(\mathbf{\boldsymbol{\nabla}}\times\widehat{\mathbf{A}}\left(\mathbf{r}\right)\right)^{2}\right\} ,
\end{equation}
and the light-matter interaction is 
\begin{equation}
\widehat{H}_{I}=-\frac{1}{\epsilon_{0}}\int\mathrm{d}v\widehat{\mathbf{D}}_{\perp}\left(\mathbf{r}\right)\cdot\widehat{\mathbf{P}}_{\perp}\left(\mathbf{r}\right).\label{eq:PZW_int}
\end{equation}
Here $\widehat{\mathbf{A}}\left(\mathbf{r}\right)$ is the vector
potential of the EM field and $\widehat{\mathbf{D}}_{\perp}\left(\mathbf{r}\right)$
is the transverse field displacement. Note that the displacement $\widehat{\mathbf{D}}_{\perp}\left(\mathbf{r}\right)$
is the momentum conjugate to the vector potential $\widehat{\mathbf{A}}\left(\mathbf{r}\right)$,
satisfying the canonical commutation relation, $[\widehat{\mathbf{D}}_{\perp}\left(\mathbf{r}\right),\widehat{\mathbf{A}}\left(\mathbf{r}^{\prime}\right)]=i\hbar\delta^{\perp}\left(\mathbf{r}-\mathbf{r}^{\prime}\right)$.
We denote the polarization operator of the subsystem as $\widehat{\mathbf{P}}$
and use the Helmholtz decomposition expression ($\widehat{\mathbf{P}}=\widehat{\mathbf{P}}_{\perp}+\widehat{\mathbf{P}}_{\parallel}$)
to separate the the transverse polarization (satisfying $\boldsymbol{\nabla}\cdot\widehat{\mathbf{P}}_{\perp}=0$)
and the longitudinal polarization (satisfying $\boldsymbol{\nabla}\times\widehat{\mathbf{P}}_{\parallel}=0$).
$\widehat{H}_{s}$ is the Hamiltonian of the matter system and will
be specified below. Note that the Power-Zienau-Woolley Hamiltonian
is rigorously equivalent to the more standard Coulomb ($\widehat{\mathbf{P}}\cdot\widehat{\mathbf{A}}$)
representation of QED, but the matter field is now conveniently decomposed
into a multipolar form. That being said, in Eq.~(\ref{eq:Power-Zienau-Woolley})
we have ignored all magnetic couplings and an infinite Coulomb self
energy; we are also assuming we may ignore any relativistic dynamics
of the matter field.

For QED in the Coulomb gauge, we choose the vector potential and
the displacement following the standard canonical quantization approach:\citep{cohen-tannoudji_photons_1997}
\begin{equation}
\widehat{\mathbf{A}}\left(\mathbf{r}\right)=i\sum_{\mathrm{i}}\frac{{\cal E}_{\mathbf{\mathrm{i}}}}{\omega_{\mathbf{\mathrm{i}}}}\mathbf{s}_{\mathrm{i}}\left(\widehat{a}_{\mathrm{i}}e^{i\mathbf{k}_{\mathrm{i}}\cdot\mathbf{r}}+\widehat{a}_{\mathrm{i}}^{\dagger}e^{-i\mathbf{k}_{\mathrm{i}}\cdot\mathbf{r}}\right),\label{eq:A-operator}
\end{equation}
\begin{equation}
\widehat{\mathbf{D}}_{\perp}\left(\mathbf{r}\right)=i\epsilon_{0}\sum_{\mathrm{i}}{\cal E}_{\mathrm{i}}\mathbf{s}_{\mathrm{i}}\left(\widehat{a}_{\mathrm{i}}e^{i\mathbf{k}_{\mathrm{i}}\cdot\mathbf{r}}-\widehat{a}_{\mathrm{i}}^{\dagger}e^{-i\mathbf{k}_{\mathrm{i}}\cdot\mathbf{r}}\right).\label{eq:D_operator}
\end{equation}
Here, the matrix element ${\cal E}_{\mathrm{i}}=\sqrt{\frac{\hbar\omega_{\mathrm{i}}}{2\epsilon_{0}L^{n}}}$
is associated with the frequency $\omega_{\mathrm{i}}=c\left|\mathbf{k}_{\mathrm{i}}\right|$,
and $L^{n}$ is the volume of the $n$-dimensional space. $\mathbf{s}_{\mathrm{i}}$
is a unit vector of transverse polarization associated with the wave
vector $\mathbf{k}_{\mathrm{i}}$. $\widehat{a}_{\mathrm{i}}$ and
$\widehat{a}_{\mathbf{\mathrm{i}}}^{\dagger}$ are the destruction
and creation operators of the photon field where the index $\mathrm{i}$
designates the set $\left\{ \mathbf{k}_{\mathrm{i}},\mathbf{s}_{\mathrm{i}}\right\} $,
and satisfy the commutation relations: $\left[\widehat{a}_{\mathbf{\mathrm{i}}},\widehat{a}_{\mathrm{i}^{\prime}}^{\dagger}\right]=\delta\left(\mathbf{s}_{\mathrm{i}}-\mathbf{s}_{\mathrm{i}^{\prime}}\right)\delta\left(\mathbf{k}_{\mathrm{i}}-\mathbf{k}_{\mathrm{i}^{\prime}}\right)$.
In terms of $\widehat{a}_{\mathrm{i}}$ and $\widehat{a}_{\mathrm{i}}^{\dagger}$,
the transverse Hamiltonian of the EM field can be represented equivalently
as 
\begin{equation}
\widehat{H}_{R}=\sum\hbar\omega_{\mathrm{i}}\left(\widehat{a}_{\mathrm{i}}^{\dagger}\widehat{a}_{\mathrm{i}}+\frac{1}{2}\right).
\end{equation}
Note that $\widehat{\mathbf{A}}$ and $\widehat{\mathbf{D}}_{\perp}$
are pure EM field operators in the PZW representation.

Finally, within the Coulomb gauge, the electric and magnetic fields
can be obtained from the vector potential:
\begin{eqnarray}
\widehat{\mathbf{B}}\left(\mathbf{r}\right) & = & \mathbf{\boldsymbol{\nabla}}\times\widehat{\mathbf{A}}\left(\mathbf{r}\right),\\
\widehat{\mathbf{E}}_{\perp}\left(\mathbf{r}\right) & = & -\frac{\partial}{\partial t}\widehat{\mathbf{A}}\left(\mathbf{r}\right)=-\frac{i}{\hbar}\left[\widehat{H}_{R}+\widehat{H}_{I},\widehat{\mathbf{A}}\left(\mathbf{r}\right)\right],\label{eq:E=00003DdAdt}
\end{eqnarray}
recalling that $\mathbf{\boldsymbol{\nabla}}\cdot\widehat{\mathbf{A}}\left(\mathbf{r}\right)=0$
in the Coulomb gauge. The transverse electric field is related to
the displacement and the polarization by $\epsilon_{0}\widehat{\mathbf{E}}_{\perp}\left(\mathbf{r}\right)=\widehat{\mathbf{D}}_{\perp}\left(\mathbf{r}\right)-\widehat{\mathbf{P}}_{\perp}\left(\mathbf{r}\right)$.
Thus, these physical observables can also be expressed in terms of
$\widehat{a}_{\mathrm{i}}$ and $\widehat{a}_{\mathrm{i}}^{\dagger}$,
\begin{equation}
\widehat{\mathbf{B}}\left(\mathbf{r}\right)=i\sum_{\mathrm{i}}{\cal E}_{\mathrm{i}}\mathbf{k}_{\mathrm{i}}\times\mathbf{s}_{\mathrm{i}}\left(\widehat{a}_{\mathrm{i}}e^{i\mathbf{k}_{\mathrm{i}}\cdot\mathbf{r}}-\widehat{a}_{\mathrm{i}}^{\dagger}e^{-i\mathbf{k}_{\mathrm{i}}\cdot\mathbf{r}}\right),\label{eq:B_operator}
\end{equation}
\begin{equation}
\widehat{\mathbf{E}}_{\perp}\left(\mathbf{r}\right)=i\sum_{\mathrm{i}}{\cal E}_{\mathrm{i}}\mathbf{s}_{\mathrm{i}}\left(\widehat{a}_{\mathrm{i}}e^{i\mathbf{k}_{\mathrm{i}}\cdot\mathbf{r}}-\widehat{a}_{\mathrm{i}}^{\dagger}e^{-i\mathbf{k}_{\mathrm{i}}\cdot\mathbf{r}}\right)-\frac{1}{\epsilon_{0}}\widehat{\mathbf{P}}_{\perp}\left(\mathbf{r}\right).\label{eq:E_operator}
\end{equation}
Here, we note that $\widehat{\mathbf{E}}_{\perp}$ is \emph{not} a
pure EM field operator in the PZW representation. Instead, $\widehat{\mathbf{D}}_{\perp}\left(\mathbf{r}\right)$
is the pure EM field operator, satisfying Eq.~(\ref{eq:D_operator}),
as well as:
\begin{equation}
\widehat{\mathbf{D}}_{\perp}\left(\mathbf{r}\right)=-\epsilon_{0}\frac{\partial}{\partial t}\widehat{\mathbf{A}}\left(\mathbf{r}\right)+\widehat{\mathbf{P}}_{\perp}\left(\mathbf{r}\right).\label{eq:D-defination}
\end{equation}

Before proceeding, for readers more familiar with QED using the normal
coupling by $\widehat{\mathbf{P}}\cdot\widehat{\mathbf{A}}$ Hamiltonian,
a few more words are appropriate regarding Eqs.~(\ref{eq:D_operator}),
(\ref{eq:E=00003DdAdt}), (\ref{eq:E_operator}), and (\ref{eq:D-defination}).
Here, one may recall that, within the $\widehat{\mathbf{P}}\cdot\widehat{\mathbf{A}}$
Hamiltonian, the operator on the right hand side of Eq.~(\ref{eq:D_operator})
is associated with the transverse electric field $\epsilon_{0}\widehat{\mathbf{E}}_{\perp}$
(rather than $\widehat{\mathbf{D}}_{\perp}$).\citep{cohen-tannoudji_photons_1997}
With this apparent difference in mind, we stress that, when gaining
intuition for the PZW approach, one must never forget that the assignment
of mathematical operators for physical quantities can depend strongly
on the choice of representation and Hamiltonian. Luckily, for us in
many cases, one need not always distinguish between $\widehat{\mathbf{E}}_{\perp}$
and $\widehat{\mathbf{D}}_{\perp}$ because the transverse displacement
and electric field are the same up to a factor of $\epsilon_{0}$
($\epsilon_{0}\widehat{\mathbf{E}}_{\perp}=\widehat{\mathbf{D}}_{\perp}$)
in regions of space far away from the polarization of the subsystem
(where $\widehat{\mathbf{P}}_{\perp}\left(\mathbf{r}\right)=0$).

\subsection{Electric Dipole Hamiltonian}

In practice, for atomic problems, we often consider an electronic
system with a spatial distribution on the order of a Bohr radius interacting
with an EM field which has a wavelength much larger than the size
of the system. In this case, we can exploit the long-wavelength approximation
and recover the standard electric dipole Hamiltonian (i.e. a Göppert-Mayer
transformation\citep{cohen-tannoudji_photons_1997}):
\begin{equation}
\widehat{H}_{I}\approx-i\sum_{\mathrm{i}}{\cal E}_{\mathrm{i}}\widehat{\mathbf{d}}\cdot\mathbf{s}_{\mathrm{i}}\left(\widehat{a}_{\mathrm{i}}-\widehat{a}_{\mathrm{i}}^{\dagger}\right).\label{eq:electric-dipole-Hamiltonian}
\end{equation}
In this representation, the coupling between the atom and the photon
field is simple: one multiplies the dipole moment operator, $\mathbf{\widehat{\mathbf{d}}}=\sum_{\alpha}q_{\alpha}\mathbf{\widehat{\mathbf{r}}}_{\alpha}$,
by the electric field evaluated at the origin (where the atom is positioned).
This bi-linear electric dipole Hamiltonian is the usual starting point
for studying quantum optical effects, such as spontaneous emission.

\subsection{Quantum Theory of Spontaneous Emission}

For a quantum electrodynamics description of spontaneous emission,
we may consider a simple two-level system 
\begin{equation}
\widehat{H}_{s}=\varepsilon_{0}\left|0\right\rangle \left\langle 0\right|+\varepsilon_{1}\left|1\right\rangle \left\langle 1\right|\label{eq:H_s}
\end{equation}
which is coupled to the photon field. We assume $\varepsilon_{0}<\varepsilon_{1}$
and $\varepsilon_{1}-\varepsilon_{0}=\hbar\Omega$. The electronic
dipole moment operator takes the form of 
\begin{equation}
\widehat{\mathbf{d}}=\boldsymbol{\mu}_{01}\left(\left|0\right\rangle \left\langle 1\right|+\left|1\right\rangle \left\langle 0\right|\right),
\end{equation}
where $\boldsymbol{\mu}_{01}=\left\langle 0\right|\sum_{\alpha}q_{\alpha}\mathbf{\widehat{\mathbf{r}}}_{\alpha}\left|1\right\rangle $
is the transition dipole moment of the two states. Using Eq.~(\ref{eq:electric-dipole-Hamiltonian}),
with a dipolar approximation, the coupling between the two level system
and the photon field can be expressed as 
\begin{equation}
\widehat{H}_{I}=\sum_{\mathrm{i}}V_{\mathrm{i}}\left(\widehat{a}_{\mathrm{i}}-\widehat{a}_{\mathrm{i}}^{\dagger}\right)\left(\left|0\right\rangle \left\langle 1\right|+\left|1\right\rangle \left\langle 0\right|\right)\label{eq:ED-interaction}
\end{equation}
where the matrix element is given by $V_{\mathrm{i}}=i{\cal E}_{\mathrm{i}}\boldsymbol{\mu}_{01}\cdot\mathbf{s}_{\mathrm{i}}$.
Let us assume that the initial wavefunction for the two-level system
is $\left|\psi\left(0\right)\right\rangle =C_{0}\left|0\right\rangle +C_{1}\left|1\right\rangle $
and the reduced density matrix element is $\rho_{ij}\left(0\right)=C_{i}C_{j}^{*}$.

Based on the generalization of Weisskopf\textendash Wigner theory
(see Appendix~\ref{sec:Weisskopf=002013Wigner}), we can write down
the excited state population as 
\begin{equation}
\rho_{11}\left(t\right)=\rho_{11}\left(0\right)e^{-\kappa t},\label{eq:WW-population}
\end{equation}
assuming that $\kappa\ll\Omega/2\pi$. The coherence of the reduced
density matrix satisfies
\begin{equation}
\left|\rho_{01}\left(t\right)\right|=\left|\rho_{01}\left(0\right)\right|e^{-\frac{\kappa}{2}t}.\label{eq:WW-coherence}
\end{equation}
and the ``impurity'' of the reduced density matrix is
\begin{equation}
\begin{split}1-\eta\left(t\right) & =\mathrm{Tr}_{s}\left\{ \widehat{\rho}\left(t\right)-\widehat{\rho}^{2}\left(t\right)\right\} \\
 & =2\left|\rho_{11}\left(0\right)\right|^{2}\left(e^{-\kappa t}-e^{-2\kappa t}\right).
\end{split}
\label{eq:WW-impurity}
\end{equation}
Eq.~(\ref{eq:WW-impurity}) gives a measure of how much the matter
system appears mixed as a result of interacting with the EM environment.

The decay rate for a three-dimensional system is given by the Fermi's
golden rule (FGR) rate\citep{nitzan_chemical_2006} 
\begin{equation}
\kappa^{\text{3D}}=\frac{\left|\boldsymbol{\mu}_{01}\right|^{2}\Omega^{3}}{3\pi\hbar\epsilon_{0}c^{3}}.\label{eq:FGR_rate_3D}
\end{equation}
Similarly, for an effectively one-dimensional system, we imagine
a uniform charge distributions in the $yz$ plane and a delta function
in the $x$ direction. The effective dipole moment in 1D is defined
as $\mu_{01}^{2}=\left|\boldsymbol{\mu}_{01}\right|^{2}/L_{y}L_{z}$.
The decay rate for this effectively 1D case is 
\begin{equation}
\kappa^{\text{1D}}=\frac{\mu_{01}^{2}\Omega}{\hbar\epsilon_{0}c}.\label{eq:FGR_rate_1D}
\end{equation}
Eqs.~(\ref{eq:FGR_rate_3D}) and (\ref{eq:FGR_rate_1D}) are proven
in Ref~\onlinecite{li_mixed_2018}, as well as in Appendix~\ref{sec:Weisskopf=002013Wigner}.
Below, we will use $\kappa$ to represent the FGR rate for either
$\kappa^{\text{3D}}$ or $\kappa^{\text{1D}}$ depending on context.
Note that, in general, Fermi's golden rule is valid in the weak coupling
limit ($\kappa\ll\Omega$), which is also called the FGR regime.

We assume that the initial condition of the photon field is a vacuum,
i.e. there are no photons at $t=0$. For a given initial state of
the matter, $\left|\psi\left(0\right)\right\rangle =C_{0}\left|0\right\rangle +C_{1}\left|1\right\rangle $,
the expectation value of the observed electric field for an effectively
1D system is given by
\begin{equation}
\left\langle \mathbf{E}_{\perp}\left(x,t\right)\right\rangle =\left|C_{0}\right|\left|C_{1}\right|\times R\left(x,t\right)\sin\Omega\left(t-\left|x\right|/c\right)\label{eq:E-field-WWtheroy}
\end{equation}
where 
\begin{equation}
R\left(x,t\right)=\frac{\Omega\mu_{01}}{c\epsilon_{0}}e^{-\frac{\kappa}{2}\left(t-\frac{\left|x\right|}{c}\right)}\times\theta\left(ct-\left|x\right|\right)
\end{equation}
Note that $R\left(x,t\right)$ contains an event horizon ($\left|x\right|<ct$)
for the emitting radiation. The observed electric field represents
the \emph{coherent emission }at the frequency $\Omega$. In a coarse-grained
sense, since $\overline{\sin^{2}\Omega t}\approx\frac{1}{2}$, the
coherent emission has a magnitude given by 
\begin{equation}
\overline{\left\langle \mathbf{E}_{\perp}\left(x,t\right)\right\rangle ^{2}}=\left|C_{0}\right|^{2}\left|C_{1}\right|^{2}\times\frac{R\left(x,t\right)^{2}}{2}.\label{eq:E-field2-WWtheroy}
\end{equation}
We note that the coherent emission depends on the initial population
of the ground state $\left|C_{0}\right|^{2}$. 

The expectation value of the intensity distribution can be obtained
as 
\begin{equation}
\overline{\left\langle \mathbf{E}_{\perp}^{2}\left(x,t\right)\right\rangle }=\left|C_{1}\right|^{2}\times\frac{R\left(x,t\right)^{2}}{2},\label{eq:intensity-WWtheroy}
\end{equation}
which conserves the energy of the total system. Note that the variance
of the observed electric field (i.e. the fact that $\overline{\left\langle \mathbf{E}_{\perp}^{2}\right\rangle }\neq\overline{\left\langle \mathbf{E}_{\perp}\right\rangle ^{2}}$)
reflects a quantum mechanical feature of spontaneous emission. For
proofs of Eqs.~(\ref{eq:E-field-WWtheroy}\textendash \ref{eq:intensity-WWtheroy}),
see Appendix~\ref{sec:Weisskopf=002013Wigner}.

\section{Ehrenfest Dynamics as ansatz for quantum electrodynamics\label{sec:Ehrenfest-Dynamics}}

Ehrenfest dynamics provides a semiclassical ansatz for modeling QED
based on a mean-field approximation together with a classical EM field
and quantum matter field.\citep{li_mixed_2018} In general, a mean-field
approximation should be valid when there are no strong correlations
among different subsystems. In this section, we review the Ehrenfest
approach for treating coupled electron\textendash radiation dynamics,
specifically spontaneous emission. 

\subsection{Ehrenfest dynamics }

Within Ehrenfest dynamics, the electronic system is described by
the electronic reduced density matrix $\widehat{\rho}\left(t\right)$
while the EM fields, $\mathbf{E}\left(\mathbf{r},t\right)$ and $\mathbf{B}\left(\mathbf{r},t\right)$,
are classical. As far as dynamics are concerned, the electronic density
matrix evolves according to the Liouville equation,
\begin{equation}
\frac{\partial}{\partial t}\widehat{\rho}\left(t\right)=-\frac{i}{\hbar}\left[\widehat{H}^{\mathrm{el}},\widehat{\rho}\left(t\right)\right],\label{eq:liouville_rho}
\end{equation}
where $\widehat{H}^{\mathrm{el}}=\widehat{H}^{\mathrm{el}}\left(\mathbf{E},\mathbf{B}\right)$
is a semiclassical Hamiltonian for the quantum subsystem which depends
only parametrically on the EM fields. This semiclassical electronic
Hamiltonian $\widehat{H}^{\mathrm{el}}$ in Eq.~(\ref{eq:liouville_rho})
must approximate $\widehat{H}_{P}+\widehat{H}_{I}$ in Eq.~(\ref{eq:Power-Zienau-Woolley}),
and according to Ehrenfest dynamics, we choose\citep{mukamel_principles_1999}
\begin{equation}
\widehat{H}^{\mathrm{el}}=\widehat{H}_{s}-\int\mathrm{d}v\mathbf{E}_{\perp}\left(\mathbf{r},t\right)\cdot\widehat{\mathbf{P}}\left(\mathbf{r}\right).\label{eq:electronic_Hamiltonian}
\end{equation}

For the EM fields, dynamics are governed by Maxwell's equations 
\begin{eqnarray}
\frac{\partial}{\partial t}\mathbf{B}\left(\mathbf{r},t\right) & = & -\boldsymbol{\nabla}\times\mathbf{E}\left(\mathbf{r},t\right),\label{eq:maxwell_BE}\\
\frac{\partial}{\partial t}\mathbf{E}\left(\mathbf{r},t\right) & = & c^{2}\boldsymbol{\nabla}\times\mathbf{B}\left(\mathbf{r},t\right)-\frac{1}{\epsilon_{0}}\mathbf{J}\left(\mathbf{r},t\right),\label{eq:maxwell_EB}
\end{eqnarray}
where the average current is generated by the average polarization
of the electronic system
\begin{equation}
\mathbf{J}\left(\mathbf{r},t\right)=\frac{\partial}{\partial t}\text{Tr}_{s}\left\{ \widehat{\rho}\left(t\right)\widehat{\mathbf{P}}\left(\mathbf{r}\right)\right\} \equiv\frac{\partial}{\partial t}\mathbf{P}\left(\mathbf{r},t\right).\label{eq:polarization_mean}
\end{equation}
Here we define the average polarization (without hat) $\mathbf{P}\left(\mathbf{r},t\right)=\text{Tr}_{s}\left\{ \widehat{\rho}\left(t\right)\widehat{\mathbf{P}}\left(\mathbf{r}\right)\right\} $.
Note that Eq.~(\ref{eq:maxwell_EB}) suggests that the longitudinal
component of the classical electric field is 
\begin{equation}
\mathbf{E}_{\parallel}\left(\mathbf{r},t\right)=-\frac{1}{\epsilon_{0}}\mathbf{P}_{\parallel}\left(\mathbf{r},t\right),\label{eq:maxwell_EB_para}
\end{equation}
and the transverse component satisfies
\begin{equation}
\frac{\partial}{\partial t}\mathbf{E}_{\perp}\left(\mathbf{r},t\right)=c^{2}\boldsymbol{\nabla}\times\mathbf{B}\left(\mathbf{r},t\right)-\frac{1}{\epsilon_{0}}\mathbf{J}_{\perp}\left(\mathbf{r},t\right)\label{eq:maxwell_EB_perp}
\end{equation}
with $\mathbf{J}_{\perp}\left(\mathbf{r},t\right)=\frac{\partial}{\partial t}\mathbf{P}_{\perp}\left(\mathbf{r},t\right)$.

The total energy of the electronic system and the classical EM field
is 
\begin{equation}
\begin{split}U_{\mathrm{tot}}\left(\widehat{\rho},\mathbf{E},\mathbf{B}\right) & =\mathrm{Tr}_{s}\left(\widehat{\rho}\left(t\right)\widehat{H}_{s}\right)+\\
 & \int\mathrm{d}v\left(\frac{\epsilon_{0}}{2}\mathbf{E}_{\perp}\left(\mathbf{r},t\right)^{2}+\frac{1}{2\mu_{0}}\mathbf{B}\left(\mathbf{r},t\right)^{2}\right).
\end{split}
\label{eq:total_energy}
\end{equation}
In Eq.~(\ref{eq:total_energy}), we have replaced all quantum mechanical
operators for the EM field by their classical expectation values,
i.e. $\mathbf{\nabla}\times\widehat{\mathbf{A}}\rightarrow\mathbf{B}$
and $\widehat{\mathbf{D}}\rightarrow\mathbf{D}_{\perp}=\epsilon_{0}\mathbf{E}_{\perp}+\mathbf{P}_{\perp}$,
where $\mathbf{P}_{\perp}=\text{Tr}_{s}\left\{ \widehat{\rho}\widehat{\mathbf{P}}_{\perp}\right\} $.
One of the most important strengths of Ehrenfest dynamics is that
the total energy ($U_{\mathrm{tot}}$) is conserved (as can be shown
easily). Altogether, Ehrenfest dynamics is a self-consistent, computationally
inexpensive approach for propagating the electronic states and EM
field dynamics simultaneously.

As a sidenote, we mention that, in Eqs.~(\ref{eq:Power-Zienau-Woolley}\textendash \ref{eq:PZW_int}),
we have neglected a formally infinite self-interaction energy. If
we include such a term, we can argue that, for a single charge center,
one can write a slightly different electronic Hamiltonian (instead
of Eq.~(\ref{eq:electronic_Hamiltonian})) namely\citep{sukharev_numerical_2011}\footnote{In QED, the Coulomb interaction between particles $\alpha$ and $\beta$
can be expressed as\citep{cohen-tannoudji_photons_1997}
\[
\hat{V}_{\mathrm{Coul}}=\frac{1}{\epsilon_{0}}\int\mathrm{d}v\hat{\mathbf{P}}_{\parallel}^{\left(\alpha\right)}\left(\mathbf{r}\right)\cdot\hat{\mathbf{P}}_{\parallel}^{\left(\beta\right)}\left(\mathbf{r}\right).
\]
Consider a quantum subsystem composed of a single electron within
a semiclassical approximation. The Coulomb self-interaction energy
in Eq.~(\ref{eq:Power-Zienau-Woolley}) is
\[
\hat{V}_{\mathrm{self}}=\frac{1}{\epsilon_{0}}\int\mathrm{d}v\mathbf{P}_{\parallel}\left(\mathbf{r},t\right)\cdot\mathbf{\hat{P}}_{\parallel}\left(\mathbf{r}\right).
\]
If we add this term to the Hamiltonian in Eq.~(\ref{eq:electronic_Hamiltonian})
and substitute $\mathbf{E}_{\perp}=\mathbf{E}+\frac{1}{\epsilon_{0}}\mathbf{P}_{\parallel}$,
we find that the Coulomb self energy is canceled, yielding Eq.~(\ref{eq:electric_Hamiltonian_withself})
\[
\hat{H}^{\mathrm{el}}+\hat{V}_{\mathrm{self}}=\hat{H}_{s}-\int\mathrm{d}v\mathbf{E}\left(\mathbf{r},t\right)\cdot\mathbf{\hat{P}}\left(\mathbf{r}\right).
\]
For dynamics propagated with the semiclassical electronic Hamiltonian
in Eq.~(\ref{eq:electric_Hamiltonian_withself}), the conserved energy
becomes
\[
\begin{split}U_{\mathrm{tot}} & =\mathrm{Tr}_{s}\left(\hat{\rho}\left(t\right)H_{s}\right)+\\
 & \int\mathrm{d}v\left(\frac{\epsilon_{0}}{2}\mathbf{E}\left(\mathbf{r},t\right)^{2}+\frac{1}{2\mu_{0}}\mathbf{B}\left(\mathbf{r},t\right)^{2}\right)
\end{split}
\]
}
\begin{equation}
\widehat{H}^{\mathrm{el}}=\widehat{H}_{s}-\int\mathrm{d}v\mathbf{E}\left(\mathbf{r},t\right)\cdot\widehat{\mathbf{P}}\left(\mathbf{r}\right).\label{eq:electric_Hamiltonian_withself}
\end{equation}
All numerical results presented below are nearly identical using either
Eq.~(\ref{eq:electronic_Hamiltonian}) or Eq.~(\ref{eq:electric_Hamiltonian_withself})
for a semiclassical Hamiltonian.

\subsection{Drawbacks of Ehrenfest Dynamics: Spontaneous Emission}

For the purposes of this paper, it will now be fruitful to discuss
spontaneous emission in more detail within the context of Ehrenfest
dynamics. In the FGR regime, if we approximate the transition dipole
moment of the two level system to be a delta function at the origin
and consider again the case of no electric field at time zero, we
can show that the electric dipole coupling within Ehrenfest dynamics
satisfies the relationship
\begin{equation}
H_{01}^{\mathrm{el}}=-\hbar\kappa\mathrm{Im}\rho_{01}\label{eq:coupling_ehr}
\end{equation}
for both 1D and 3D systems. For a 1D system, this relation was derived
previously in Ref.~\onlinecite{li_mixed_2018}. For a 3D system,
this relation can be derived using Jefimenko's equation for classical
electrodynamics with a current source given by Eq.~(\ref{eq:polarization_mean})
(see Appendix~\ref{sec:Instantaneous-Decay-rate}). 

With Eq.~(\ref{eq:coupling_ehr}), we can convert the Liouville
equation (Eq.~(\ref{eq:liouville_rho})) for Ehrenfest dynamics into
a set of self-consistent, non-linear equations of motion for the electronic
subsystem. To be precise, let $\widehat{H}^{\mathrm{el}}=\left(\begin{array}{cc}
0 & H_{01}^{\mathrm{el}}\\
H_{10}^{\mathrm{el}} & \Omega
\end{array}\right)$ and substitute Eq.~(\ref{eq:coupling_ehr}) for $H_{01}^{\mathrm{el}}=H_{10}^{\mathrm{el}}$.
Now, the commutator in Eq.~(\ref{eq:liouville_rho}) yields:
\begin{eqnarray}
\frac{\partial\rho_{11}}{\partial t} & = & -2\kappa\left(\mathrm{Im}\rho_{01}\right)^{2},\label{eq:drho_11dt}\\
\frac{\partial\rho_{01}}{\partial t} & = & i\Omega\rho_{01}+i\kappa\mathrm{Im}\rho_{01}\left(\rho_{11}-\rho_{00}\right).\label{eq:drho_01dt}
\end{eqnarray}
In the FGR regime, because $\kappa\ll\Omega$, we can approximate
the coherence $\rho_{01}\approx\left|\rho_{01}\right|e^{i\Omega t}$
for a time $\tau$ satisfying $2\pi/\Omega\ll\tau\ll1/\kappa$ so
that $\left(\mathrm{Im}\rho_{01}\right)^{2}\approx\left|\rho_{01}\right|^{2}\sin^{2}\Omega t$.
We may then define an instantaneous decay rate $k_{Eh}\left(t\right)$
for $\rho_{11}$, satisfying $\frac{\partial}{\partial t}\rho_{11}=-k_{Eh}\left(t\right)\rho_{11}$,
where 
\begin{equation}
k_{Eh}\left(t\right)=2\kappa\frac{\left|\rho_{01}\right|^{2}}{\rho_{11}}\sin^{2}\Omega t,\label{eq:Ehrenfest_rate-3D}
\end{equation}
so long as $\rho_{11}\neq0$. (Note that $k_{Eh}=0$ if $\rho_{11}=0$.)
Note also that $\rho_{11}$ does not change much within the time scale
$\tau$. To monitor the population decay in a coarse-grained sense,
we can perform a moving average over $\tau$ and denote the average
decay rate as
\begin{equation}
\overline{k_{Eh}}\left(t\right)=\frac{1}{\tau}\int_{t}^{t+\tau}dt^{\prime}k_{Eh}\left(t^{\prime}\right)=\kappa\frac{\left|\rho_{01}\right|^{2}}{\rho_{11}};\label{eq:tau_averaged-3D}
\end{equation}
here we have used $\overline{\sin^{2}\Omega t}\approx\frac{1}{2}$. 

This analysis quantifies Ehrenfest's failure to capture spontaneous
emission: Eq.~(\ref{eq:tau_averaged-3D}) demonstrates that Ehrenfest
dynamics yields a non-exponential decay and, when $\rho_{00}=0$,
Ehrenfest dynamics does not predict any spontaneous emission. Interestingly,
the Ehrenfest decay rate ends up being the correct spontaneous emission
rate multiplied by the lower state population at time $t$. 

Now we turn our attention to the coherence of the density matrix
$\left|\rho_{01}\right|$. From Eq.~(\ref{eq:drho_01dt}), we can
evaluate the change of the coherence:
\begin{equation}
\frac{\partial}{\partial t}\left|\rho_{01}\right|^{2}=-2\kappa\left(\mathrm{Im}\rho_{01}\right)^{2}\left(\rho_{00}-\rho_{11}\right).
\end{equation}
In analogy to our approach above for FGR dynamics, we can define an
instantaneous ``dephasing'' rate, $\gamma_{Eh}\left(t\right)$,
for $\left|\rho_{01}\right|$, satisfying $\frac{\partial}{\partial t}\left|\rho_{01}\right|=-\gamma_{Eh}\left(t\right)\left|\rho_{01}\right|$,
where
\begin{equation}
\gamma{}_{Eh}\left(t\right)=\kappa\left(\rho_{00}-\rho_{11}\right)\sin^{2}\Omega t,
\end{equation}
so long as $\rho_{01}=0$. (Note that $\gamma_{Eh}=0$ if $\rho_{01}=0$.)
We can now perform a moving average over $\tau$ and denote the average
rate in a coarse-grained sense:
\begin{equation}
\overline{\gamma_{Eh}}\left(t\right)=\frac{1}{\tau}\int_{t}^{t+\tau}dt^{\prime}\gamma_{Eh}\left(t\right)=\frac{\kappa}{2}\left(\rho_{00}-\rho_{11}\right).\label{eq:average_gamma}
\end{equation}
Apparently, the average dephasing rate (Eq.~(\ref{eq:average_gamma}))
is proportional to the instantaneous population difference $\left(\rho_{00}-\rho_{11}\right)$
of the system. Note that this Ehrenfest ``dephasing'' rate can be
negative, such that the value of $\left|\rho_{01}\right|$ can grow
exponentially with time. This analysis leads to another drawback of
Ehrenfest dynamics: for the case of an isolated two-level system interacting
with a vacuum EM field, when $\rho_{00}<\rho_{11}$, there is an unphyscial
increase of the coherence ($\left|\rho_{01}\right|$) with respect
to time. This increase does not agree with Eq.~(\ref{eq:WW-coherence}).

Regarding the purity of the reduced density matrix, one can easily
show that the purity is conserved within Ehrenfest dynamics, i.e.
\begin{equation}
\frac{\partial}{\partial t}\mathrm{Tr}\left\{ \rho^{2}\right\} =0.
\end{equation}
If we consider a system initialized to be in a pure state, the density
matrix will stay as a pure state within Ehrenfest dynamics, i.e.$\left|\rho_{01}\right|^{2}=\rho_{00}\rho_{11}$,
and we find Eq.~(\ref{eq:tau_averaged-3D}) can be written as
\begin{equation}
\overline{k_{Eh}}\left(t\right)=\kappa\rho_{00}.\label{eq:kEh_pure}
\end{equation}
This Ehrenfest purity conservation does not agree with Eq.~(\ref{eq:WW-impurity}).

\section{Ehrenfest+R Method\label{sec:Ehrenfest+R-method}}

Given the failure of Ehrenfest dynamics to capture spontaneous emission
fully as described above, we now propose an \emph{ad~hoc} \emph{Ehrenfest+R}
method for ensuring that the dynamics of quantum subsystem in vacuum
do agree with FGR decay. Our approach is straightforward: we will
enforce an additional relaxation pathway on top of Ehrenfest dynamics
such that the total Ehrenfest+R emission should agree with the true
spontaneous decay rate. We will benchmark this Ehrenfest+R approach
in the context of a two-level system in 1D or 3D space. Note that
the classical radiation field is at zero temperature, so we may exclude
all thermal transitions from $\left|0\right\rangle $ to $\left|1\right\rangle $.
We begin by motivating our choice of an \emph{ad~hoc} algorithm.
In Sec.~\ref{subsec:Step-by-step-Algorithm-of}, we provide a step-by-step
outline so that the reader can easily reproduce our algorithm and
data.

\subsection{The Quantum Subsystem\label{subsec:The-Quantum-Subsystem}}

\subsubsection{Liouville equation}

As far as the quantum subsystem is concerned, in order to recover
the FGR rate of the population in the excited state and the correct
dephasing rate, we will include an additional relaxation (``+R'')
term on top of the Liouville equation,

\begin{equation}
\frac{\partial\widehat{\rho}}{\partial t}=\doublehat{{\cal L}_{\hspace{0.1pt}}}_{Eh}\widehat{\rho}+\doublehat{{\cal L}_{\hspace{0.1pt}}}_{R}\widehat{\rho},\label{eq:Liouville+R}
\end{equation}
where the super-operator 
\begin{equation}
\doublehat{{\cal L}_{\hspace{0.1pt}}}_{Eh}\widehat{\rho}=-\frac{i}{\hbar}\left[\widehat{H}^{\mathrm{el}},\widehat{\rho}\right]
\end{equation}
accounts for Ehrenfest dynamics (Eq.~(\ref{eq:liouville_rho})) and
the super-operator $\doublehat{{\cal L}_{\hspace{0.1pt}}}_{R}$ enforces
relaxation. For a relaxation pathway from state $a$ to state $b$,
the super-operator affects only $\rho_{ij}$ for $i,j\in\left\{ a,b\right\} $.
We choose the diagonal elements of the super-operator to be 
\begin{equation}
\left[\doublehat{{\cal L}_{\hspace{0.1pt}}}_{R}\widehat{\rho}\right]_{aa}=-\left[\doublehat{{\cal L}_{\hspace{0.1pt}}}_{R}\widehat{\rho}\right]_{bb}=-k_{R}\rho_{aa},\label{eq:Liouville+R_diagonal}
\end{equation}
and the the off-diagonal elements to be
\begin{equation}
\left[\doublehat{{\cal L}_{\hspace{0.1pt}}}_{R}\widehat{\rho}\right]_{ab}=\left[\doublehat{{\cal L}_{\hspace{0.1pt}}}_{R}\widehat{\rho}\right]_{ba}^{*}=-\gamma_{R}\rho_{ab}.\label{eq:Liouville+R_offdiagonal}
\end{equation}
Specifically, for a two level system, the super-operator can be written
as
\begin{equation}
\doublehat{{\cal L}_{\hspace{0.1pt}}}_{R}\widehat{\rho}=\left(\begin{array}{cc}
+k_{R}\rho_{11} & -\gamma_{R}\rho_{01}\\
-\gamma_{R}\rho_{10} & -k_{R}\rho_{11}
\end{array}\right)\label{eq:Liouville+R_matrix}
\end{equation}

The +R relaxation rate in Eq.~(\ref{eq:Liouville+R_matrix}) is chosen
as
\begin{equation}
k_{R}\equiv2\kappa\left(1-\frac{\left|\rho_{01}\right|^{2}}{\rho_{11}}\right)\text{Im}\left[\frac{\rho_{01}}{\left|\rho_{01}\right|}e^{i\phi}\right]^{2},\label{eq:k_R}
\end{equation}
where $\kappa$ is the FGR rate ($k_{R}=0$ if $\rho_{11}=0$). Eq.~(\ref{eq:k_R})
is similar to Eq.~(\ref{eq:Ehrenfest_rate-3D}) but with an arbitrary
phase $\phi\in\left(0,2\pi\right)$. Averaging over a time scale $\tau$
(defined in Eq.~(\ref{eq:tau_averaged-3D})), we find
\begin{equation}
\overline{k_{R}}=\kappa\left(1-\frac{\left|\rho_{01}\right|^{2}}{\rho_{11}}\right).
\end{equation}
Thus, the average total population decay rate predicted by Eq.~(\ref{eq:Liouville+R})
is 

\begin{equation}
\kappa=\overline{k_{Eh}}+\overline{k_{R}}.
\end{equation}
In other words, Eqs.~(\ref{eq:Liouville+R}\textendash \ref{eq:k_R})
should recover the true FGR rate of the excited state decay by correcting
Ehrenfest dynamics. 

The +R dephasing rate $\gamma_{R}$ in Eq.~(\ref{eq:Liouville+R_matrix})
is chosen to be
\begin{equation}
\gamma_{R}\equiv\frac{\kappa}{2}\left(1-\rho_{00}+\rho_{11}\right)\label{eq:gamma_R}
\end{equation}
Together with the dephasing rate of Ehrenfest dynamics $\overline{\gamma_{Eh}}$
given in Eq.~(\ref{eq:average_gamma}), the total dephasing rate
of Eq.~(\ref{eq:Liouville+R}) is 
\begin{equation}
\frac{\kappa}{2}=\overline{\gamma_{Eh}}+\gamma_{R}.
\end{equation}
Note that $\gamma_{R}$ is always positive. The additional dephasing
should eliminate the unphysical increase of $\left|\rho_{01}\right|$
within Ehrenfest dynamics and recover the correct result for spontaneous
emission.

The phase $\phi$ in Eq.~(\ref{eq:k_R}) can be chosen arbitrarily
without affecting the total decay rate in a coarse-grained sense (i.e.
if we perform a moving average over $\tau$). In what follows, we
will run multiple trajectories (indexed by $\ell\in N_{\text{traj }}$)
with $\phi^{\ell}$ chosen randomly. The choice of a random $\phi^{\ell}$
allows us effectively to introduce decoherence within the EM field,
so that we may represent the time/phase uncertainty of the emitted
light as an ensemble of classical fields. Each individual trajectory
still carries a pure electronic wavefunction. Note that a random phase
does not affect the FGR decay rate of the quantum subsystem.

Before finishing up this subsection, a few words are now appropriate
about how Ehrenfest+R dynamics are different from the more standard
Maxwell\textendash Bloch equations, whereby one introduces phenomenological
damping of the electronic density matrix. (Indeed, this will be a
topic of future discussion for another paper\citep{li_ehrenfest+r_2018}).
Within such a comparison, we note that, when solving the Maxwell\textendash Bloch
equations for the electronic subsystem, one must take great care to
separate the effects of incoming EM fields from the effect of self-interaction.
Such a separation is required to avoid double counting of all electronic
relaxation, and several techniques have been proposed over the years.\citep{neuhauser_molecular_2007,lopata_nonlinear_2009,lopata_multiscale_2009}
Furthermore, once such a separation has been achieved, one must construct
a robust algorithm to transfer all energy lost by electronic relaxation
into energy of the EM field. By contrast, for the case of Ehrenfest+R
dyanmics, we do not require any separation between incoming EM and
self-interaction EM fields, and we avoid double counting by insisting
that the +R relaxation rate must itself depend on the population on
the upper state\textemdash though this leads to nonlinear matrix elements;
see Eqs.~(\ref{eq:k_R}) and (\ref{eq:gamma_R}). Energy conservation
can be achieved by properly rescaling the EM fields. 

In the end, in seeking to capture light-matter interactions and fluorescence
correctly, the Ehrenfest+R approach eliminates one problem (the separation
of self-interacting fields) but creates another problem (solving nonlinear
Schrodinger equations). Now, from our perspective, given the subtle
problems that inevitably arise with any quantum-classical algorithm,\citep{deinega_self-interaction-free_2014}
the usefulness of a semiclassical electrodynamics approach (including
Ehrenfest+R dynamics) can only be assessed by rigorously benchmarking
the algorithm over a host of different model problems. And so, in
the present paper (Paper I) and the following paper (Paper II\citep{chen_ehrenfest+r_2018-2}),
we will perform such benchmarks. Furthermore, in a companion paper,
we will make direct comparisons to more standard Maxwell-Bloch approaches
(where we also discuss energy conservation at length).

\subsubsection{Practical Implementation}

Formally, for an infinitesimal time step $dt$, the electronic density
matrix can be evolved with a two-step propagation scheme:
\begin{equation}
\widehat{\rho}\left(t+dt\right)=e^{\widehat{\widehat{{\cal L}_{\hspace{0.1pt}}}}_{R}dt}e^{\widehat{\widehat{{\cal L}_{\hspace{0.1pt}}}}_{Eh}dt}\widehat{\rho}\left(t\right).
\end{equation}
Here, the propagator
\begin{equation}
e^{\widehat{\widehat{{\cal L}_{\hspace{0.1pt}}}}_{Eh}dt}\widehat{\rho}\equiv e^{i\widehat{H}^{\mathrm{el}}dt/\hbar}\widehat{\rho}e^{-i\widehat{H}^{\mathrm{el}}dt/\hbar}
\end{equation}
carries out standard propagation of the Liouville equation with the
electronic Hamiltonian given by Eq.~(\ref{eq:electronic_Hamiltonian}).
The propagator
\begin{equation}
e^{\widehat{\widehat{{\cal L}_{\hspace{0.1pt}}}}_{R}dt}\widehat{\rho}\equiv\left(\begin{array}{cc}
1-e^{-k_{R}dt}\rho_{11} & e^{-\gamma_{R}dt}\rho_{01}\\
e^{-\gamma_{R}dt}\rho_{10} & e^{-k_{R}dt}\rho_{11}
\end{array}\right)
\end{equation}
 implements the additional +R relaxation from Eqs.~(\ref{eq:Liouville+R_matrix})
with a population relaxation rate $k_{R}$ given by Eq.~(\ref{eq:k_R})
and a dephasing rate $\gamma_{R}$ given by Eq.~(\ref{eq:gamma_R}). 

In practice, we will work below with the wavefunction $\left|\psi\right\rangle $,
rather than the density matrix $\widehat{\rho}=\left|\psi\right\rangle \left\langle \psi\right|$.
For each time step $dt$, the wavefunction is evolved with a two-step
propagation scheme:
\begin{equation}
\left|\psi\left(t+dt\right)\right\rangle =e^{i\widehat{\Phi}\left[\gamma_{R}\right]}\widehat{{\cal T}}_{0\leftarrow1}\left[k_{R}\right]\cdot e^{-i\widehat{H}^{\mathrm{el}}dt/\hbar}\left|\psi\left(t\right)\right\rangle .\label{eq:EhrenfestPropagation}
\end{equation}
The operator $e^{-i\widehat{H}^{\mathrm{el}}dt/\hbar}$ carries out
standard propagation of the Schr\"odinger equation with the electronic
Hamiltonian given by Eq.~(\ref{eq:electronic_Hamiltonian}). The
quantum transition operator $\widehat{{\cal T}}_{0\leftarrow1}\left[k_{R}\right]$
implements the additional +R population relaxation from Eqs.~(\ref{eq:Liouville+R_matrix}),
(\ref{eq:k_R}) and (\ref{eq:gamma_R}). Explicitly, the transition
operator is defined by 
\begin{equation}
\left(\begin{array}{c}
c_{0}^{\prime}\\
c_{1}^{\prime}
\end{array}\right)=\widehat{{\cal T}}_{0\leftarrow1}\left[k_{R}\right]\left(\begin{array}{c}
c_{0}\\
c_{1}
\end{array}\right)\label{eq:TransitonOperator}
\end{equation}
where
\begin{equation}
\begin{split}c_{1}^{\prime} & =c_{1}e^{-k_{R}dt/2}\\
 & \approx\frac{c_{1}}{\left|c_{1}\right|}\sqrt{\left|c_{1}\right|^{2}-k_{R}\left|c_{1}\right|^{2}dt}
\end{split}
,
\end{equation}
and if $\left|c_{0}\right|\neq0$,
\begin{equation}
\begin{split}c_{0}^{\prime} & =c_{0}\sqrt{1+\frac{\left|c_{1}\right|^{2}}{\left|c_{0}\right|^{2}}\left(1-e^{-k_{R}dt}\right)}\\
 & \approx\frac{c_{0}}{\left|c_{0}\right|}\sqrt{\left|c_{0}\right|^{2}+k_{R}\left|c_{1}\right|^{2}dt}
\end{split}
.
\end{equation}

Note that, if the subsystem happens to begin purely on the excited
state (i.e. $\widehat{\rho}=\left|1\right\rangle \left\langle 1\right|$
or $\left|c_{0}\right|=0$), there is an undetermined phase in the
wavefunction representation. In other words, we can write say $\left|\psi\right\rangle =e^{i\theta}\left|1\right\rangle $
and choose $\theta$ randomly. In this case, the transition operator
is defined as 
\begin{eqnarray}
c_{1}^{\prime} & = & e^{i\theta}e^{-\kappa dt/2}\approx e^{i\theta}\sqrt{1-\kappa dt},\\
c_{0}^{\prime} & = & \sqrt{1-e^{-\kappa dt}}\approx\sqrt{\kappa dt}.
\end{eqnarray}
As emphasized in Ref.~\onlinecite{li_mixed_2018} and Sec.~\ref{sec:Ehrenfest-Dynamics},
for these initial conditions, $\overline{k_{Eh}}=0$ and $\overline{k_{R}}=\kappa$
so that the +R relaxation must account for all of the required spontaneous
decay.

Finally, we introduce a stochastic random phase operator defined by
\begin{equation}
e^{i\widehat{\Phi}\left[\gamma_{R}\right]}=\begin{cases}
\left(\begin{array}{cc}
e^{i\Phi_{0}} & 0\\
0 & e^{i\Phi_{1}}
\end{array}\right) & \text{if RN}<\gamma_{R}dt\\
\widehat{1} & \text{otherwise}
\end{cases}
\end{equation}
where $\text{RN}\in\left[0,1\right]$ is a random number and $\Phi_{0},\Phi_{1}\in\left[0,2\pi\right]$
are random phases. This stochastic random phase operator enforces
the additional dephasing $\gamma_{R}$. That is, within time interval
$dt$, one reduces the ensemble average coherence $\left\langle c_{0}^{\prime}c_{1}^{\prime*}\right\rangle $
by an amount of $\left\langle c_{0}^{\prime}c_{1}^{\prime*}\right\rangle \times\gamma_{R}dt$
\textendash even though each individual trajectory still carries a
pure wavefunction. Put differently, the average coherence decays following
an inhomogeneous Poisson processes with instantaneous decay rate $\gamma_{R}$.
In practice, as shown in Paper II, it would appear much more robust
to set $\Phi_{1}=0$, and give a nonzero phase only to the ground
state ($\Phi_{0}\neq0$).

\subsubsection{Energy Conservation}

While Ehrenfest dynamics conserves the total energy of the quantum
subsystem together with the EM field, our proposed extra +R relaxation
changes the energy of the quantum subsystem $U_{s}=\text{Tr}\left\{ \widehat{\rho}\widehat{H}_{s}\right\} $
by an additional amount (relative to Ehrenfest dynamics):
\begin{eqnarray}
 &  & \frac{\partial U_{s}^{Eh+R}}{\partial t}-\frac{\partial U_{s}^{Eh}}{\partial t}\nonumber \\
 &  & \ =\text{Tr}\left\{ \widehat{H}_{s}\left(\doublehat{{\cal L}_{\hspace{0.1pt}}}_{Eh}+\doublehat{{\cal L}_{\hspace{0.1pt}}}_{R}\right)\widehat{\rho}\right\} -\text{Tr}\left\{ \widehat{H}_{s}\doublehat{{\cal L}_{\hspace{0.1pt}}}_{Eh}\widehat{\rho}\right\} \nonumber \\
 &  & \ =-\Omega k_{R}\rho_{11}
\end{eqnarray}
Thus, during a time step $dt$, the change in energy for the radiation
field is 
\begin{equation}
\delta U_{R}=\Omega k_{R}\rho_{11}dt.\label{eq:EnergyChange}
\end{equation}
For the Ehrenfest+R approach to enforce the energy conservation, this
energy loss must flow into the EM field in the form of light emission.
In other words, we must rescale the $\mathbf{E}$ and $\mathbf{B}$
fields.

\subsection{The Classical EM fields\label{subsec:The-Classical-EM}}

At every time step, with the +R correction of the quantum wavefunction,
we will rescale the Ehrenfest EM field ($\mathbf{E}_{Eh}$ and $\mathbf{B}_{Eh}$)
for each trajectory ($\ell$) as follows:

\begin{equation}
\mathbf{E}_{Eh+R}^{\ell}=\mathbf{E}_{Eh}^{\ell}+\text{\ensuremath{\alpha^{\ell}}}\delta\mathbf{E}_{R},\label{eq:rescaling_operator_E}
\end{equation}
\begin{equation}
\mathbf{B}_{Eh+R}^{\ell}=\mathbf{B}_{Eh}^{\ell}+\text{\ensuremath{\beta^{\ell}}}\delta\mathbf{B}_{R},\label{eq:rescaling_operator_B}
\end{equation}
or, in matrix notation,
\begin{equation}
\left(\begin{array}{c}
\mathbf{E}_{Eh+R}^{\ell}\\
\mathbf{B}_{Eh+R}^{\ell}
\end{array}\right)={\cal R}\left[\delta U_{R}^{\ell}\right]\left(\begin{array}{c}
\mathbf{E}_{Eh}^{\ell}\\
\mathbf{B}_{Eh}^{\ell}
\end{array}\right).\label{eq:rescaling_operator}
\end{equation}
Here, the coefficients $\text{\ensuremath{\alpha^{\ell}}}$ and $\text{\ensuremath{\beta^{\ell}}}$
depend on the random phase $\phi^{\ell}$ from Sec.~\ref{subsec:The-Quantum-Subsystem}.
In choosing the rescaling function ${\cal R}\left[\delta U_{R}^{\ell}\right]$,
there are several requirements:
\begin{enumerate}
\item [(a)]$\delta\mathbf{E}_{R}$ and $\delta\mathbf{B}_{R}$ must be
transverse fields.
\item [(b)]Since the +R correction enforces the FGR rate, it is crucial
that the rescaled EM field does not interfere with propagating the
quantum subsystem. Therefore, the spatial distribution of $\delta\mathbf{E}_{R}$
and $\delta\mathbf{B}_{R}$ must be located outside of the polarization
distribution. In other words, $\int\mathrm{d}v\widehat{\mathbf{P}}\cdot\delta\mathbf{E}_{R}\approx0$,
ensuring the electronic Hamiltonian, Eq.~(\ref{eq:electronic_Hamiltonian}),
does not change much after we rescale the classical EM field.
\item [(c)]The magnitude of $\beta\delta\mathbf{B}_{R}$ must be equal
to $1/c$ times the magnitude of $\alpha\delta\mathbf{E}_{R}$ for
all $\mathbf{r}$ in space so that the emission light propagates only
in one direction. 
\item [(d)]The directional energy flow must be outward, i.e. the Poynting
vector, $\mathbf{S}=\frac{1}{\mu_{0}}\mathbf{E}_{Eh+R}\times\mathbf{B}_{Eh+R}$
must have $\mathbf{S}\left(\mathbf{r}\right)\cdot\hat{\mathbf{r}}>0$
for all $\mathbf{r}$ (assuming the light is emanating from the origin).
\item [(e)]On average, we must have energy conservation, i.e. the energy
increase of the classical EM field must be equal to the energy loss
of the quantum subsystem described in Eq.~(\ref{eq:EnergyChange}).
\end{enumerate}
Unfortunately, it is very difficult to satisfy all of these requirements
concurrently, especially (c), (d), and (e). Nevertheless, we will
make an ansatz below which we believe will be robust.

Given a polarization distribution $\mathbf{P}$, the rescaling functions
for our ansatz are picked to be of the form

\begin{eqnarray}
\delta\mathbf{E}_{R} & = & \boldsymbol{\nabla}\times\boldsymbol{\nabla}\times\mathbf{P}-g\mathbf{P}_{\bot},\label{eq:rescaling_E}\\
\delta\mathbf{B}_{R} & = & -\boldsymbol{\nabla}\times\mathbf{P}-h\left(\boldsymbol{\nabla}\times\right)^{3}\mathbf{P},\label{eq:rescaling_B}
\end{eqnarray}
where $g$ and $h$ are chosen to best accommodate requirements (b)\textendash (d).
Note that Eqs.~(\ref{eq:rescaling_E}) and (\ref{eq:rescaling_B})
are both transverse fields. Eqs.~(\ref{eq:rescaling_E}) and (\ref{eq:rescaling_B})
arise naturally by iterating Maxwell's equations to low order. Since
the average current has the same spatial distribution as $\mathbf{P}$,
the $\mathbf{E}$ field derived from Maxwell's equations must be a
linear combination of $\mathbf{P}$ and even order derivatives of
$\mathbf{P}$. Vice versa, the $\mathbf{B}$ field must a linear combination
of the odd derivatives of $\mathbf{P}$.\footnote{Formally, the rescaling direction in Eqs.~(\ref{eq:rescaling_E})
and (\ref{eq:rescaling_B}) are motivated by a comparison of the electrodynamical
quantum\textendash classical Liouville equation (QCLE) and Ehrenfest
dynamics in the framework of mixed quantum-classical theory (to be
published).} In 3D space, we simply choose $g=h=0$, but the dynamics in 1D are
more complicated. (In Appendix~\ref{sec:Justify-the-rescaling},
we show numerically that $\boldsymbol{\nabla}\times\boldsymbol{\nabla}\times\mathbf{P}$
and $-\boldsymbol{\nabla}\times\mathbf{P}$ are good directions of
the emanated $\mathbf{E}$ and $\mathbf{B}$ fields in 3D. For a 1D
geometry, we choose $g$ and $h$ to minimize the spatial overlap
of both $\delta\mathbf{E}_{R}\cdot\mathbf{P}$ and $\delta\mathbf{B}_{R}\cdot\mathbf{P}$.
See Appendix~\ref{sec:Justify-the-rescaling}.)

For a Ehrenfest+R trajectory (labeled by $\ell$), the parameters
$\text{\ensuremath{\alpha^{\ell}}}$ and $\text{\ensuremath{\beta^{\ell}}}$
are chosen to be

\begin{equation}
\text{\ensuremath{\alpha^{\ell}}}=\sqrt{\frac{cdt}{\Lambda}\frac{\delta U_{R}^{\ell}}{\epsilon_{0}\int\mathrm{d}v\left|\delta\mathbf{E}_{R}\right|^{2}}}\times\text{sgn}\left(\text{Im}\left[\rho_{01}e^{i\phi^{\ell}}\right]\right)\label{eq:alpha_ell-sic}
\end{equation}
\begin{equation}
\text{\ensuremath{\beta^{\ell}}}=\sqrt{\frac{cdt}{\Lambda}\frac{\mu_{0}\delta U_{R}^{\ell}}{\int\mathrm{d}v\left|\delta\mathbf{B}_{R}\right|^{2}}}\times\text{sgn}\left(\text{Im}\left[\rho_{01}e^{i\phi^{\ell}}\right]\right)\label{eq:beta_ell-sic}
\end{equation}
where $\Lambda$ is the self-interference length determined by 
\begin{equation}
\Lambda=\frac{2\pi^{2}\left|\delta\widetilde{\mathbf{E}}_{R}\left(0\right)\right|^{2}}{\int\mathrm{d}x\left|\delta\mathbf{E}_{R}\right|^{2}}+\frac{2\pi^{2}\left|\delta\widetilde{\boldsymbol{B}}_{R}\left(0\right)\right|^{2}}{\int\mathrm{d}x\left|\delta\mathbf{B}_{R}\right|^{2}}.
\end{equation}
Here, $\delta\widetilde{\mathbf{E}}_{R}$ and $\delta\widetilde{\boldsymbol{B}}_{R}$
are the Fourier components of the rescaling fields $\delta\mathbf{E}_{R}$
and $\delta\mathbf{B}_{R}$. For $\mathbf{P}$ in the form of a Gaussian
distribution (e.g. $\left|\mathbf{P}\right|\sim e^{-ax^{2}}$ in a
1D system), we find that the self-interference length is always $\Lambda^{\text{1D}}=\frac{2}{3}\sqrt{\frac{2\pi}{a}}$.
By construction, Eqs.~(\ref{eq:alpha_ell-sic}) and (\ref{eq:beta_ell-sic})
should conserve energy only on average, i.e. an individual trajectory
with a random phase $\phi^{\ell}$ may not conserve energy, but the
ensemble energy should satisfy energy conservation (see Appendix~\ref{sec:Derivation-of-the-rescaling}).

\subsection{Step-by-step Algorithm of Ehrenfest+R method\label{subsec:Step-by-step-Algorithm-of}}

Here we give a detailed step-by-step outline of the Ehrenfest+R method.
For now, we restrict ourselves to the case of two electronic states.
Given a polarization $\mathbf{P}\left(\mathbf{r}\right)$ between
the electronic states, before starting an Ehrenfest+R trajectory,
we precompute the FGR rate $\kappa$ (Eq.~(\ref{eq:FGR_rate_3D})
or Eq.~(\ref{eq:FGR_rate_1D})) and a self-interference length $\Lambda$
(see Appendix~\ref{sec:Derivation-of-the-rescaling}). At this point,
we can initialize an Ehrenfest+R trajectory $\ell$ with a random
phase $\phi^{\ell}$. For time step $dt$, 
\begin{enumerate}
\item Propagate the wavefunction by $\left|\psi_{Eh}\left(t+dt\right)\right\rangle =e^{-i\widehat{H}^{\mathrm{el}}dt/\hbar}\left|\psi\left(t\right)\right\rangle $
and the EM field by Maxwell equations, Eqs.~(\ref{eq:maxwell_BE})
and (\ref{eq:maxwell_EB}). Here, we denote the EM field as $\mathbf{E}_{Eh}^{\ell}\left(t+dt\right)$
and $\mathbf{B}_{Eh}^{\ell}\left(t+dt\right)$ and $\widehat{H}^{\mathrm{el}}$
is defined by Eq.~(\ref{eq:electronic_Hamiltonian}).
\item Calculate the +R relaxation rate $k_{R}^{\ell}$ (Eq.~(\ref{eq:k_R})),
the +R dephasing rate $\gamma_{R}^{\ell}$ (Eq.~(\ref{eq:gamma_R})),
and energy change $\delta U_{R}^{\ell}$ (Eq.~(\ref{eq:EnergyChange})).
\item Apply the transition operator $\left|\psi\left(t+dt\right)\right\rangle =\widehat{{\cal T}}_{0\leftarrow1}\left[k_{R}^{\ell},\gamma_{R}^{\ell}\right]\left|\psi_{Eh}\left(t+dt\right)\right\rangle $
(Eq.~(\ref{eq:TransitonOperator})). Draw a random number $r\in\left(0,1\right)$.
If $r<\gamma_{R}^{\ell}dt$, draw another two random numbers $\Phi_{0},\Phi_{1}\in\left(0,2\pi\right)$
and apply $e^{i\Phi\left[\gamma_{R}^{\ell}\right]}.$
\item Calculate $\text{\ensuremath{\alpha^{\ell}}}$ and $\text{\ensuremath{\beta^{\ell}}}$
according to Eq.~(\ref{eq:alpha_ell-sic}) and Eq.~(\ref{eq:beta_ell-sic})
and then rescale the EM field by $\left(\begin{array}{c}
\mathbf{E}^{\ell}\left(t+dt\right)\\
\mathbf{B}^{\ell}\left(t+dt\right)
\end{array}\right)={\cal R}\left[\delta U_{R}^{\ell}\right]\left(\begin{array}{c}
\mathbf{E}_{Eh}^{\ell}\left(t+dt\right)\\
\mathbf{B}_{Eh}^{\ell}\left(t+dt\right)
\end{array}\right)$ according to Eq.~(\ref{eq:rescaling_operator_E}\textendash \ref{eq:rescaling_operator}).
\item Apply absorbing boundary conditions if the classical EM field reaches
the end of the spatial grid.
\end{enumerate}

\section{Results: Spontaneous Emission\label{sec:Results}}

\begin{figure}
\begin{centering}
\includegraphics{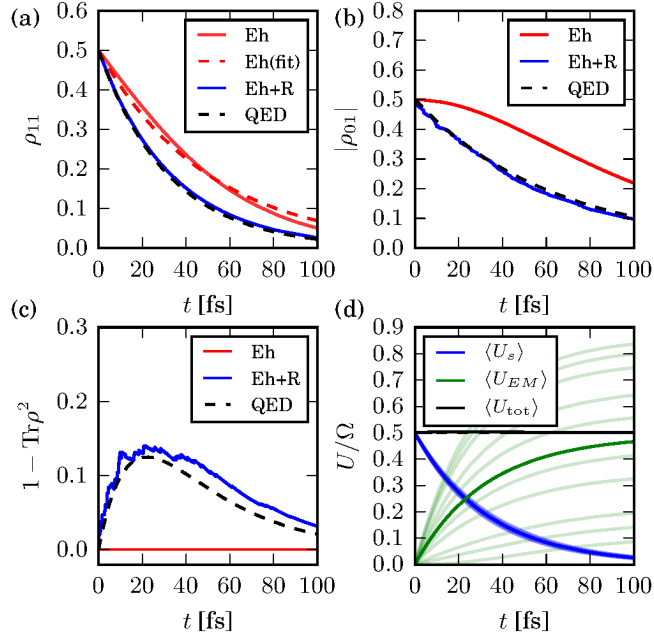}
\par\end{centering}
\caption{(a) Population of the excited state as a function of time. The black
dashed line indicates the FGR decay ($e^{-\kappa t}$). The red solid
line is the standard Ehrenfest dynamics and the red dashed line is
an exponential fit of the data. The blue solid line is Ehrenfest+R
dynamics. (b) Coherence of the reduced density matrix as a function
of time. The black dashed line indicates a decay at the true dephasing
rate ($e^{-\kappa t/2}$). The red solid line is the standard Ehrenfest
dynamics and the blue solid line is Ehrenfest+R dynamics. (c) Impurity
of the reduced density matrix as a function of time. The black dashed
line is the correct QED theoretical result given by Eq.~(\ref{eq:WW-impurity}).
The blue solid line is Ehrenfest+R dynamics. Note that the electronic
state remain a pure state ($1-\mathrm{Tr}\left\{ \rho^{2}\right\} =0$
for all time) within the standard Ehrenfest dynamics (red solid line).
(d) Energy as a function of time. The average energy of the two level
system is plotted in blue lines and the average energy of the EM field
is plotted in green lines. The dim lines are data from individual
trajectories. The solid black line is the average total energy (which
is effectively a constant). The initial state is $\left|\psi\right\rangle =\sqrt{\frac{1}{2}}\left|0\right\rangle +\sqrt{\frac{1}{2}}\left|1\right\rangle $
for all panels. The Ehrenfest+R dynamics data are averaged over $N_{\text{traj}}=200$
trajectories. \label{fig:population-dynamics-and} }
\end{figure}

As a test for our proposed Ehrenfest+R ansatz, we study spontaneous
emission of a two-level system in vacuum for 1D and 3D systems. We
assume the system lies in the FGR regime and the polarization distribution
is relatively small in space so that the long-wavelength approximation
is valid. For a two-level system with energy difference $\varepsilon_{1}-\varepsilon_{0}=\hbar\Omega$,
we consider two types of initial conditions $\left|\psi\left(0\right)\right\rangle $
with distinct behaviors:
\begin{enumerate}
\item [\#1]A superposition state with a fixed relative phase, i.e. $\left|\psi\left(0\right)\right\rangle =C_{0}\left|0\right\rangle +C_{1}\left|1\right\rangle $
where $\left|C_{0}\right|^{2}+\left|C_{1}\right|^{2}=1$ and $\left|C_{0}\right|\neq0$,
$\left|C_{1}\right|\neq1$:
\begin{itemize}
\item The upper state population $\rho_{11}\left(t\right)$ should decay
according to the FGR rate $\kappa$, and the coherence $\left|\rho_{01}\left(t\right)\right|$
should decay at the dephasing rate $\frac{\kappa}{2}$.
\item According to Eqs.~(\ref{eq:E-field-WWtheroy})\textendash (\ref{eq:intensity-WWtheroy}),
the electric field $\left\langle \mathbf{E}\right\rangle $ should
exhibit coherent emission at frequency $\Omega$. 
\item The averaged intensity $\left\langle \mathbf{E}^{2}\right\rangle $
should not equal the coherent emission $\left\langle \mathbf{E}\right\rangle ^{2}$,
i.e. $\left\langle \mathbf{E}^{2}\right\rangle -\left\langle \mathbf{E}\right\rangle ^{2}\neq0$. 
\end{itemize}
\item [\#2] A pure state with a random phase, i.e. $\widehat{\rho}\left(0\right)=\left|1\right\rangle \left\langle 1\right|$,
which corresponds to $\left|\psi\left(0\right)\right\rangle =e^{i\theta}\left|1\right\rangle $
where $\theta$ is a random phase:
\begin{itemize}
\item The upper state population $\rho_{11}\left(t\right)$ should still
decay according to the FGR rate, and the coherence $\left|\rho_{01}\left(t\right)\right|$
must remain zero.
\item The electric field of each individual trajectory should oscillate
at frequency $\Omega$, but the phases of different trajectories should
cancel out\textemdash so that the ensemble average of the electric
field becomes zero, i.e. $\left\langle \mathbf{E}\right\rangle =0$. 
\item The averaged intensity should not vanish, i.e. $\left\langle \mathbf{E}^{2}\right\rangle \neq0$.
\end{itemize}
\end{enumerate}
Model problems \#1 and \#2 capture key features when simulating spontaneous
emission and can be considered critical tests for the proposed Ehrenfest+R
approach. The parameters for our simulation are as follows. The energy
difference of the two levels system is $\hbar\Omega=16.46\ \text{eV}$.
The transition dipole moment is $\mu_{01}=11282\ \text{C/nm/mol}$.

\begin{figure}
\centering{}\includegraphics{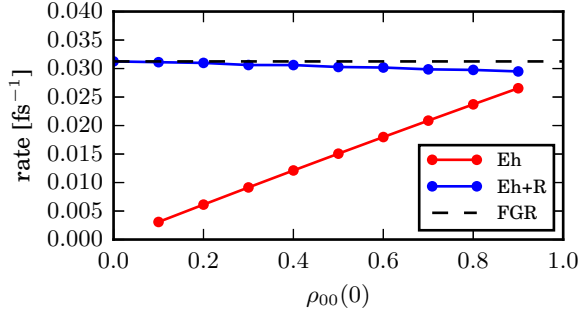}\caption{Spontaneous decay rates extracted from excited state population dynamics
for different initial states. As a function of the initial ground
state population $\rho_{00}$, we plot the exponential decay rates
for both the standard Ehrenfest method (red) and Ehrenfest+R method
(blue). The black dashed line indicates the FGR rate. Note that, for
all cases, Ehrenfest+R dynamics recover the true FGR spontaneous emission
rate.\label{fig:decay-rates}}
\end{figure}
For a 1D geometry, we consider a polarization distribution of the
form:
\begin{equation}
\mathbf{P}^{\text{1D}}\left(x\right)=\mu_{01}\sqrt{\frac{a}{\pi}}e^{-ax^{2}}\hat{\mathbf{z}},\label{eq:P_1D}
\end{equation}
with $a=1/2\sigma^{2}$ and $\sigma=3.0\ \text{nm}$. According to
Eq.~(\ref{eq:P_1D}), the polarization is in the $z$ direction varying
along the $x$ direction. For this polarization, the self-interference
length is $\Lambda^{\text{1D}}\approx7.0\ \text{nm}$. (As a reminder,
$\Lambda^{\text{1D}}=\frac{2}{3}\sqrt{\frac{2\pi}{a}}=2.363\sigma$.)
We use the rescaling function derived in Appendix.~\ref{sec:Justify-the-rescaling}:
\begin{eqnarray}
\delta\mathbf{E}_{R}^{\text{1D}}\left(x\right) & = & -\mu_{01}\sqrt{\frac{a}{\pi}}4a^{2}x^{2}e^{-ax^{2}}\hat{\mathbf{z}},\label{eq:dE_1D-1}\\
\delta\mathbf{B}_{R}^{\text{1D}}\left(x\right) & = & \mu_{01}\sqrt{\frac{a}{\pi}}\frac{4}{3}a^{2}x^{3}e^{-ax^{2}}\hat{\mathbf{y}}.\label{eq:dB_1D-1}
\end{eqnarray}

For a 3D geometry, we again assume the polarization is only in the
$z$ direction, now of the form
\begin{equation}
\mathbf{P}^{\text{3D}}\left(\mathbf{r}\right)=\hat{\mathbf{z}}\mu_{01}\frac{2a^{3/2}}{\pi^{3/2}}e^{-ar^{2}},
\end{equation}
where we use the same parameters for $a$ and $\mu_{01}$ as for the
1D geometry. The rescaling field in 3D is chosen to be:
\begin{eqnarray}
\delta\mathbf{E}_{R}^{\text{3D}}\left(\mathbf{r}\right) & = & \boldsymbol{\nabla}\times\boldsymbol{\nabla}\times\mathbf{P}^{\text{3D}}\left(\mathbf{r}\right),\label{eq:dE_3D}\\
\delta\mathbf{B}_{R}^{\text{3D}}\left(\mathbf{r}\right) & = & -\boldsymbol{\nabla}\times\mathbf{P}^{\text{3D}}\left(\mathbf{r}\right).\label{eq:dB_3D}
\end{eqnarray}
The self-interference length can be obtained numerically as $\Lambda^{\text{3D}}\approx0.6\ \text{nm}$.\footnote{Note that the self-interference length strongly depends on dimensionality
and is much smaller in 3D than in 1D.}

Our simulation is propagated using Cartesian coordinates with $dx=0.1$
for 1D and $dx=dy=dz=0.3\ \text{nm}$ for 3D. The time step is $dt=10^{-3}\ \text{fs}$.
Without loss of generality, the random phase $\phi^{\ell}$ for Ehrenfest+R
trajectories is chosen from an evenly space distribution, i.e. $\phi^{\ell}=2\pi j/N_{\text{traj}}$
for $j=1,\cdots,N_{\text{traj}}$. 

\begin{figure*}
\begin{centering}
\includegraphics{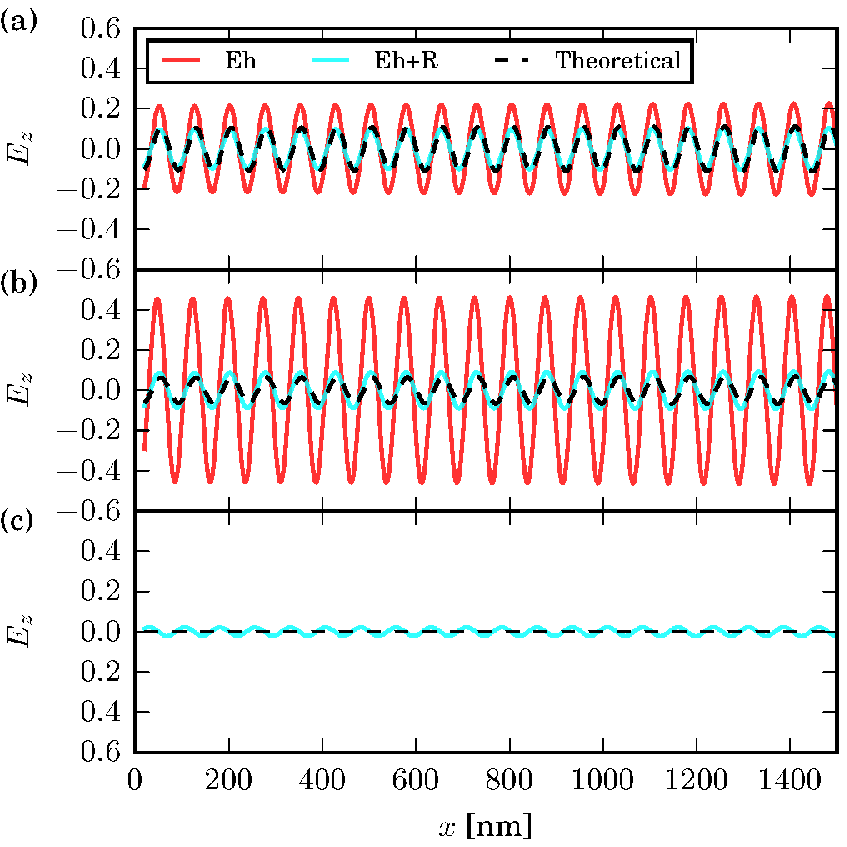}~\includegraphics{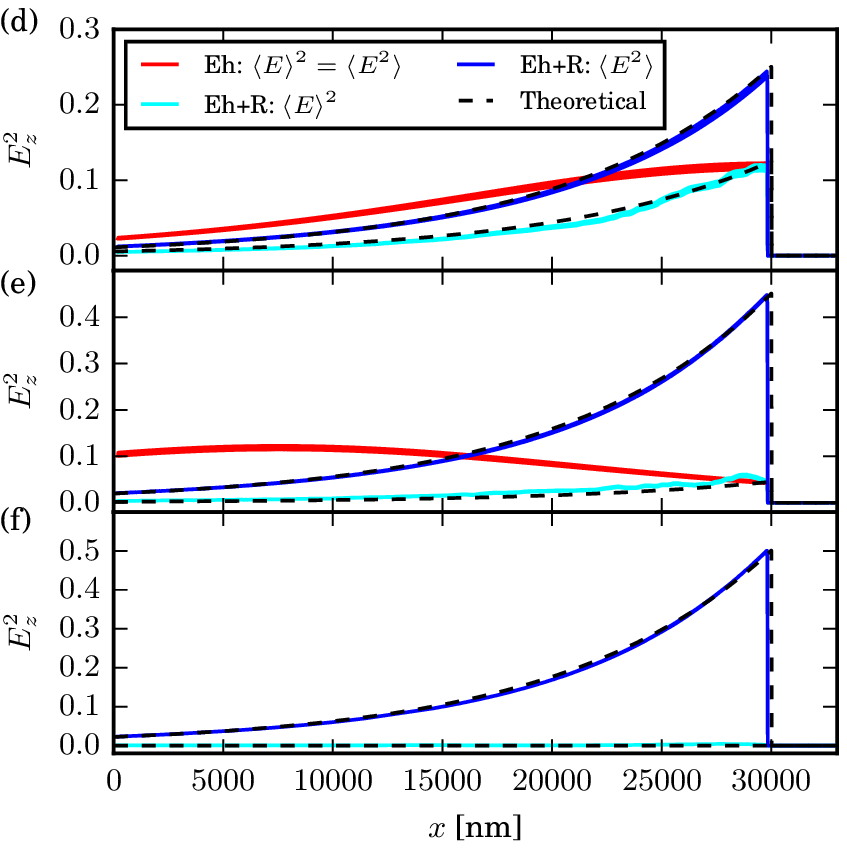}
\par\end{centering}
\caption{The electric field produced for spontaneous emission as a function
of $x$ at $t=100\ \text{fs}$. The initial population on the excited
state is $\rho_{11}\left(0\right)=0.5$ for (a), (d), $\rho_{11}\left(0\right)=0.9$
for (b), (e), and $\rho_{11}\left(0\right)=1$ for (c), (f). Left
panels are the electric field in the $z$ direction $\left\langle E_{z}\right\rangle $
in units of $\mu_{01}\Omega/\epsilon_{0}c$, where the black dashed
lines are the theoretical results (see Eq.~(\ref{eq:E-field-WWtheroy}).)
The solid lines are calculated by standard Ehrenfest (red) and by
Ehrenfest+R (cyan) dynamics. Right panels are the intensity ($\left\langle E_{z}^{2}\right\rangle $)
and the magnitude of the coherent emission ($\left\langle E_{z}\right\rangle ^{2}$)
in units of $\left(\mu_{01}\Omega/\epsilon_{0}c\right)^{2}$, where
the black dashed lines are Eq.~(\ref{eq:E-field2-WWtheroy}) and
Eq.~(\ref{eq:intensity-WWtheroy}). On the right panels, we perform
a moving average over $c\tau=720\ \text{nm}$ (10 oscillations) to
show the coarse-grained behavior. The solid lines are $\left\langle E_{z}^{2}\right\rangle =\left\langle E_{z}\right\rangle ^{2}$
calculated by standard Ehrenfest dynamics (red), and $\left\langle E_{z}^{2}\right\rangle $
(blue) and $\left\langle E_{z}\right\rangle ^{2}$ (cyan) calculated
by Ehrenfest+R approach. The event horizon can be observed at $x=ct=30000\ \text{nm}$.
$N_{\text{traj}}=200$. Note that Ehrenfest+R recovers all observables
quantitatively, whereas Ehrenfest dynamics are accurate only when
$\rho_{11}\left(0\right)\ll1$. Note also that Ehrenfest dynamics
predicts no emission when $\rho_{00}\left(0\right)=0$ (f). \label{fig:Radiation-field-intensity}}
\end{figure*}

\subsection{Spontaneous decay rate}

Our first focus is an initially coherent state with $\rho_{00}\left(0\right)=\rho_{11}\left(0\right)=0.5$.
We plot the upper state population and the decay rate of a 1D system
($e^{-\kappa t}$) in Fig~\ref{fig:population-dynamics-and}(a).
As shown in Ref.~\onlinecite{li_mixed_2018} and summarized in Sec.~\ref{sec:Ehrenfest-Dynamics}
above, standard Ehrenfest dynamics does not agree with the FGR decay
and cannot be fit to an exponential decay. With Ehrenfest+R dynamics,
however, we can quantitatively correct the errors of Ehrenfest dynamics
and recover the full spontaneous decay rate accurately. Furthermore,
in Fig.~\ref{fig:population-dynamics-and}(b), we plot the coherence
$\left|\rho_{01}\right|$ of the 1D system. At early times where the
system is not far from initial state ($\rho_{00}\approx\rho_{11}\approx0.5$),
we find that the coherence of Ehrenfest dynamics remain a constant
of time, i.e. $\gamma_{Eh}=0$ as Eq.~(\ref{eq:average_gamma}) suggested.
By contrast Ehrenfest+R dynamics recover the correct dephasing rate
($\approx e^{-\kappa t/2}$). Finally, with an accurate evaluation
of the population and coherence, it is not surprising that Ehrenfest+R
recover the correct impuriy ($1-\mathrm{Tr}\left\{ \rho^{2}\right\} $)
in Fig.~\ref{fig:population-dynamics-and}(c).

Regarding energy conservation, individual Ehrenfest+R trajectories
do not conserve energy by design. While the energy loss of the quantum
system is roughly the same for every trajectory, the emitted EM energy
fluctuates and is not equal to the corresponding quantum energy loss
(see Fig.~\ref{fig:population-dynamics-and}(d)). However, an ensemble
of trajectories does converse energy on average.

In Fig.~\ref{fig:decay-rates}, for all initial conditions, we plot
decay rates extracted from excited state population dynamics for a
short time ($t<10\ \text{fs}$). As shown in Eq.~(\ref{eq:kEh_pure}),
the Ehrenfest decay rate is proportional to the lower state population.
However, even though Ehrenfest dynamics fails to predict the correct
decay rate as a function of initial condition, the decay rate extracted
from Ehrenfest+R dynamics agrees very well with the FGR decay rate
for all initial conditions. Note that, for the extreme case $\rho_{00}\left(0\right)=0$,
Ehrenfest dynamics does not predict any population decay.

\subsection{Emission Fields in 1D}

We now turn our attention to the coherent emission and the intensity
of the EM field. We start by considering a 1D geometry. According
to Eq.~(\ref{eq:E-field-WWtheroy}), for a given time $t$, the electric
field of spontaneous emission can be expressed as a function of $x$
and shows oscillatory behavior proportional to $\sin\Omega\left(t-\left|x\right|/c\right)$
for short times. Also, an event horizon is observed at $\left|x\right|=ct$,
i.e. no electric field should be observed for $\left|x\right|>ct$
because of causality. 

We find that the electric field obtained by an individual Ehrenfest+R
trajectory shows the correct oscillations at frequency $\Omega$ with
an additional phase shift. For an initially coherent state, the ensemble
average of Ehrenfest+R trajectories agrees with Eq.~(\ref{eq:E-field-WWtheroy})
very well (see Figs.~\ref{fig:Radiation-field-intensity}(a) and
\ref{fig:Radiation-field-intensity}(b) for two cases with different
initial conditions.) When the initial state is exclusively the excited
state, the ensemble average of Ehrenfest+R trajectories vanishes by
phase cancellation and we recover $\left\langle \mathbf{E}\right\rangle =0$
(see Fig.~\ref{fig:Radiation-field-intensity}(c)).

Now we compare the emission intensity $\overline{\left\langle \mathbf{E}^{2}\right\rangle }$
and the magnitude of the coherent emission $\overline{\left\langle \mathbf{E}\right\rangle ^{2}}$.
On the right panels of Fig.~\ref{fig:Radiation-field-intensity},
we plot the coarse-grained behavior of Ehrenfest+R trajectories. We
show that Ehrenfest+R can accurately recover the spatial distribution
of both $\overline{\left\langle \mathbf{E}^{2}\right\rangle }$ and
$\overline{\left\langle \mathbf{E}\right\rangle ^{2}}$, as well as
the event horizon. Note that in Fig.~\ref{fig:Radiation-field-intensity},
the electric field and the intensity at large $x$ corresponds to
emission at earlier times. If we start with a coherent initial state,
the relative proportion of coherent emission is given by $\overline{\left\langle \mathbf{E}\right\rangle ^{2}}/\overline{\left\langle \mathbf{E}^{2}\right\rangle }=\rho_{00}\left(0\right)$,
see Eqs.~(\ref{eq:E-field-WWtheroy}) and (\ref{eq:E-field2-WWtheroy}).
For $\rho_{11}\left(0\right)=0.5$, the coherent emission is responsible
for 50\% of the total energy emission at early times ($x\sim ct=3\times10^{4}\ \text{nm}$),
and the coherent emission dominates later ($x\sim0$). Obviously,
if we begin with a wavefunction prepared exclusively on the excited
state, there is no coherent emission due to phase cancellation among
Ehrenfest+R trajectories. In the end, using an ensemble of trajectories
with random phases $\phi^{\ell}$, Ehrenfest+R is effectively able
to introduce some quantum decoherence among the classical trajectories
and can recover both $\overline{\left\langle \mathbf{E}^{2}\right\rangle }$
and $\overline{\left\langle \mathbf{E}^{2}\right\rangle }$ .

This behavior of Ehrenfest+R dynamics should be contrasted with the
behavior of standard Ehrenfest dynamics, where we run only one trajectory
and we observe only coherent emission with $\left\langle \mathbf{E}^{2}\right\rangle =\left\langle \mathbf{E}\right\rangle ^{2}$.
Although the coherent emission obtained by standard Ehrenfest dynamics
is close to the quantum result when $\rho_{11}\left(0\right)$ is
small (see Fig.~\ref{fig:Radiation-field-intensity}(a)), the magnitude
of the coherent emission is incorrect in general. The electric field
does oscillate at the correct frequency.

\subsection{Emission Fields in 3D}

\begin{figure}
\begin{centering}
\includegraphics{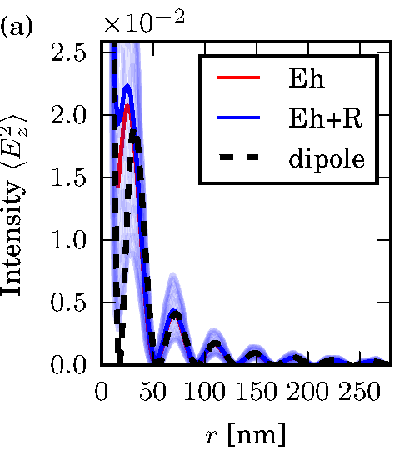}~\includegraphics{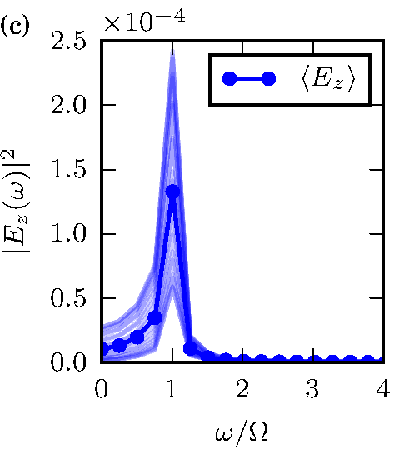}
\par\end{centering}
\begin{centering}
\includegraphics[clip]{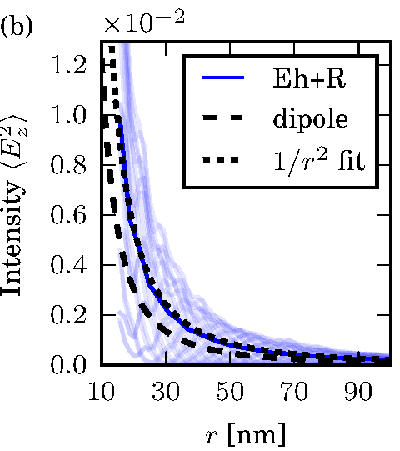}~\includegraphics{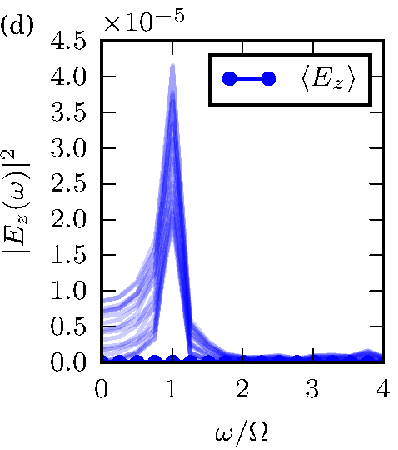}
\par\end{centering}
\caption{Spontaneous emission intensity calculated by Ehrenfest+R dynamics
as a function of radius $r$ at $t=1.0\ \text{fs}$ for the initial
population (a) $\rho_{11}=0.5$, and (b)$\rho_{11}=1.0$. The polar
angle is $\theta=\frac{\pi}{2}$ and the intensity is plotted in units
of $\mu_{0}\Omega^{4}\mu_{01}^{2}/32\pi^{2}c^{2}$.The right panels
are the corresponding spectrum of the electric field in the $z$ direction.
The dim lines are data from individual trajectories and the solid
circles are the average data. Note that, in (d), there is a phase
cancellation and, even though all EM fields have non-zero Fourier
transform components around $\omega=\Omega$, the net average EM field
is zero. The black dashed line is the theoretical energy flux. The
self-interference length is $\Lambda^{\text{3D}}\approx0.6\ \text{nm}$.
\label{fig:3D}}
\end{figure}

For a 3D geometry, for reasons of computational cost, we propagate
the dynamics of spontaneous emission for short-times only ($t<1.0\ \text{fs}$).
Our results are similar to the 1D case and are plotted in Fig.~\ref{fig:3D}.
For a coherent initial state (Fig.~\ref{fig:3D} (a), (c)), each
Ehrenfest+R trajectory yields an electric field and EM intensity oscillating
at frequency $\Omega$, and these features are retained by the ensemble
average. For the case of dynamics initiated from the excited state
only (Fig.~\ref{fig:3D} (b), (d)), each trajectory still oscillates
at frequency $\Omega$, but the average electric field is actually
zero ($\left\langle \mathbf{E}\right\rangle =0$).

In Fig.~\ref{fig:3D}, we also compare our result versus the well-known
classical Poynting flux of electric dipole radiation. In Fig.~\ref{fig:3D}
(a), our reference is
\begin{equation}
I\left(\mathbf{r},t\right)=\frac{\mu_{0}}{c^{2}}\frac{\Omega^{4}\mu_{01}^{2}}{16\pi^{2}}\frac{\sin^{2}\theta}{r^{2}}\sin^{2}\Omega\left(t-\frac{r}{c}\right),
\end{equation}
and, in Fig.~\ref{fig:3D} (b), our reference is the mean electromagnetic
energy flux
\begin{equation}
\overline{I}\left(\mathbf{r}\right)=\frac{\mu_{0}}{c^{2}}\frac{\Omega^{4}\mu_{01}^{2}}{32\pi^{2}}\frac{\sin^{2}\theta}{r^{2}}.
\end{equation}
In general, Ehrenfest+R dynamics yields a similar distribution as
the classical dipole radiation. When initiated from a coherent state,
both methods behave as $\sin^{2}\Omega\left(t-\frac{r}{c}\right)$;
when initiated from the excited state, Ehrenfest+R method shows $1/r^{2}$
dependence for $\left\langle \mathbf{E}^{2}\right\rangle $ while
Ehrenfest dynamics does not yield any emission (not shown in the plot.)
However, we note that the intensity of the Ehrenfest+R results is
slightly larger than that of classical dipole radiation. This difference
is attributed to the fact that the classical dipole radiation includes
only coherent emission, which is captured by standard Ehrenfest dynamics.
By contrast, Ehrenfest+R dynamics can also yield so-called incoherent
emission ($\left\langle \mathbf{E}^{2}\right\rangle -\left\langle \mathbf{E}\right\rangle ^{2}\neq0$),
which is effectively a quantum mechanical feature with no classical
analogue. 

\section{Conclusions and Future work\label{sec:Conclusions}}

In this work, we have proposed a heuristic, new semiclassical approach
to quantum electrodynamics, based on Ehrenfest dynamics and designed
to capture spontaneous emission correctly. Our ansatz is to enforce
extra electronic relaxation while also rescaling the EM field in the
direction $\delta\mathbf{E}_{R}=\boldsymbol{\nabla}\times\boldsymbol{\nabla}\times\mathbf{P}$
and $\delta\mathbf{B}_{R}=-\boldsymbol{\nabla}\times\mathbf{P}$.
Our results suggest that this Ehrenfest+R approach can indeed recover
the correct FGR decay rate for a two-level system. More importantly,
both intensity and coherent emission can be accurately captured by
Ehrenfest+R dynamics, where an ensemble of classical trajectories
effectively simulates the statistical variations of a quantum electrodynamics
field. Obviously, our approach here is not unique; a more standard
approach would be to explicitly model the EM vacuum fluctuations with
a set of harmonic oscillators. Nevertheless, by avoiding the inclusion
of high frequency oscillator modes, our ansatz eliminates any possibility
of artificial zero point energy loss or other anomalies from quasi-classical
dynamics.\citep{peslherbe_analysis_1994,brieuc_zero-point_2016}

As far as computational cost is concerned, one Ehrenfest+R trajectory
costs roughly the same amount as one standard Ehrenfest trajectory,
and all dynamics are numerically stable. Implementation of Ehrenfest+R
dynamics is easy to parallelize and incorporate within sophisticated
numerical packages for classical electromagnetics (e.g. FDTD\citep{taflove_advances_1998}).

Given the promising results presented above for Ehrenfest+R, we can
foresee many interesting applications. First, we would like to include
nuclear degrees of freedom within the quantum subsystem to explicitly
address the role of dephasing in spontaneous and stimulated emission.
Second, we would like to study more than two states. For instance,
a three-level system with an incoming EM field can be employed for
studying inelastic light scattering processes, such as Raman spectroscopy.
This will be the focus of paper II. Third, we would also like to model
multiple spatial separated quantum emitters, such as resonance energy
transfer.

At the same time, many questions remain and need to be addressed:
\begin{enumerate}
\item The current prescription for Ehrenfest+R approach is fundamentally
based on enforcing the FGR rate. However, in many physical situations,
such as molecules in a resonant cavity or near a metal surface, the
decay rate of the quantum subsystem can be modified by interactions
with environmental degrees of freedom. How should we modify the Ehrenfest+R
approach to account for each environment?
\item For a quantum subsystem interacting with a strong incoming field,
including the well-known Mollow triplet phenomenon\citep{mollow_power_1969}
and other multi-photon processes, EM field quantization can lead to
complicated emission spectra involving frequencies best described
with dressed states. Can these effectively quantum features be captured
by Ehrenfest+R dynamics?
\item Finally, and most importantly, it remains to test how the approach
presented here behaves when there are many quantum subsystems interacting,
leading to coherent effects (i.e. plasmonic excitations). Can our
approach simulate these fascinating experiments? Can our approach
simulate these fascinating experiments? How will other nonadiabatic
dynamics methods based on Ehrenfest dynamics (e.g. PLDM,\citep{huo_consistent_2012}
PBME,\citep{kim_quantum-classical_2008} and SQC\citep{miller_classical/semiclassical_1978})
behave?
\end{enumerate}
These questions will be investigated in the future.

\section*{Acknowledgment}

J.E.S. acknowledges start up funding from the University of Pennsylvania.
The research of AN is supported by the Israel-U.S. Binational Science
Foundation, the German Research Foundation (DFG TH 820/11-1), the
U.S. National Science Foundation (Grant No. CHE1665291), and the University
of Pennsylvania. M.S. would also like to acknowledge financial support
by the Air Force Office of Scientific Research under Grant No. FA9550-15-1-0189
and Binational Science Foundation under Grant No. 2014113. We thank
Kirk McDonald for very interesting discussions related to the calculation
in Appendix~\ref{sec:Instantaneous-Decay-rate}. 

\appendix

\section{Generalized Weisskopf\textendash Wigner Theory of Spontaneous Emission\label{sec:Weisskopf=002013Wigner}}

Consider the electric dipole Hamiltonian given by Eq.~(\ref{eq:ED-interaction}).
For comparison with semiclassical dynamics in Sec.~\ref{sec:Results}
we will now derive the exact population dynamics and the emission
EM field of a two level system in vacuum based on Weisskopf\textendash Wigner
theory and a retarded Green's function approach.

\subsection{Dressed state representation}

Let $\left|0,\cdots,1_{\mathrm{k}},\cdots,0\right\rangle $ be a state
of the EM field with one photon of mode $\omega_{\mathrm{k}}$, as
expressed in a Fock space representation. Let us denote the vacuum
state as $\left|\left\{ 0\right\} \right\rangle $. For a system composed
of an atom interacting with the EM field, the dressed state representation
has the following basis (including up to a single photon per mode)\citep{scully_quantum_1997,nitzan_chemical_2006}
\begin{eqnarray}
\left|j;\mathrm{k}\right\rangle  & = & \left|j\right\rangle \left|0,\cdots,1_{\mathrm{k}},\cdots,0\right\rangle \\
\left|j;0\right\rangle  & = & \left|j\right\rangle \left|\left\{ 0\right\} \right\rangle 
\end{eqnarray}
Here $\left|j\right\rangle =\left|0\right\rangle ,\left|1\right\rangle $
are the wavefunctions for the two level system. For such a setup,
the total wavefunction in the dressed state representation must be
of the form:
\begin{equation}
\begin{split}\left|\psi\left(t\right)\right\rangle = & C_{00}\left(t\right)\left|0;0\right\rangle +C_{10}\left(t\right)\left|1;0\right\rangle +\\
 & \sum_{\mathrm{k}}C_{0\mathrm{k}}\left(t\right)\left|0;\mathrm{k}\right\rangle +\sum_{\mathrm{k}}C_{1\mathrm{k}}\left(t\right)\left|1;\mathrm{k}\right\rangle .
\end{split}
\label{eq:WW-wavefunction}
\end{equation}
For spontaneous emission, let the initial wavefunction of the two-level
system in vacuum be written as 
\begin{equation}
\left|\psi\left(0\right)\right\rangle =C_{0}\left|0;0\right\rangle +C_{1}\left|1;0\right\rangle \label{eq:WW-initialstate}
\end{equation}
with $\left|C_{0}\right|^{2}+\left|C_{1}\right|^{2}=1$. We would
like to propagate $\left|\psi\left(0\right)\right\rangle $ and calculate
$\left|\psi\left(t\right)\right\rangle $ as a function of time. We
emphasize that, in Eqs.~(\ref{eq:WW-wavefunction}) and (\ref{eq:WW-initialstate}),
the Hilbert space is restricted to one photon states. 

For visualization purpose, it is helpful to write down the electric
dipole Hamiltonian explicitly in matrix form in the dressed state
representation, 
\begin{eqnarray}
{\cal H} & = & {\cal H}_{0}+{\cal V}\\
 &  & \begin{array}{cccc}
\ \ \ \ \ \ \left\{ \left|0;\mathrm{k}\right\rangle \right\}  & \ \ \ \ \left|0;0\right\rangle  & \ \ \ \left|1;0\right\rangle  & \ \ \ \ \ \left\{ \left|1;\mathrm{k}\right\rangle \right\} \end{array}\nonumber \\
 & = & \left(\begin{array}{cccc}
\left\{ \left[\varepsilon_{0}+\hbar\omega_{\mathrm{k}}\right]\right\}  & 0 & \left[\left\{ V_{\mathrm{k}}\right\} \right]^{\dagger} & 0\\
0 & \varepsilon_{0} & 0 & \left[\left\{ V_{\mathrm{k}}\right\} \right]\\
\left[\left\{ V_{\mathrm{k}}\right\} \right] & 0 & \varepsilon_{1} & 0\\
0 & \left[\left\{ V_{\mathrm{k}}\right\} \right]^{\dagger} & 0 & \left\{ \left[\varepsilon_{1}+\hbar\omega_{\mathrm{k}}\right]\right\} 
\end{array}\right)\nonumber 
\end{eqnarray}
Here the set $\left\{ \left[\varepsilon_{j}+\hbar\omega_{\mathrm{k}}\right]\right\} $
is an infinite set of matrices with exclusively diagonal elements
$\varepsilon_{j}+\hbar\omega_{\mathrm{k}}$ for $j=0,1$. $\left[V_{\mathrm{k}}\right]$
is an infinite row with corresponding elements 
\begin{equation}
V_{\mathrm{k}}=i\boldsymbol{\mu}_{01}\cdot\mathbf{s}_{\mathrm{k}}\sqrt{\frac{\hbar\omega_{\mathrm{k}}}{2\epsilon_{0}L^{n}}}
\end{equation}
between the vacuum state $\left|\left\{ 0\right\} \right\rangle $
and a one-photon state with mode $\omega_{\mathrm{k}}$. Let us denote
the diagonal part of the matrix as the unperturbed Hamiltonian ${\cal H}_{0}$
and the off-diagonal part as the coupling Hamilton ${\cal V}$. Note
that the two quantum states in vacuum ($\left|0;0\right\rangle $
and $\left|1;0\right\rangle $) are coupled to two different continuous
manifolds $\left\{ \left|1;\mathrm{k}\right\rangle \right\} $ and
$\left\{ \left|0;\mathrm{k}\right\rangle \right\} $, respectively.

Given that $\varepsilon_{0}<\varepsilon_{1}$, the $\left\{ \left|0;\mathrm{k}\right\rangle \right\} $
manifold will always include a quantum state that is energetically
resonant with the $\left|1;0\right\rangle $ state. However, the the
$\left\{ \left|1;\mathrm{k}\right\rangle \right\} $ manifold will
always be off-resonant with $\left|0;0\right\rangle $ for all $\mathrm{k}$.
Therefore, as the lowest order approximation, we can assume 
\begin{equation}
C_{1\mathrm{k}}\left(t\right)\approx0,\label{eq:C1k}
\end{equation}
 and 
\begin{equation}
C_{00}\left(t\right)\approx C_{0}e^{-i\varepsilon_{0}t/\hbar}.\label{eq:C00}
\end{equation}
Eqs.~(\ref{eq:C1k}) and (\ref{eq:C00}) are known as the rotating
wave approximation (RWA).

\subsection{Retarded Green's function formulation}

We employ a retarded Green's function formulation\citep{nitzan_chemical_2006}
to obtain the time evolution of $C_{10}\left(t\right)$ and $C_{0\mathrm{k}}\left(t\right)$.
The retarded Green's operators are ${\cal G}\left(\varepsilon\right)=\left[\varepsilon-{\cal H}+i\eta\right]^{-1}$
for the full Hamiltonian and ${\cal G}_{0}\left(\varepsilon\right)=\left[\varepsilon-{\cal H}_{0}+i\eta\right]^{-1}$
for the unperturbed Hamiltonian where $\eta$ is a positive small
quantity ($\eta\rightarrow0^{+}$). Using Dyson's identity ${\cal G}={\cal G}_{0}+{\cal G}_{0}{\cal V}{\cal G}={\cal G}_{0}+{\cal G}{\cal V}{\cal G}_{0}$,
we can obtain the retarded Green's function in a self-consistent expression
\begin{eqnarray}
{\cal G}_{10,10}\left(\varepsilon\right) & = & \frac{1}{\varepsilon-\varepsilon_{1}+i\eta+\frac{i}{2}\Gamma\left(\varepsilon\right)},\\
{\cal G}_{0\mathrm{k},10}\left(\varepsilon\right) & = & \frac{V_{\mathrm{k}}}{\varepsilon-\varepsilon_{0}-\hbar\omega_{\mathrm{k}}+i\eta}{\cal G}_{10,10}\left(\varepsilon\right),
\end{eqnarray}
where the self energy is $\Gamma\left(\varepsilon\right)=2i\sum_{\mathrm{k}}\left|V_{\mathrm{k}}\right|^{2}/\left(\varepsilon-\varepsilon_{0}-\hbar\omega_{\mathrm{k}}+i\eta\right)$.
The self energy can be evaluated by a Cauchy integral identity (ignoring
the principle value part). For 1D, we can consider a dipole moment
$\mu_{01}$ and use the density of states of a 1D system to obtain
the self energy as
\begin{eqnarray*}
\Gamma^{\text{1D}}\left(\varepsilon\right) & = & 2i\frac{L}{2\pi}\sum_{s}\int dk\frac{\mu_{01}^{2}\mathcal{E}_{\mathrm{k}}^{2}}{\varepsilon-\varepsilon_{0}-\hbar\omega_{\mathrm{k}}+i\eta}\\
 & = & i\frac{\mu_{01}^{2}}{2\pi\epsilon_{0}\hbar c}\left[-i\pi\left(\varepsilon-\varepsilon_{0}\right)\right]\\
 & = & \frac{\mu_{01}^{2}}{\epsilon_{0}\hbar c}\left(\varepsilon-\varepsilon_{0}\right)
\end{eqnarray*}
Here, ${\cal E}_{\mathrm{k}}=\sqrt{\frac{\hbar\omega_{\mathrm{k}}}{2\epsilon_{0}L}}$.
For 3D, we consider a dipole moment $\boldsymbol{\mu}_{01}=\mu_{01}\hat{\mathbf{z}}$
so that $\boldsymbol{\mu}_{01}\cdot\mathbf{s}_{\mathrm{k}}=\mu_{01}\sin\theta$
and the self energy is
\begin{eqnarray*}
\Gamma^{\text{3D}}\left(\varepsilon\right) & = & 4\pi i\left(\frac{L}{2\pi}\right)^{3}\int_{0}^{\pi}\sin^{3}\theta\mathrm{d}\theta\int_{0}^{\infty}k^{2}\mathrm{d}k\times\\
 &  & \ \frac{\mu_{01}^{2}\mathcal{E}_{k}^{2}}{\varepsilon-\varepsilon_{0}-\hbar\omega_{\mathbf{k}}+i\eta}\\
 & = & i\frac{\mu_{01}^{2}}{3\pi^{2}\epsilon_{0}\hbar^{3}c^{3}}\left[-i\pi\left(\varepsilon-\varepsilon_{0}\right)^{3}\right]\\
 & = & \frac{\mu_{01}^{2}}{3\pi\epsilon_{0}\hbar^{3}c^{3}}\left(\varepsilon-\varepsilon_{0}\right)^{3}
\end{eqnarray*}
Here, ${\cal E}_{\mathrm{k}}=\sqrt{\frac{\hbar\omega_{\mathrm{k}}}{2\epsilon_{0}L^{3}}}$
and we have used the identity $\int_{0}^{\pi}\sin^{3}\theta\mathrm{d}\theta=\frac{4}{3}$.
Note that the $\varepsilon$ dependence of the self energy will result
in a non-exponential decay. In the FGR regime, since all dynamics
can be extracted from Fourier transforms of the Green's function,
and the Green's operators ${\cal G}\left(\varepsilon\right)$ are
expected to have a single pole near $\varepsilon=\varepsilon_{1}$
that will dominate all Cauchy integrals, we approximate the self energy
by the value $\Gamma\left(\varepsilon\right)\approx\Gamma\left(\varepsilon_{1}\right)$
\begin{eqnarray}
\Gamma^{\text{1D}}\left(\varepsilon\right) & \approx & \hbar\kappa^{\text{1D}}=\frac{\mu_{01}^{2}\Omega}{\epsilon_{0}c},\\
\Gamma^{\text{3D}}\left(\varepsilon\right) & \approx & \hbar\kappa^{\text{3D}}=\frac{\mu_{01}^{2}\Omega^{3}}{3\pi\epsilon_{0}c^{3}}.
\end{eqnarray}
In the following, we will use $\kappa$ to represent either $\kappa^{\text{1D}}$
or $\kappa^{\text{3D}}$ and $\Gamma=\hbar\kappa$ depending on context.
Finally, the retarded Green's function is approximated as 
\begin{eqnarray}
{\cal G}_{10,10}\left(\varepsilon\right) & \approx & \frac{1}{\varepsilon-\varepsilon_{1}+i\eta+\frac{i}{2}\Gamma},\\
{\cal G}_{0\mathrm{k},10}\left(\varepsilon\right) & \approx & \frac{V_{\mathrm{k}}}{\varepsilon-\varepsilon_{0}-\hbar\omega_{\mathrm{k}}+i\eta}{\cal G}_{10,10}\left(\varepsilon\right).
\end{eqnarray}

The total wavefunction can then be obtained by the Fourier transform
of the Green's function 
\begin{equation}
\left|\psi\left(t\right)\right\rangle =-\frac{1}{2\pi i}\int_{-\infty}^{\infty}d\varepsilon e^{-i\left(\varepsilon+i\eta\right)t/\hbar}{\cal G}\left(\varepsilon\right)\left|\psi\left(0\right)\right\rangle 
\end{equation}
 with Cauchy integral:
\begin{equation}
C_{10}\left(t\right)=C_{1}e^{-i\frac{\varepsilon_{1}}{\hbar}t-\frac{\kappa}{2}t},
\end{equation}
\begin{equation}
C_{0\mathrm{k}}\left(t\right)=\frac{C_{1}V_{\mathrm{k}}/\hbar}{\omega_{\mathrm{k}}-\Omega+i\frac{\kappa}{2}}\left[e^{-i\left(\frac{\varepsilon_{0}}{\hbar}+\omega_{\mathrm{k}}\right)t}-e^{-i\frac{\varepsilon_{1}}{\hbar}t-\frac{\kappa}{2}t}\right].
\end{equation}
The reduced density matrix of the electronic system is defined by
taking trace over the photon modes of the total density matrix, $\rho\left(t\right)=\mathrm{Tr}_{\mathrm{photon}}\left\{ \left|\psi\left(t\right)\right\rangle \left\langle \psi\left(t\right)\right|\right\} $.
The reduced density matrix element can be evaluated by
\begin{equation}
\begin{split}\rho_{ij}\left(t\right)= & \left\langle i;0\left|\psi\left(t\right)\right\rangle \right.\left.\left\langle \psi\left(t\right)\right|j;0\right\rangle +\\
 & \sum_{\mathrm{k}}\left\langle i;\mathrm{k}\left|\psi\left(t\right)\right\rangle \right.\left.\left\langle \psi\left(t\right)\right|j;\mathrm{k}\right\rangle .
\end{split}
\end{equation}
As must be the case, the population of the excited state decays as
\begin{equation}
\rho_{11}\left(t\right)=\left|C_{10}\left(t\right)\right|^{2}=\left|C_{1}\right|^{2}e^{-\kappa t},\label{eq:WW-population-decay}
\end{equation}
and the coherence (the off-diagonal element) is 
\begin{equation}
\rho_{01}\left(t\right)=C_{00}\left(t\right)C_{10}^{*}\left(t\right)=C_{0}C_{1}^{*}e^{i\Omega t-\frac{\kappa}{2}t}.\label{eq:WW-coherence-decay}
\end{equation}
Here, since we do note include pure dephasing, the total dephasing
rate of the system is half of the population decay rate ($\frac{\kappa}{2}$).
The purity of electronic quantum state is a scalar defined as

\begin{equation}
\eta=\mathrm{Tr}\left\{ \rho^{2}\right\} =1-2\left|C_{1}\right|^{4}\left(e^{-\kappa t}-e^{-2\kappa t}\right).\label{eq:WW-purity}
\end{equation}

\subsection{Radiation Field Observables in 1D}

While Eq.~(\ref{eq:WW-population-decay}) expresses the standard
FGR decay of the electronic excited state, in Sec.~\ref{sec:Results}
our primary interest is in the dynamics of the EM field. To that end,
we now calculate the expectation value of the radiation intensity
$\left\langle \widehat{\mathbf{E}}_{\perp}\left(x,t\right)^{2}\right\rangle $
and the observed electric field $\left\langle \widehat{\mathbf{E}}_{\perp}\left(x,t\right)\right\rangle $
using the electric field operator (Eq.~(\ref{eq:E_operator})) for
a 1D system. Eq.~(\ref{eq:E_operator}) suggests that the $\left\{ \left|0;\mathrm{k}\right\rangle \right\} $
manifold is coupled to the $\left|0;0\right\rangle $ state and the
$\left\{ \left|1;\mathrm{k}\right\rangle \right\} $ manifold is coupled
to the $\left|1;0\right\rangle $ state. Since $C_{1\mathrm{k}}\left(t\right)\approx0$,
the expectation value can be expressed as
\begin{equation}
\left\langle \widehat{\mathbf{E}}_{\perp}\left(x,t\right)\right\rangle =\sum_{\mathrm{k}}i{\cal E}_{\mathrm{k}}e^{ikx}C_{00}^{*}\left(t\right)C_{0\mathrm{k}}\left(t\right)+c.c.
\end{equation}
where ${\cal E}_{\mathrm{k}}=\sqrt{\frac{\hbar\omega_{\mathrm{k}}}{2\epsilon_{0}L^{n}}}$.
By plugging in the density of states for a 1D system, we have 
\begin{equation}
\begin{split}\left\langle \widehat{\mathbf{E}}_{\perp}\left(x,t\right)\right\rangle = & C_{0}^{*}C_{1}\frac{\mu_{01}}{4\pi\epsilon_{0}c}\int d\omega\frac{\omega}{\omega-\Omega+i\frac{\kappa}{2}}\times\\
 & \left\{ e^{-i\Omega t-\frac{\kappa}{2}t+i\omega x/c}-e^{-i\omega t+i\omega x/c}\right\} +c.c.
\end{split}
\end{equation}
Then we use a Cauchy integral to carry out the integration over $\omega$
\begin{equation}
\begin{split}\left\langle \widehat{\mathbf{E}}_{\perp}\left(x,t\right)\right\rangle  & =\left|C_{0}\right|\left|C_{1}\right|\frac{\mu_{01}}{c\epsilon_{0}}e^{-\frac{\kappa}{2}\left(t-\frac{\left|x\right|}{c}\right)}\theta\left(ct-\left|x\right|\right)\times\\
 & \left\{ \Omega\sin\Omega\left(t-\frac{\left|x\right|}{c}\right)+\frac{\kappa}{2}\cos\Omega\left(t-\frac{\left|x\right|}{c}\right)\right\} 
\end{split}
\end{equation}
where the step function $\theta$ appears because of the Cauchy integral
and we will drop the $\frac{\kappa}{2}$ term since $\kappa\ll\Omega$.
Therefore, we obtain the expectation value of the electric field in
a 1D system as
\begin{equation}
\left\langle \widehat{\mathbf{E}}_{\perp}\left(x,t\right)\right\rangle =\left|C_{0}\right|\left|C_{1}\right|\times R\left(x,t\right)\sin\Omega\left(t-\frac{\left|x\right|}{c}\right)
\end{equation}
where the spatial distribution function is given by 
\begin{equation}
R\left(x,t\right)=\frac{\Omega\mu_{01}}{c\epsilon_{0}}e^{-\frac{\kappa}{2}\left(t-\frac{\left|x\right|}{c}\right)}\times\theta\left(ct-\left|x\right|\right).
\end{equation}
For a given time $t$, we find that $\left\langle \widehat{\mathbf{E}}_{\perp}\left(x,t\right)\right\rangle $
oscillates in space at frequency $\Omega/c$ and the event horizon
can be observed at $\left|x\right|=ct$. The magnitude of the electric
field can be estimated by $\left\langle \widehat{\mathbf{E}}_{\perp}\left(x,t\right)\right\rangle ^{2}.$
If we calculate a coarse-grained average over a short time $\tau$,
satisfying $2\pi/\Omega\ll\tau\ll1/\kappa$, we obtain 
\begin{eqnarray}
\overline{\left\langle \widehat{\mathbf{E}}_{\perp}\left(x,t\right)\right\rangle ^{2}} & = & \frac{1}{\tau}\int_{t}^{t+\tau}dt^{\prime}\left\langle \widehat{\mathbf{E}}_{\perp}\left(x,t\right)\right\rangle ^{2}\label{eq:WW-<E>2}\\
 & = & \left|C_{0}\right|^{2}\left|C_{1}\right|^{2}\times\frac{R\left(x,t\right)^{2}}{2},
\end{eqnarray}
In Eq.~(\ref{eq:WW-<E>2}), we have approximated $\overline{\sin^{2}\Omega t}\approx\frac{1}{2}$.
Within the time scale $\tau$, the population does not change much
and the coherence is just a rapid oscillation.

Beyond $\left\langle \widehat{\mathbf{E}}_{\perp}\right\rangle ^{2}$,
it is standard to evaluate $\left\langle \widehat{\mathbf{E}}_{\perp}^{2}\right\rangle $,
so as to better understand the nature of the quantum fluctuations
of the EM field. According to Eq.~(\ref{eq:E_operator}), the $\widehat{\mathbf{E}}_{\perp}^{2}$
operator includes couplings only within the manifolds $\left\{ \left|0;\mathrm{k}\right\rangle \right\} $
and $\left\{ \left|1;\mathrm{k}\right\rangle \right\} $. Since $\left\{ \left|1;\mathrm{k}\right\rangle \right\} $
is the off-resonant manifold, we will ignore this contribution. Therefore,
following the same procedure as above, we can obtain the expectation
value for the radiation intensity by

\begin{equation}
\left\langle \widehat{\mathbf{E}}_{\perp}^{2}\left(x,t\right)\right\rangle =2\sum_{k,k^{\prime}}\mathcal{E}_{k}\mathcal{E}_{k^{\prime}}\cos\left[\left(k-k^{\prime}\right)x\right]C_{0\mathrm{k}}^{*}\left(t\right)C_{0\mathrm{k}^{\prime}}\left(t\right)
\end{equation}
where we ignore the vacuum fluctuations of the radiation field. We
then calculate a coarse-grained average over a short time $\tau$,
\begin{equation}
\overline{\left\langle \widehat{\mathbf{E}}_{\perp}^{2}\left(x,t\right)\right\rangle }=\left|C_{1}\right|^{2}\times\frac{R\left(x,t\right)^{2}}{2}.\label{eq:Intensity}
\end{equation}
Note that the equation
\begin{equation}
\left\langle \widehat{\mathbf{E}}_{\perp}\left(x,t\right)\right\rangle ^{2}=\left|C_{0}\right|^{2}\left\langle \widehat{\mathbf{E}}_{\perp}^{2}\left(x,t\right)\right\rangle 
\end{equation}
establishes a simple relationship between $\left\langle \widehat{\mathbf{E}}_{\perp}^{2}\right\rangle $
and $\left\langle \widehat{\mathbf{E}}_{\perp}\right\rangle ^{2}$.

\section{Derivation of the electric dipole coupling in Ehrenfest dynamics\label{sec:Instantaneous-Decay-rate}}

To derive the electric dipole coupling of the semiclassical electronic
Hamiltonian (Eq.~(\ref{eq:electronic_Hamiltonian})), we need a solution
to Maxwell's equation Eqs.~(\ref{eq:maxwell_BE}\textendash \ref{eq:maxwell_EB})
with the source given by the average polarization and the average
current (Eq.~(\ref{eq:polarization_mean})). Here, we will consider
a polarization distribution idealized as a delta function at the origin
and derive the electric dipole coupling within Ehrenfest dynamics.

In a 3D system, Jefimenko's equations give a general expression for
the classical EM field due to an arbitrary charge and current density,
taking into account the retardation of the field. The retarded electric
field in the frequency domain is given by \citep{mcdonald_relation_1997,panofsky_classical_2005}
\begin{equation}
\mathbf{E}_{\omega}\left(\mathbf{r}\right)=\frac{1}{4\pi\epsilon_{0}}\int\mathrm{d}v^{\prime}e^{iks}\left\{ \frac{\rho_{\omega}^{\prime}\hat{\mathbf{s}}}{s^{2}}-ik\frac{\rho_{\omega}^{\prime}\hat{\mathbf{s}}}{s}+ik\frac{\mathbf{J}_{\omega}^{\prime}}{cs}\right\} 
\end{equation}
where $\mathbf{s}=\mathbf{r}-\mathbf{r}^{\prime}$, $s=\left|\mathbf{r}-\mathbf{r}^{\prime}\right|$,
$\hat{\mathbf{s}}=\mathbf{s}/s$, and $\omega=ck$. Here, we denote
the Fourier transform of a time-dependent function $f\left(t\right)$
as $f_{\omega}=\frac{1}{2\pi}\int f\left(t\right)e^{i\omega t}\mathrm{d}t$
for convenience. According to the definition of bound charge ($\rho=-\boldsymbol{\nabla}\cdot\mathbf{P}$)
and the continuity equation ($\dot{\rho}+\boldsymbol{\nabla}\cdot\mathbf{J}=0$,
transformed to Fourier space as $-i\omega\rho_{\omega}+\boldsymbol{\nabla}\cdot\mathbf{J}_{\omega}=0$),
the retarded field can be written as
\begin{equation}
\begin{split}\mathbf{E}_{\omega}\left(\mathbf{r}\right)= & \frac{1}{4\pi\epsilon_{0}}\int\mathrm{d}v^{\prime}e^{iks}\times\\
 & \left\{ -\frac{\boldsymbol{\nabla}^{\prime}\cdot\mathbf{P}_{\omega}\left(\mathbf{r}^{\prime}\right)}{s^{2}}\hat{\mathbf{s}}-\frac{\boldsymbol{\nabla}^{\prime}\cdot\mathbf{J}_{\omega}\left(\mathbf{r}^{\prime}\right)}{cs}\hat{\mathbf{s}}+\frac{ik\mathbf{J}_{\omega}\left(\mathbf{r}^{\prime}\right)}{cs}\right\} .
\end{split}
\end{equation}
Now, given the polarization operator $\widehat{\mathbf{P}}\left(\mathbf{r}\right)=\boldsymbol{\xi}\left(\mathbf{r}\right)\left(\left|0\right\rangle \left\langle 1\right|+\left|1\right\rangle \left\langle 0\right|\right)$,
the average polarization ($\mathbf{P}\left(\mathbf{r},t\right)=\text{Tr}_{s}\left\{ \widehat{\rho}\left(t\right)\widehat{\mathbf{P}}\left(\mathbf{r}\right)\right\} $)
can be expressed in the frequency domain as 
\begin{equation}
\mathbf{P}_{\omega}\left(\mathbf{r}\right)=2{\cal R}_{\omega}\boldsymbol{\xi}\left(\mathbf{r}\right),
\end{equation}
where we define ${\cal R}_{\omega}=\left(\mathrm{Re}\rho_{01}\right)_{\omega}$.
The average current ($\mathbf{J}\left(\mathbf{r},t\right)=\frac{\partial}{\partial t}\mathbf{P}\left(\mathbf{r},t\right)$)
can be obtained by taking the time derivative of $\mathbf{P}\left(\mathbf{r},t\right)=\int\mathbf{P}_{\omega}\left(\mathbf{r}\right)e^{-i\omega t}\mathrm{d}\omega$:
\begin{equation}
\mathbf{J}\left(\mathbf{r},t\right)=\int-i\omega\mathbf{P}_{\omega}\left(\mathbf{r}\right)e^{-i\omega t}\mathrm{d}\omega,
\end{equation}
or, in Fourier space,
\begin{equation}
\mathbf{J}_{\omega}\left(\mathbf{r}\right)=-i2\omega{\cal R}_{\omega}\boldsymbol{\xi}\left(\mathbf{r}\right).
\end{equation}
Alternatively, according to Liouville equation for the reduced density
matrix $\widehat{\rho}\left(t\right)$ (Eq.~(\ref{eq:liouville_rho})),
the average current can be expressed in terms of 
\begin{equation}
\mathbf{J}_{\omega}\left(\mathbf{r}\right)=-2\Omega{\cal I}_{\omega}\boldsymbol{\xi}\left(\mathbf{r}\right)
\end{equation}
where ${\cal I}_{\omega}=\left(\mathrm{Im}\rho_{01}\right)_{\omega}$.

We would like to calculate the electric dipole coupling:
\begin{equation}
H_{01}^{\mathrm{el}}\left(t\right)=-\int\mathrm{d}\omega e^{-i\omega t}\int\mathrm{d}v\mathbf{E}_{\omega}\left(\mathbf{r}\right)\cdot\boldsymbol{\xi}\left(\mathbf{r}\right)\label{eq:ehr_coupling_fft}
\end{equation}
where the spatial integration is \begin{widetext}
\begin{equation}
\int\mathrm{d}v\mathbf{E}_{\omega}\left(\mathbf{r}\right)\cdot\boldsymbol{\xi}\left(\mathbf{r}\right)=\frac{{\cal R}_{\omega}}{2\pi\epsilon_{0}}\int\mathrm{d}v\int\mathrm{d}v^{\prime}e^{iks}\left\{ -\frac{\boldsymbol{\nabla}^{\prime}\cdot\boldsymbol{\xi}\left(\mathbf{r}^{\prime}\right)}{s^{2}}\xi_{s}\left(\mathbf{r}\right)+i\omega\frac{\boldsymbol{\nabla}^{\prime}\cdot\boldsymbol{\xi}\left(\mathbf{r}^{\prime}\right)}{cs}\xi_{s}\left(\mathbf{r}\right)+\frac{\omega^{2}\boldsymbol{\xi}\left(\mathbf{r}\right)\cdot\boldsymbol{\xi}\left(\mathbf{r}^{\prime}\right)}{c^{2}s}\right\} ,\label{eq:Jefimenko-integral-1}
\end{equation}
and $\xi_{s}\left(\mathbf{r}\right)=\boldsymbol{\xi}\left(\mathbf{r}\right)\cdot\hat{\mathbf{s}}$.
The spatial integration can be carried out using integration by parts
and eliminating boundary contributions:
\begin{eqnarray*}
-\int\mathrm{d}v^{\prime}e^{iks}\frac{\boldsymbol{\nabla}^{\prime}\cdot\boldsymbol{\xi}\left(\mathbf{r}^{\prime}\right)}{s^{2}}\xi_{s}\left(\mathbf{r}\right) & = & \int\mathrm{d}v^{\prime}\boldsymbol{\xi}\left(\mathbf{r}^{\prime}\right)\cdot\boldsymbol{\nabla}^{\prime}\frac{\xi_{s}\left(\mathbf{r}\right)e^{iks}}{s^{2}}\\
 & = & \int\mathrm{d}v^{\prime}e^{iks}\left[-\frac{ik}{s^{2}}\xi_{s}\left(\mathbf{r}\right)\xi_{s}\left(\mathbf{r}^{\prime}\right)+\frac{2}{s^{3}}\xi_{s}\left(\mathbf{r}\right)\xi_{s}\left(\mathbf{r}^{\prime}\right)+\frac{1}{s^{2}}\left(\boldsymbol{\xi}\left(\mathbf{r}^{\prime}\right)\cdot\boldsymbol{\nabla}^{\prime}\right)\xi_{s}\left(\mathbf{r}\right)\right]
\end{eqnarray*}
\begin{eqnarray*}
\int\mathrm{d}v^{\prime}e^{iks}i\omega\frac{\boldsymbol{\nabla}^{\prime}\cdot\boldsymbol{\xi}\left(\mathbf{r}^{\prime}\right)}{cs}\xi_{s}\left(\mathbf{r}\right) & = & -\int\mathrm{d}v^{\prime}ik\boldsymbol{\xi}\left(\mathbf{r}^{\prime}\right)\cdot\boldsymbol{\nabla}^{\prime}\frac{\xi_{s}\left(\mathbf{r}\right)e^{iks}}{s}\\
 & = & \int\mathrm{d}v^{\prime}e^{iks}\left[-\frac{k^{2}}{s}\xi_{s}\left(\mathbf{r}\right)\xi_{s}\left(\mathbf{r}^{\prime}\right)-\frac{ik}{s^{2}}\xi_{s}\left(\mathbf{r}\right)\xi_{s}\left(\mathbf{r}^{\prime}\right)-\frac{ik}{s}\left(\boldsymbol{\xi}\left(\mathbf{r}^{\prime}\right)\cdot\boldsymbol{\nabla}^{\prime}\right)\xi_{s}\left(\mathbf{r}\right)\right]
\end{eqnarray*}
Here, we have used the identity $\boldsymbol{\nabla}^{\prime}s=-\hat{\mathbf{s}}$.
Now, Eq.~(\ref{eq:Jefimenko-integral-1}) becomes
\begin{equation}
\begin{split}\int\mathrm{d}v\mathbf{E}_{\omega}\left(\mathbf{r}\right)\cdot\boldsymbol{\xi}\left(\mathbf{r}\right)= & \frac{{\cal R}_{\omega}}{2\pi\epsilon_{0}}\int\mathrm{d}v\int\mathrm{d}v^{\prime}e^{iks}\\
 & \left\{ \left(-2\frac{ik}{s^{2}}+\frac{2}{s^{3}}-\frac{k^{2}}{s}\right)\xi_{s}\left(\mathbf{r}\right)\xi_{s}\left(\mathbf{r}^{\prime}\right)+\left(\frac{1}{s^{2}}-\frac{ik}{s}\right)\left(\boldsymbol{\xi}\left(\mathbf{r}^{\prime}\right)\cdot\boldsymbol{\nabla}^{\prime}\right)\xi_{s}\left(\mathbf{r}\right)+\frac{k^{2}}{s}\boldsymbol{\xi}\left(\mathbf{r}\right)\cdot\boldsymbol{\xi}\left(\mathbf{r}^{\prime}\right)\right\} .
\end{split}
\label{eq:Jefimenko-integral-2}
\end{equation}
Explicitly, in Cartesian coordinates, let $\mathbf{s}=s_{x}\hat{\mathbf{x}}+s_{y}\hat{\mathbf{y}}+s_{z}\hat{\mathbf{z}}$,
so we can evaluate 
\begin{equation}
\left(\boldsymbol{\xi}\left(\mathbf{r}^{\prime}\right)\cdot\boldsymbol{\nabla}^{\prime}\right)\xi_{s}\left(\mathbf{r}\right)=-\frac{1}{s}\boldsymbol{\xi}\left(\mathbf{r}\right)\cdot\boldsymbol{\xi}\left(\mathbf{r}^{\prime}\right)+\frac{1}{s^{3}}\left[\xi_{x}\left(\mathbf{r}\right)\xi_{x}\left(\mathbf{r}^{\prime}\right)s_{x}^{2}+\xi_{y}\left(\mathbf{r}\right)\xi_{y}\left(\mathbf{r}^{\prime}\right)s_{y}^{2}+\xi_{z}\left(\mathbf{r}\right)\xi_{z}\left(\mathbf{r}^{\prime}\right)s_{z}^{2}\right]\label{eq:Jefimenko-identity-1}
\end{equation}
\end{widetext}

Let us now assume that the source distribution is a delta function
at the origin without dependence on either $\theta$ or $\phi$, and
polarized in the $z$ direction:
\begin{equation}
\boldsymbol{\xi}\left(\mathbf{r}^{\prime}\right)=\mu_{01}\delta^{3}\left(r^{\prime}\right)\hat{\mathbf{z}},
\end{equation}
where $\delta^{3}\left(r^{\prime}\right)$ is a 3D delta function
and $r^{\prime}=\left|\mathbf{r}^{\prime}\right|$. Because we integrate
over $\mathbf{r}$ and $\mathbf{r}^{\prime}$ in Eq.~(\ref{eq:Jefimenko-integral-2}),
we need only consider $\mathbf{r}\approx\mathbf{r}^{\prime}\approx0$
in the above integral, and so we can approximate
\begin{equation}
\boldsymbol{\xi}\left(\mathbf{r}\right)=\mu_{01}\delta^{3}\left(\left|\mathbf{r}^{\prime}+\mathbf{s}\right|\right)\hat{\mathbf{z}}\approx\mu_{01}\delta^{3}\left(\left|\mathbf{s}\right|\right)\hat{\mathbf{z}}=\mu_{01}\delta^{3}\left(s\right)\hat{\mathbf{z}}.
\end{equation}
Now we transform the integral by $\int\mathrm{d}v\int\mathrm{d}v^{\prime}\rightarrow\int\mathrm{d}v^{\prime}\int\mathrm{d}s\mathrm{d}\theta\mathrm{d}\phi s^{2}\sin\theta$
and use , 
\begin{eqnarray}
\xi_{s}\left(\mathbf{r}\right) & \approx & \mu_{01}\delta^{3}\left(s\right)\cos\theta,\\
\xi_{s}\left(\mathbf{r}^{\prime}\right) & = & \mu_{01}\delta^{3}\left(r^{\prime}\right)\cos\theta,
\end{eqnarray}
and by Eq.~(\ref{eq:Jefimenko-identity-1})
\begin{equation}
\left(\boldsymbol{\xi}\left(\mathbf{r}^{\prime}\right)\cdot\boldsymbol{\nabla}^{\prime}\right)\xi_{s}\left(\mathbf{r}\right)\approx-\mu_{01}^{2}\delta^{3}\left(r^{\prime}\right)\delta^{3}\left(s\right)\frac{\sin^{2}\theta}{s}.
\end{equation}
Then Eq.~(\ref{eq:Jefimenko-integral-2}) turns into\begin{widetext}
\begin{equation}
\begin{split}\int\mathrm{d}v\mathbf{E}_{\omega}\left(\mathbf{r}\right)\cdot\boldsymbol{\xi}\left(\mathbf{r}\right)= & \frac{\mu_{01}^{2}{\cal R}_{\omega}}{2\pi\epsilon_{0}}\int\mathrm{d}v^{\prime}\int\mathrm{d}s\mathrm{d}\theta\mathrm{d}\phi s^{2}\sin\theta e^{iks}\delta^{3}\left(r^{\prime}\right)\delta^{3}\left(s\right)\\
 & \left\{ \left(-\frac{ik}{s^{2}}+\frac{2}{s^{3}}-\frac{k^{2}}{s}-\frac{ik}{s^{2}}\right)\cos^{2}\theta-\left(\frac{1}{s^{3}}-\frac{ik}{s^{2}}\right)\sin^{2}\theta+\frac{k^{2}}{s}\right\} .
\end{split}
\end{equation}
\end{widetext}Now we transform the 3D $\delta$-function to a 1D
$\delta$-function: $\delta^{3}\left(s\right)=\frac{1}{2\pi s^{2}}\delta\left(s\right)$,
and use $\int\mathrm{d}\mathbf{r}^{\prime}\delta^{3}\left(r^{\prime}\right)=1$.
After carrying out the $\theta$ and $\phi$ integration in spherical
coordinates using $\int_{0}^{\pi}\mathrm{d}\theta\sin\theta\cos^{2}\theta=\frac{2}{3}$,
$\int_{0}^{\pi}\mathrm{d}\theta\sin^{3}\theta=\frac{4}{3}$, and $\int_{0}^{\pi}\mathrm{d}\theta\sin\theta=2$,
we obtain
\begin{equation}
\int\mathrm{d}v\mathbf{E}_{\omega}\left(\mathbf{r}\right)\cdot\boldsymbol{\xi}\left(\mathbf{r}\right)=\frac{2\mu_{01}^{2}{\cal R}_{\omega}k^{2}}{3\pi\epsilon_{0}}\int_{0}^{\infty}ds\delta\left(s\right)\frac{e^{iks}}{s}\label{eq:ehr_coupling_1/s}
\end{equation}
where all of the $1/s^{2}$ and $1/s^{3}$ terms cancel. The radial
integration of Eq.~(\ref{eq:ehr_coupling_1/s}) gives
\begin{eqnarray}
\int_{0}^{\infty}ds\delta\left(s\right)\frac{e^{iks}}{s} & = & \int_{0}^{\infty}ds\delta\left(s\right)\left(\frac{\cos ks}{s}+i\frac{\sin ks}{s}\right),\nonumber \\
 & = & \lim_{\eta\rightarrow0}\frac{1}{\eta}+i\frac{k}{2}\label{eq:radial_integration}
\end{eqnarray}
where the real part of the integral is infinite but does not depend
on $k$. When plugging into Eq.~(\ref{eq:ehr_coupling_fft}), this
real part turns out to be $\lim_{\eta\rightarrow0}\frac{1}{\eta}\delta\left(t\right)$
which represents a self-interaction at $t=0$, and will be ignored. 

At this point, we can plug Eqs.~(\ref{eq:ehr_coupling_1/s}) and
(\ref{eq:radial_integration}) into Eq.~(\ref{eq:ehr_coupling_fft})
and use $ik^{3}{\cal R}_{\omega}=\dddot{{\cal R}}_{\omega}/c^{3}$
to obtain the electric dipole coupling
\begin{equation}
H_{01}^{\mathrm{el}}\left(t\right)=-\frac{\mu_{01}^{2}}{3\pi\epsilon_{0}c^{3}}\dddot{{\cal R}}\left(t\right)
\end{equation}
The presence of a \emph{third} derivative of $\mathrm{Re}\rho_{01}\left(t\right)$
is reminiscent of the Abraham\textendash Lorentz force in classical
electrodynamics.\citep{griffiths_introduction_2014} Finally, we
approximate $\dddot{{\cal R}}\approx\Omega^{3}{\cal I}$, and conclude
\begin{equation}
H_{01}^{\mathrm{el}}\left(t\right)=-\frac{\mu_{01}^{2}\Omega^{3}}{3\pi\epsilon_{0}c^{3}}{\cal I}\left(t\right)=-\hbar\kappa^{\text{3D}}\mathrm{Im}\rho_{01}\left(t\right).
\end{equation}

\section{The direction of the rescaling field\label{sec:Justify-the-rescaling}}

\subsection{The 3D case}

Here, we provide numerical proof that $\delta\mathbf{E}_{R}=\boldsymbol{\nabla}\times\boldsymbol{\nabla}\times\text{\ensuremath{\mathbf{P}}}$
and $\delta\mathbf{B}_{R}=-\boldsymbol{\nabla}\times\text{\ensuremath{\mathbf{P}}}$
are reasonable rescaling directions for spontaneous emission. To do
so, we run Ehrenfest dynamics for the 3D system in Sec.~\ref{sec:Results}.
We calculate the overlap of the Ehrenfest EM field arising from the
origin (where$\mathbf{P}^{\text{3D}}\neq0$) with $\boldsymbol{\nabla}\times\boldsymbol{\nabla}\times\mathbf{P}^{\text{3D}}$and
$-\boldsymbol{\nabla}\times\mathbf{P}^{\text{3D}}$. To be precise,
consider a spherical shell outside of the region of $\mathbf{P}^{\text{3D}}\left(\mathbf{r}\right)$.
We calculate the normalized overlap estimation in this region defined
as 
\begin{equation}
\left(\mathbf{E}_{Eh}|\delta\mathbf{E}_{R}\right)=\frac{\int_{\circledcirc}\mathrm{d}v\mathbf{E}_{Eh}\cdot\delta\mathbf{E}_{R}}{\sqrt{\int_{\circledcirc}\mathrm{d}v\left|\mathbf{E}_{Eh}\right|^{2}\int_{\circledcirc}\mathrm{d}v\left|\delta\mathbf{E}_{R}\right|^{2}}}
\end{equation}
where $\int_{\circledcirc}\mathrm{d}v$ denote the integral within
the spherical shell. If our intuition is correct, the overlap should
be large and oscillatory as the emanated wave propagates out into
free space.

\begin{figure}
\begin{centering}
\par\end{centering}
\begin{centering}
\includegraphics{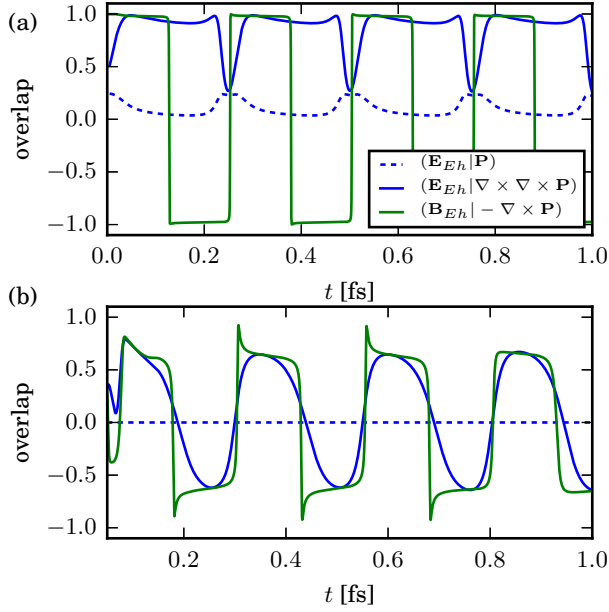}
\par\end{centering}
\begin{centering}
\par\end{centering}
\caption{The normalized overlap of the EM field of Ehrenfest dynamics a function
of time in 3D space. The initial state of Ehrenfest dynamics is $\left|\psi\right\rangle =\sqrt{\frac{1}{2}}\left|0\right\rangle +\sqrt{\frac{1}{2}}\left|1\right\rangle $.
The blue dashed lines are is the overlap of $\left(\mathbf{E}_{Eh}|\text{\ensuremath{\mathbf{P}}}\right)$,
the blue solid line is the overlap of $\left(\mathbf{E}_{Eh}|\boldsymbol{\nabla}\times\boldsymbol{\nabla}\times\text{\ensuremath{\mathbf{P}}}\right)$,
and the green line is the overlap of $\left(\mathbf{B}_{Eh}|-\boldsymbol{\nabla}\times\text{\ensuremath{\mathbf{P}}}\right)$.
The shell radius are (a) $6.0-7.5\ \text{nm}$ and (b) $30.0-31.5\ \text{nm}$.
Note the large overlap between $\mathbf{E}_{Eh}$ and $\mathbf{B}_{Eh}$
fields with $\boldsymbol{\nabla}\times\boldsymbol{\nabla}\times\text{\ensuremath{\mathbf{P}}}$
and $-\boldsymbol{\nabla}\times\text{\ensuremath{\mathbf{P}}}$. Overall
this data suggest that $\boldsymbol{\nabla}\times\boldsymbol{\nabla}\times\text{\ensuremath{\mathbf{P}}}$
and $-\boldsymbol{\nabla}\times\text{\ensuremath{\mathbf{P}}}$ should
be the leading order contributions to the rescaled $\mathbf{E}$ and
$\mathbf{B}$ fields respectively. \label{fig:overlap-self-interference }}
\end{figure}

In Fig.~(\ref{fig:overlap-self-interference }), we plot the normalized
overlap for short times. We consider a Gaussian distribution of width
about $3\ \text{nm}$. The overlap of magnetic fields exhibit an oscillatory
behavior in the near and far field. However, the overlap of electric
field shows similar behavior only in the far field. This distortion
is attributed to the fact that the electric field behaves in a more
complicated fashion in the near field. Despite this difference, we
find that, when the emission field begins to enter the vacuum ($t<0.05\ \text{fs}$),
$\left(\mathbf{E}_{Eh}|\left(\boldsymbol{\nabla}\times\right)^{2}\text{\ensuremath{\mathbf{P}}}^{\text{3D}}\right)$
and $\left(\mathbf{B}_{Eh}|-\boldsymbol{\nabla}\times\text{\ensuremath{\mathbf{P}}}^{\text{3D}}\right)$
account for more than $90\%$ of the emission field in the near field.
Thus, this data then strongly suggests that the leading order contributions
to the rescaling field should in fact be in the direction of $\delta\mathbf{E}_{R}=\left(\boldsymbol{\nabla}\times\right)^{2}\text{\ensuremath{\mathbf{P}}}^{\text{3D}}$
for the electric field and $\delta\mathbf{B}_{R}=-\boldsymbol{\nabla}\times\text{\ensuremath{\mathbf{P}}}^{\text{3D}}$
for the magnetic field. 

\subsection{The 1D case}

Interestingly, the analysis above is less straightforward in 1D. Here
we consider a polarization distribution given by Eq.~(\ref{eq:P_1D})
and the width of Gaussian distribution is assumed to be much smaller
than the wavelength ($\frac{1}{\sqrt{a}}\ll\frac{2\pi c}{\Omega}$).
Compared against the 3D case, $\boldsymbol{\nabla}\times\boldsymbol{\nabla}\times\mathbf{P}^{\text{1D}}$
and $-\boldsymbol{\nabla}\times\mathbf{P}^{\text{1D}}$ overlap strongly
with $\mathbf{P}^{\text{1D}}$ and this overlap cannot be ignored.\cite{Note3}
For instance, for a 1D system, this overlap can lead to unwanted EM
fields propagating back to the origin. 

To circumvent this issue, we can simply add additional transverse
fields\footnote{Note that, in 1D, $\mathbf{P}^{\text{1D}}$ is always transverse.}
to the rescaling field:
\begin{eqnarray}
\delta\mathbf{E}_{R} & = & \boldsymbol{\nabla}\times\boldsymbol{\nabla}\times\mathbf{P}^{\text{1D}}-g\mathbf{P}^{\text{1D}},\label{eq:deltaE_R_1D}\\
\delta\mathbf{B}_{R} & = & -\boldsymbol{\nabla}\times\mathbf{P}^{\text{1D}}-h\left(\boldsymbol{\nabla}\times\right)^{3}\mathbf{P}^{\text{1D}},\label{eq:deltaB_R_1D}
\end{eqnarray}
where the coefficients $g$ and $h$ are determined by 
\begin{eqnarray}
\delta\mathbf{E}_{R}\left(x=0\right) & = & 0,\label{eq:g-condition}\\
\boldsymbol{\nabla}\times\delta\mathbf{B}_{R}\left(x=0\right) & = & 0.\label{eq:h-condition}
\end{eqnarray}
In the end, using Eqs.~(\ref{eq:g-condition}) and (\ref{eq:h-condition}),
we find $g=2a$ and $h=1/6a$ and the rescaling field is
\begin{eqnarray}
\delta\mathbf{E}_{R}\left(x\right) & = & -\mu_{01}\sqrt{\frac{a}{\pi}}4a^{2}x^{2}e^{-ax^{2}}\hat{\mathbf{z}},\label{eq:dE_1D}\\
\delta\mathbf{B}_{R}\left(x\right) & = & \mu_{01}\sqrt{\frac{a}{\pi}}\frac{4}{3}a^{2}x^{3}e^{-ax^{2}}\hat{\mathbf{y}}.\label{eq:dB_1D}
\end{eqnarray}
Note that all $e^{-ax^{2}}$ and $xe^{-ax^{2}}$ terms have been canceled
out by our choice of $g$ and $h$.

\section{Derivation of the rescaling factors $\alpha^{\ell}$ and $\beta^{\ell}$\label{sec:Derivation-of-the-rescaling}}

Here we discuss the details of EM field rescaling and energy conservation. 

\subsection{Each trajectory cannot conserve energy}

In an ideal world, one would like to enforce energy conservation for
every trajectory, much in the same way as Tully's FSSH algorithm operates.\citep{tully_molecular_1990,tully_perspective:_2012}
Thus, every time an electron is forced to relax, one would like to
insert a corresponding increase in the energy of the EM field so as
to satisfy conservation of energy:
\begin{equation}
\begin{split}\delta U_{R} & =\frac{\epsilon_{0}}{2}\int\mathrm{d}v\left(2\mathbf{E}_{Eh}\cdot\alpha\delta\mathbf{E}_{R}+\left|\alpha\delta\mathbf{E}_{R}\right|^{2}\right)\\
 & +\frac{1}{2\mu_{0}}\int\mathrm{d}v\left(2\mathbf{B}_{Eh}\cdot\beta\delta\mathbf{B}_{R}+\left|\beta\delta\mathbf{B}_{R}\right|^{2}\right).
\end{split}
\label{eq:all_energy_conservation}
\end{equation}
And given requirement (c) in Sec.~\ref{subsec:The-Classical-EM},
Eq.~(\ref{eq:all_energy_conservation}) implies two independent quadratic
equations: 
\begin{eqnarray}
\frac{\delta U_{R}}{2} & = & \frac{\epsilon_{0}}{2}\int\mathrm{d}v\left(2\mathbf{E}_{Eh}\cdot\alpha\delta\mathbf{E}_{R}+\left|\alpha\delta\mathbf{E}_{R}\right|^{2}\right),\label{eq:micro_energy_conserve_E}\\
 & = & \frac{1}{2\mu_{0}}\int\mathrm{d}v\left(2\mathbf{B}_{Eh}\cdot\beta\delta\mathbf{B}_{R}+\left|\beta\delta\mathbf{B}_{R}\right|^{2}\right).\label{eq:micro_energy_conserve_B}
\end{eqnarray}
Now, if $\alpha$ and $\beta$ are chosen to have well-defined signs
(e.g. in Tully's FSSH model, the sign for velocity rescaling is chosen
to minimize the change of momentum), we will necessarily find that
$\left\langle \mathbf{E}^{2}\right\rangle =\left\langle \mathbf{E}\right\rangle ^{2}$
and $\left\langle \mathbf{B}^{2}\right\rangle =\left\langle \mathbf{B}\right\rangle ^{2}$\textemdash{}
which we know to be incorrect (see Appendix~\ref{sec:Weisskopf=002013Wigner}).
Thus, it is inevitable that either we sample trajectories over which
$\alpha$ and $\beta$ have different phases or that $\alpha$ and
$\beta$ are dynamically assigned random phases within one trajectory.
In the latter case, we will necessarily obtain large discontinuities
in the $E$ and $B$ fields and the wrong emission intensity. After
all, solving Eqs.~(\ref{eq:micro_energy_conserve_E}) and (\ref{eq:micro_energy_conserve_B})
for $\alpha$ and $\beta$ must lead to two solutions with opposite
sign since $\int\mathrm{d}v\left|\delta\mathbf{E}_{R}\right|^{2}>0$,
$\int\mathrm{d}v\left|\delta\mathbf{B}_{R}\right|^{2}>0$, and $\delta U_{R}>0$.
Thus, the only way forward is to sample over trajectories where $\alpha$
and $\beta$ have different phases.

Given that $\delta U_{R}$ can be defined with a random phase $\phi$
(see Eq.~(\ref{eq:k_R}) and Eq.~(\ref{eq:EnergyChange}))
\begin{equation}
\delta U_{R}=\Omega\kappa\rho_{11}\left(1-\rho_{00}\right)\text{Im}\left[\frac{\rho_{01}}{\left|\rho_{01}\right|}e^{i\phi}\right]^{2}dt,
\end{equation}
it would seem natural to apply the following sign convention: 
\begin{equation}
\text{sgn}\left(\alpha\right)=\text{sgn}\left(\beta\right)=\text{sgn}\left(\text{Im}\left[\rho_{01}e^{i\phi}\right]\right).\label{eq:sign_alpha}
\end{equation}
This convention can achieve two goals. First, it ensures that the
Poynting vector of the rescaled field will be usually outward, away
from the polarization. Second, it ensures that we will not introduce
any artificial frequency into the EM field (because $\rho_{01}$ is
rotating at frequency $\Omega$). Nevertheless, even with these two
points in its favor, this convention is still unworkable.

\begin{figure}
\centering{}\includegraphics{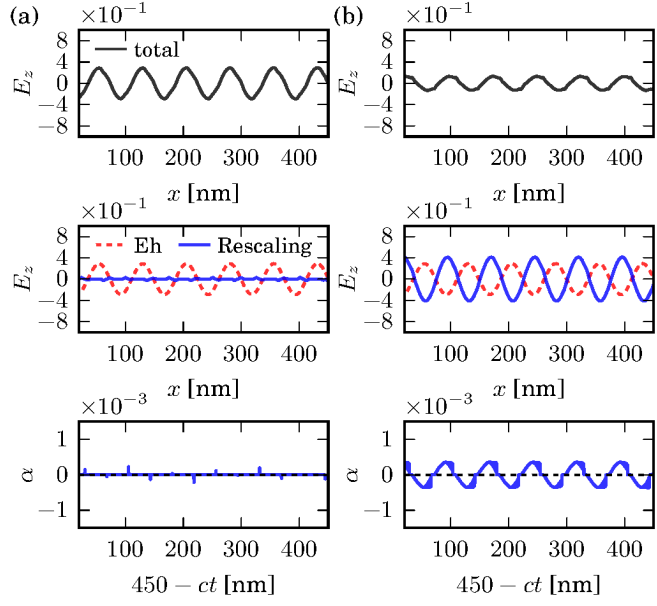}\caption{The hypothetical electric field that would result from enforcing energy
conservation for individual trajectories in a 1D system. We consider
two different phases of the rescaling field in two columns: (a) $\phi=0+\phi_{0}$
and (b) $\phi=\pi+\phi_{0}$, where $\phi_{0}=-\sqrt{3}\sigma\Omega/c=-0.13\pi$.
In the upper two panels of each column, we plot the total electric
field ($\mathbf{E}_{Eh}$), Ehrenfest component ($\mathbf{E}_{Eh}$)
and the rescaling component ($\alpha\delta\mathbf{E}_{R}$) at $t=1.5\text{ fs}$
as a function of $x$. The initial condition is $\rho_{11}=0.1$.
The dashed red lines are the Ehrenfest component ($\mathbf{E}_{Eh}$)
and the solid blue lines are the rescaling fields ($\alpha\delta\mathbf{E}_{R}$).
The lower panel of each column is the calculated $\alpha$ coefficient
along the trajectory as a function of $450-ct\ \text{nm}$ (as determined
by Eq.~(\ref{eq:micro_energy_conserve_E})). While the EM field looks
physical in (a), note that the rescaling field in (b) is completely
phase-inverted relative to the Ehrenfest component and cannot be physical.
In the end, applying energy conservation for each trajectory would
result in absurdly large changes in the EM field, even when spontaneous
emission should not be important. We believe this approach is not
reasonable for a semiclassical ansatz. \label{fig:electric-field-energy-conserved}}
\end{figure}

Consider the case where the initial electronic state is barely excited
($\rho_{11}=0.1$). In this case, Ehrenfest dynamics should be very
accurate and the effects of spontaneous emission should be very minor.
However, one will find bizarre behavior as a function of the random
phase $\phi$. On the one hand, if the rescaling field is in-phase
(i.e. $\phi=\phi_{0}$ in Fig.~\ref{fig:electric-field-energy-conserved}(a)),
we will find a slightly large, coherent outgoing electric field. On
the other hand, if the rescaling field is out of phase (e.g. $\phi=\phi_{0}+\pi$
in Fig.~\ref{fig:electric-field-energy-conserved}(b)), we will find
a large, completely inverted EM field. To understand why this inversion
is obviously unphysical, consider the extreme case where spontaneous
emission is very weak. How can a weak emission possibly lead to the
inversion of the entire EM field that was previously emitted long
ago? And to make things worse, how would this hypothetical approach
behave with an external incoming EM field; would that external EM
field also be inverted? Ultimately, averaging over a set of random
phases would not yield the correct total EM field. In this case, rescaling
the EM field leads to results that are qualitatively worse than no
correction at all. 

\sout{}

\subsection{An ensemble of trajectories can conserve energy}

In the end, our intuition is that one cannot capture the essence of
spontaneous emission by enforcing energy conservation for each trajectory;
instead, energy conservation can be enforced only on average. Note
that this ansatz agrees with a host of work modeling nuclear quantum
effects with interacting trajectories designed to reproduce the Wigner
distribution.\citep{donoso_quantum_2001,donoso_simulation_2003} For
the reader uncomfortable with this approach, we emphasize that true
spontaneous emission requires quantum (not classical) photons (bosons);
this is not the same problem as the FSSH problem, where one is dealing
with a classical nuclei (bosons). 

Now, in order to enforce energy conservation on average, imagine
that we run $N$ trajectories (indexed by $\ell$), and for each trajectory,
the EM field is written as the pure Ehrenfest EM field plus a sum
of $N_{\text{traj}}$ rescaling fields from each retarded time step
$jdt$:
\begin{eqnarray}
\mathbf{E}_{Eh+R}^{\ell}\left(t\right) & = & \mathbf{E}_{Eh}^{\ell}\left(t\right)+\sum_{j=0}^{n}\text{\ensuremath{\alpha_{j}^{\ell}}}\delta\mathbf{E}_{R}\left(t-jdt\right),\\
\mathbf{B}_{Eh+R}^{\ell}\left(t\right) & = & \mathbf{B}_{Eh}^{\ell}\left(t\right)+\sum_{j=0}^{n}\text{\ensuremath{\beta_{j}^{\ell}}}\delta\mathbf{B}_{R}\left(t-jdt\right).
\end{eqnarray}
Here $\delta\mathbf{E}_{R}\left(t-jdt\right)$ and $\delta\mathbf{B}_{R}\left(t-jdt\right)$
are the rescaling fields that were created at time $jdt$ and have
been propagated for a time $t-jdt$ according to Maxwell's equations.
For the last time step ($t=ndt$), energy conservation must satisfy
the following condition:
\begin{eqnarray}
\left\langle \delta U_{R}^{\ell}\right\rangle  & = & \frac{1}{N_{\text{traj}}^{2}}\sum_{\ell,\ell^{\prime}}\left\{ \frac{\epsilon_{0}}{2}\int\mathrm{d}v\mathbf{E}_{Eh}^{\ell}\left(t\right)\cdot\alpha_{n}^{\ell^{\prime}}\delta\mathbf{E}_{R}\right.\nonumber \\
 &  & +\frac{\epsilon_{0}}{2}\sum_{j=0}^{n-1}\int\mathrm{d}v\text{\ensuremath{\alpha_{j}^{\ell}}}\delta\mathbf{E}_{R}\left(t-jdt\right)\cdot\alpha_{n}^{\ell^{\prime}}\delta\mathbf{E}_{R}\nonumber \\
 &  & +\frac{\epsilon_{0}}{2}\int\mathrm{d}v\alpha_{n}^{\ell}\delta\mathbf{E}_{R}\cdot\alpha_{n}^{\ell^{\prime}}\delta\mathbf{E}_{R}\nonumber \\
 &  & +\frac{1}{2\mu_{0}}\int\mathrm{d}v\mathbf{B}_{Eh}^{\ell}\left(t\right)\cdot\beta_{n}^{\ell^{\prime}}\delta\mathbf{B}_{R}\nonumber \\
 &  & +\frac{1}{2\mu_{0}}\sum_{j=0}^{n-1}\int\mathrm{d}v\text{\ensuremath{\beta_{j}^{\ell}}}\delta\mathbf{B}_{R}\left(t-jdt\right)\cdot\beta_{n}^{\ell^{\prime}}\delta\mathbf{B}_{R}\nonumber \\
 &  & \left.+\frac{1}{2\mu_{0}}\int\mathrm{d}v\beta_{n}^{\ell}\delta\mathbf{B}_{R}\cdot\beta_{n}^{\ell^{\prime}}\delta\mathbf{B}_{R}\right\} .\label{eq:average_energy_conservation}
\end{eqnarray}
Now, let us assume that the phases of $\alpha_{n}^{\ell^{\prime}}$
and $\beta_{n}^{\ell^{\prime}}$ are random (i.e. we will enforce
Eq.~(\ref{eq:sign_alpha})), so that on average
\begin{equation}
\sum_{\ell,\ell^{\prime}}\mathbf{E}_{Eh}^{\ell}\left(t\right)\cdot\alpha_{n}^{\ell^{\prime}}\delta\mathbf{E}_{R}=\sum_{\ell,\ell^{\prime}}\mathbf{B}_{Eh}^{\ell}\left(t\right)\cdot\beta_{n}^{\ell^{\prime}}\delta\mathbf{B}_{R}=0.\label{eq:Er-R_cancellation}
\end{equation}
Furthermore there should also complete phase cancellation between
trajectories, e.g. for all $j$,
\begin{equation}
\begin{split}\sum_{\ell,\ell^{\prime}}\text{\ensuremath{\alpha_{j}^{\ell}}} & \delta\mathbf{E}_{R}\left(t-jdt\right)\cdot\alpha_{n}^{\ell^{\prime}}\delta\mathbf{E}_{R}\\
 & =2N_{\text{traj}}\sum_{\ell}\text{\ensuremath{\alpha_{j}^{\ell}}}\alpha_{n}^{\ell}\delta\mathbf{E}_{R}\left(t-jdt\right)\cdot\delta\mathbf{E}_{R},
\end{split}
\end{equation}
and
\begin{eqnarray}
\sum_{\ell,\ell^{\prime}}\alpha_{n}^{\ell}\delta\mathbf{E}_{R}\cdot\alpha_{n}^{\ell^{\prime}}\delta\mathbf{E}_{R} & = & N_{\text{traj}}\sum_{\ell}\left|\alpha_{n}^{\ell}\delta\mathbf{E}_{R}\right|^{2}.
\end{eqnarray}
Then, Eq.~(\ref{eq:average_energy_conservation}) becomes an equation
that must be enforced for each trajectory:
\begin{eqnarray}
\delta U_{R}^{\ell} & = & \frac{\epsilon_{0}}{2}\int\mathrm{d}v\sum_{j=0}^{n-1}2\text{\ensuremath{\alpha_{j}^{\ell}}}\delta\mathbf{E}_{R}\left(t-jdt\right)\cdot\alpha_{n}^{\ell}\delta\mathbf{E}_{R}\nonumber \\
 &  & +\frac{\epsilon_{0}}{2}\int\mathrm{d}v\left|\alpha_{n}^{\ell}\delta\mathbf{E}_{R}\right|^{2}\nonumber \\
 &  & +\frac{1}{2\mu_{0}}\int\mathrm{d}v\sum_{j=0}^{n-1}2\text{\ensuremath{\beta_{j}^{\ell}}}\delta\mathbf{B}_{R}\left(t-jdt\right)\cdot\beta_{n}^{\ell}\delta\mathbf{B}_{R}\nonumber \\
 &  & +\frac{1}{2\mu_{0}}\int\mathrm{d}v\left|\beta_{n}^{\ell}\delta\mathbf{B}_{R}\right|^{2}\label{eq:one_energy_conservation}
\end{eqnarray}
While Eq.~(\ref{eq:one_energy_conservation}) might appear daunting,
we emphasize that we never solve this equation in practice. Instead,
we will now make a simple approximation to convert this complicated
equations (with memory) into a simple, Markovian quadratic equation.

\subsection{Overlaps with previous rescaling fields cause self-interference}

\begin{figure}
\begin{centering}
\par\end{centering}
\centering{}\includegraphics[scale=0.8]{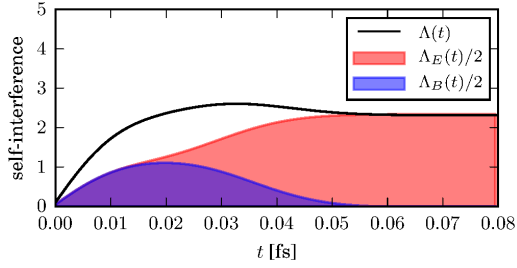}\caption{The self-interference length as a function of time for a 1D system.
The polarization distribution is given by Eq.~(\ref{eq:P_1D}) and
the spatial distribution of the rescaling fields are given by Eqs.~(\ref{eq:dE_1D})
and (\ref{eq:dB_1D}). Note that $\Lambda_{B}\left(t\right)$ is non-zero
only for a short time. \label{fig:self-interference}}
\end{figure}
Although the cross terms between the pure Ehrenfest field and the
rescaling fields will be eliminated by phase cancellation (Eq.~(\ref{eq:Er-R_cancellation})),
the rescaling fields at the current time step ($j=n$) will have a
non-vanishing cross term with the rescaling field from previous times
($j<n$). Given a polarization distribution that is small in space
and EM fields propagating freely at the speed of light, the relevant
cross term is the overlap $\int\mathrm{d}v\delta\mathbf{E}_{R}\left(t-jdt\right)\cdot\delta\mathbf{E}_{R}$
and $\int\mathrm{d}v\delta\mathbf{B}_{R}\left(t-jdt\right)\cdot\delta\mathbf{B}_{R}$
for small $t-jdt$. At this point, we presume that 
\begin{equation}
\alpha_{j}^{\ell}\approx\alpha_{n}^{\ell},\ \beta_{j}^{\ell}\approx\beta_{n}^{\ell}\label{eq:alphaisalpha}
\end{equation}
does not change much for a short, local time period and simplify Eq.~(\ref{eq:one_energy_conservation})
as 
\begin{eqnarray}
\delta U_{R}^{\ell} & = & \frac{\Lambda_{E}\left(t\right)}{cdt}\frac{\epsilon_{0}}{2}\int\mathrm{d}v\left|\alpha_{n}^{\ell}\delta\mathbf{E}_{R}\right|^{2}\nonumber \\
 &  & +\frac{\Lambda_{B}\left(t\right)}{cdt}\frac{1}{2\mu_{0}}\int\mathrm{d}v\left|\beta_{n}^{\ell}\delta\mathbf{B}_{R}\right|^{2}.\label{eq:average_energy_conservation_canceled}
\end{eqnarray}
Here we define the self-interference lengths $\Lambda_{E}\left(t\right)$
and $\Lambda_{B}\left(t\right)$ for $\delta\mathbf{E}_{R}$ and $\delta\mathbf{B}_{R}$
respectively as:
\begin{eqnarray}
\frac{\Lambda_{E}\left(t\right)}{cdt} & = & 1+\sum_{j=0}^{n-1}\frac{2\int\mathrm{d}v\delta\mathbf{E}_{R}\left(t-jdt\right)\cdot\delta\mathbf{E}_{R}}{\int\mathrm{d}v\left|\delta\mathbf{E}_{R}\right|^{2}},\label{eq:self-inter-length-Et}\\
\frac{\Lambda_{B}\left(t\right)}{cdt} & = & 1+\sum_{j=0}^{n-1}\frac{2\int\mathrm{d}v\delta\mathbf{B}_{R}\left(t-jdt\right)\cdot\delta\mathbf{B}_{R}}{\int\mathrm{d}v\left|\delta\mathbf{B}_{R}\right|^{2}}.\label{eq:self-inter-length-Bt}
\end{eqnarray}
Note that Eq.~(\ref{eq:alphaisalpha}) should hold when the time
that a rescaling field overlaps with $\delta\mathbf{E}_{R}$ or $\delta\mathbf{B}_{R}$
is much smaller than the oscillating period of the EM field, i.e.
$\sigma/c\ll2\pi/\Omega$. Given $\sigma\sim O\left(1\ \text{nm}\right)$,
this condition should be roughly $\Omega\ll10^{18}\ \text{Hz}$, i.e.
this assumption should be valid as long as the photon energy is not
in a high frequency X-ray regime. Finally, we recall that the $\delta\mathbf{E}_{R}$
and $\delta\mathbf{B}_{R}$ rescaling fields must carry equal energy
density (i.e. $\frac{\epsilon_{0}}{2}\int\mathrm{d}v\left|\alpha_{n}^{\ell}\delta\mathbf{E}_{R}\right|^{2}=\frac{1}{2\mu_{0}}\int\mathrm{d}v\left|\beta_{n}^{\ell}\delta\mathbf{B}_{R}\right|^{2}$),
so that energy conservation (Eq.~(\ref{eq:average_energy_conservation_canceled}))
can be further simplified:
\begin{eqnarray}
\delta U_{R}^{\ell} & = & \frac{\Lambda\left(t\right)}{cdt}\frac{\epsilon_{0}}{2}\int\mathrm{d}v\left|\alpha_{n}^{\ell}\delta\mathbf{E}_{R}\right|^{2}\\
 & = & \frac{\Lambda\left(t\right)}{cdt}\frac{1}{2\mu_{0}}\int\mathrm{d}v\left|\beta_{n}^{\ell}\delta\mathbf{B}_{R}\right|^{2}
\end{eqnarray}
where $\Lambda\left(t\right)=\left(\Lambda_{E}\left(t\right)+\Lambda_{B}\left(t\right)\right)/2$
is the average self-interference length. For an infinitesimal time
step $dt$, we can write $dt\sum_{j=0}^{n-1}=\int_{0}^{t}\mathrm{d}t^{\prime}$
for $t^{\prime}=t-jdt$ and the self-interference length becomes:
\begin{equation}
\begin{split}\Lambda\left(t\right)= & \frac{c\int_{0}^{t}\mathrm{d}t^{\prime}\int\mathrm{d}v\delta\mathbf{E}_{R}\left(t^{\prime}\right)\cdot\delta\mathbf{E}_{R}}{\int\mathrm{d}v\left|\delta\mathbf{E}_{R}\right|^{2}}+\\
 & \frac{c\int_{0}^{t}\mathrm{d}t^{\prime}\int\mathrm{d}v\delta\mathbf{B}_{R}\left(t^{\prime}\right)\cdot\delta\mathbf{B}_{R}}{\int\mathrm{d}v\left|\delta\mathbf{B}_{R}\right|^{2}}.
\end{split}
\label{eq:self-interference_average}
\end{equation}

At this point, to evaluate the overlap of the current rescaling field
(at time $t$) with previous rescaling fields (created at time $jdt$,
and propagated for $t^{\prime}=t-jdt$), we suppose that the rescaling
fields propagate freely according to Maxwell's equations
\begin{eqnarray}
\frac{\partial}{\partial t}\delta\mathbf{B}_{R}\left(\mathbf{r},t\right) & = & -\boldsymbol{\nabla}\times\delta\mathbf{E}_{R}\left(\mathbf{r},t\right),\\
\frac{\partial}{\partial t}\delta\mathbf{E}_{R}\left(\mathbf{r},t\right) & = & c^{2}\boldsymbol{\nabla}\times\delta\mathbf{B}_{R}\left(\mathbf{r},t\right).
\end{eqnarray}
We expand in Fourier space $\delta\mathbf{E}_{R}\left(\mathbf{r},t\right)=\int\mathrm{d}k^{n}\delta\widetilde{\mathbf{E}}_{R}\left(\mathbf{k},t\right)e^{i\mathbf{k}\cdot\mathbf{r}}$
and $\delta\mathbf{B}_{R}\left(\mathbf{r},t\right)=\int\mathrm{d}k^{n}\delta\widetilde{\mathbf{B}}_{R}\left(\mathbf{k},t\right)e^{i\mathbf{k}\cdot\mathbf{r}}$
and find the relevant equations of motion:
\begin{eqnarray}
\frac{\partial}{\partial t}\delta\widetilde{\mathbf{B}}_{R}\left(\mathbf{k},t\right) & = & i\mathbf{k}\times\delta\widetilde{\mathbf{E}}_{R}\left(\mathbf{k},t\right),\label{eq:free_field_Ek}\\
\frac{\partial}{\partial t}\delta\widetilde{\mathbf{E}}_{R}\left(\mathbf{k},t\right) & = & -ic^{2}\mathbf{k}\times\delta\widetilde{\mathbf{B}}_{R}\left(\mathbf{k},t\right).\label{eq:free_field_Bk}
\end{eqnarray}
Here, without loss of generality, we let $\mathbf{k}=k\hat{\mathbf{x}}$,
$\delta\widetilde{\mathbf{E}}_{R}\left(\mathbf{k},t\right)=\delta\widetilde{E}_{R}\left(k,t\right)\hat{\mathbf{z}}$
and $\delta\widetilde{\mathbf{B}}_{R}\left(\mathbf{k},t\right)=\delta\widetilde{B}_{R}\left(k,t\right)\left(-\hat{\mathbf{y}}\right)$.
For an arbitrary initial condition given by $\delta\widetilde{E}_{R}\left(k\right)$
and $\delta\widetilde{B}_{R}\left(k\right)$, the general solution
of Eqs.~(\ref{eq:free_field_Ek}) and (\ref{eq:free_field_Bk}) is
(with $\omega=ck$)
\begin{eqnarray}
\delta\widetilde{E}_{R}\left(k,t\right) & = & \delta\widetilde{E}_{R}\left(k\right)\cos\omega t+ic\delta\widetilde{B}_{R}\left(k\right)\sin\omega t,\label{eq:free_field_Ekt}\\
\delta\widetilde{B}_{R}\left(k,t\right) & = & \delta\widetilde{B}_{R}\left(k\right)\cos\omega t+\frac{i}{c}\delta\widetilde{E}_{R}\left(k\right)\sin\omega t.\label{eq:free_field_Bkt}
\end{eqnarray}
With this general solution for free propagation, we can evaluate the
total overlap in the Fourier space by\begin{widetext}
\begin{equation}
\int\mathrm{d}v\delta\mathbf{E}_{R}\left(t^{\prime}\right)\cdot\delta\mathbf{E}_{R}=2\pi\int_{-\infty}^{\infty}\mathrm{d}k\biggl[\delta\widetilde{E}_{R}\left(k\right)\cos\omega t^{\prime}+ic\delta\widetilde{B}_{R}\left(k\right)\sin\omega t^{\prime}\biggr]\delta\widetilde{E}_{R}\left(-k\right),\label{eq:overlap_E}
\end{equation}
\begin{equation}
\int\mathrm{d}v\delta\mathbf{B}_{R}\left(t^{\prime}\right)\cdot\delta\mathbf{B}_{R}=2\pi\int_{-\infty}^{\infty}\mathrm{d}k\left[\delta\widetilde{B}_{R}\left(k\right)\cos\omega t+\frac{i}{c}\delta\widetilde{E}_{R}\left(k\right)\sin\omega t\right]\delta\widetilde{B}_{R}\left(-k\right).\label{eq:overlap_B}
\end{equation}
\end{widetext}Here we have used $\int_{-\infty}^{\infty}\mathrm{d}xe^{i\left(k+k^{\prime}\right)x}=2\pi\delta\left(k+k^{\prime}\right)$.
We now plug Eqs.~(\ref{eq:overlap_E}) and (\ref{eq:overlap_B})
back into Eq.~(\ref{eq:self-interference_average}), so that the
time integration of the overlap becomes
\begin{eqnarray}
\int_{0}^{t}\mathrm{d}t^{\prime}\cos\omega t^{\prime} & = & \frac{1}{2}\int_{-t}^{t}\mathrm{d}t^{\prime}e^{i\omega t^{\prime}},\label{eq:coswt_integral}\\
\int_{0}^{t}\mathrm{d}t^{\prime}\sin\omega t^{\prime} & = & \frac{1-\cos kct}{kc}.\label{eq:sinwt_integral}
\end{eqnarray}
Note that the cross terms (the second terms of Eqs.~(\ref{eq:overlap_E})
and (\ref{eq:overlap_B})) become zero after we carry out $\int_{-\infty}^{\infty}\mathrm{d}k$
with Eq.~(\ref{eq:sinwt_integral}) using a Cauchy integral. We now
assume that the rescaling field overlaps with only a short history
of itself, so that the time integral of the overlap must reach a constant
in a reasonably short period of time. With this assumption in mind,
we can approximate $\Lambda\equiv\Lambda\left(t\rightarrow\infty\right)$
for all time, so that Eq.~(\ref{eq:coswt_integral}) becomes 
\begin{equation}
\int_{0}^{\infty}\mathrm{d}t^{\prime}\cos\omega t^{\prime}=\frac{\pi}{c}\delta\left(k\right).\label{eq:coswt_integral_infty}
\end{equation}
Therefore, the self-interference length turns out to be 
\begin{equation}
\Lambda=\frac{2\pi^{2}\left|\delta\widetilde{E}_{R}\left(0\right)\right|^{2}}{\int\mathrm{d}x\left|\delta E_{R}\right|^{2}}+\frac{2\pi^{2}\left|\delta\widetilde{B}_{R}\left(0\right)\right|^{2}}{\int\mathrm{d}x\left|\delta B_{R}\right|^{2}}.\label{eq:self-interference_average-1}
\end{equation}
As a practical matter for a Gaussian polarization distribution in
1D, we use the rescaling fields derived in Appendix~\ref{sec:Justify-the-rescaling}
(Eq.~(\ref{eq:dE_1D}) and (\ref{eq:dB_1D})) and find an analytical
expression for the self-interference length given by
\begin{equation}
\Lambda^{\text{1D}}=\frac{2}{3}\sqrt{\frac{2\pi}{a}}.
\end{equation}
In this particular 1D case, $\left|\delta\widetilde{B}_{R}\left(0\right)\right|^{2}=0$
and the overlap of the $\delta B_{R}$ field is canceled out for long
time (see Fig.~\ref{fig:self-interference} blue area) since $\delta B_{R}\left(x\right)$
is an odd spatial function.

Thus, in the end, $\alpha_{n}^{\ell}$ and $\beta_{n}^{\ell}$ can
be determined by 
\begin{equation}
\text{\ensuremath{\alpha_{n}^{\ell}}}=\sqrt{\frac{cdt}{\Lambda}\frac{\delta U_{R}^{\ell}}{\epsilon_{0}\int\mathrm{d}v\left|\delta\mathbf{E}_{R}\right|^{2}}}\times\text{sgn}\left(\text{Im}\left[\rho_{01}e^{i\phi^{\ell}}\right]\right),\label{eq:alpha_ell-1}
\end{equation}
\begin{equation}
\text{\ensuremath{\beta_{n}^{\ell}}}=\sqrt{\frac{cdt}{\Lambda}\frac{\mu_{0}\delta U_{R}^{\ell}}{\int\mathrm{d}v\left|\delta\mathbf{B}_{R}\right|^{2}}}\times\text{sgn}\left(\text{Im}\left[\rho_{01}e^{i\phi^{\ell}}\right]\right).\label{eq:beta_ell-1}
\end{equation}
We have now justified Eqs.~(\ref{eq:alpha_ell-sic}\textendash \ref{eq:beta_ell-sic})
in the main body of the text.

\bibliographystyle{apsrev4-1}
\bibliography{MyLibrary}

\begin{thebibliography}{59}%
\makeatletter
\providecommand \@ifxundefined [1]{%
 \@ifx{#1\undefined}
}%
\providecommand \@ifnum [1]{%
 \ifnum #1\expandafter \@firstoftwo
 \else \expandafter \@secondoftwo
 \fi
}%
\providecommand \@ifx [1]{%
 \ifx #1\expandafter \@firstoftwo
 \else \expandafter \@secondoftwo
 \fi
}%
\providecommand \natexlab [1]{#1}%
\providecommand \enquote  [1]{``#1''}%
\providecommand \bibnamefont  [1]{#1}%
\providecommand \bibfnamefont [1]{#1}%
\providecommand \citenamefont [1]{#1}%
\providecommand \href@noop [0]{\@secondoftwo}%
\providecommand \href [0]{\begingroup \@sanitize@url \@href}%
\providecommand \@href[1]{\@@startlink{#1}\@@href}%
\providecommand \@@href[1]{\endgroup#1\@@endlink}%
\providecommand \@sanitize@url [0]{\catcode `\\12\catcode `\$12\catcode
  `\&12\catcode `\#12\catcode `\^12\catcode `\_12\catcode `\%12\relax}%
\providecommand \@@startlink[1]{}%
\providecommand \@@endlink[0]{}%
\providecommand \url  [0]{\begingroup\@sanitize@url \@url }%
\providecommand \@url [1]{\endgroup\@href {#1}{\urlprefix }}%
\providecommand \urlprefix  [0]{URL }%
\providecommand \Eprint [0]{\href }%
\providecommand \doibase [0]{http://dx.doi.org/}%
\providecommand \selectlanguage [0]{\@gobble}%
\providecommand \bibinfo  [0]{\@secondoftwo}%
\providecommand \bibfield  [0]{\@secondoftwo}%
\providecommand \translation [1]{[#1]}%
\providecommand \BibitemOpen [0]{}%
\providecommand \bibitemStop [0]{}%
\providecommand \bibitemNoStop [0]{.\EOS\space}%
\providecommand \EOS [0]{\spacefactor3000\relax}%
\providecommand \BibitemShut  [1]{\csname bibitem#1\endcsname}%
\let\auto@bib@innerbib\@empty
\bibitem [{\citenamefont {Thompson}(1998)}]{thompson_nonlinear_1998}%
  \BibitemOpen
  \bibfield  {author} {\bibinfo {author} {\bibfnamefont {R.~J.}\ \bibnamefont
  {Thompson}},\ }\href {\doibase 10.1103/PhysRevA.57.3084} {\bibfield
  {journal} {\bibinfo  {journal} {Phys. Rev. A}\ }\textbf {\bibinfo {volume}
  {57}},\ \bibinfo {pages} {3084} (\bibinfo {year} {1998})}\BibitemShut
  {NoStop}%
\bibitem [{\citenamefont {Solano}\ \emph {et~al.}(2003)\citenamefont {Solano},
  \citenamefont {Agarwal},\ and\ \citenamefont
  {Walther}}]{solano_strong-driving-assisted_2003}%
  \BibitemOpen
  \bibfield  {author} {\bibinfo {author} {\bibfnamefont {E.}~\bibnamefont
  {Solano}}, \bibinfo {author} {\bibfnamefont {G.~S.}\ \bibnamefont {Agarwal}},
  \ and\ \bibinfo {author} {\bibfnamefont {H.}~\bibnamefont {Walther}},\ }\href
  {\doibase 10.1103/PhysRevLett.90.027903} {\bibfield  {journal} {\bibinfo
  {journal} {Phys. Rev. Lett.}\ }\textbf {\bibinfo {volume} {90}},\ \bibinfo
  {pages} {027903} (\bibinfo {year} {2003})}\BibitemShut {NoStop}%
\bibitem [{\citenamefont {Fink}\ \emph {et~al.}(2008)\citenamefont {Fink},
  \citenamefont {G{\"o}ppl}, \citenamefont {Baur}, \citenamefont {Bianchetti},
  \citenamefont {Leek}, \citenamefont {Blais},\ and\ \citenamefont
  {Wallraff}}]{fink_climbing_2008}%
  \BibitemOpen
  \bibfield  {author} {\bibinfo {author} {\bibfnamefont {J.~M.}\ \bibnamefont
  {Fink}}, \bibinfo {author} {\bibfnamefont {M.}~\bibnamefont {G{\"o}ppl}},
  \bibinfo {author} {\bibfnamefont {M.}~\bibnamefont {Baur}}, \bibinfo {author}
  {\bibfnamefont {R.}~\bibnamefont {Bianchetti}}, \bibinfo {author}
  {\bibfnamefont {P.~J.}\ \bibnamefont {Leek}}, \bibinfo {author}
  {\bibfnamefont {A.}~\bibnamefont {Blais}}, \ and\ \bibinfo {author}
  {\bibfnamefont {A.}~\bibnamefont {Wallraff}},\ }\href {\doibase
  10.1038/nature07112} {\bibfield  {journal} {\bibinfo  {journal} {Nature}\
  }\textbf {\bibinfo {volume} {454}},\ \bibinfo {pages} {315} (\bibinfo {year}
  {2008})}\BibitemShut {NoStop}%
\bibitem [{\citenamefont {Gibbs}\ \emph {et~al.}(2011)\citenamefont {Gibbs},
  \citenamefont {Khitrova},\ and\ \citenamefont
  {Koch}}]{gibbs_excitonpolariton_2011}%
  \BibitemOpen
  \bibfield  {author} {\bibinfo {author} {\bibfnamefont {H.~M.}\ \bibnamefont
  {Gibbs}}, \bibinfo {author} {\bibfnamefont {G.}~\bibnamefont {Khitrova}}, \
  and\ \bibinfo {author} {\bibfnamefont {S.~W.}\ \bibnamefont {Koch}},\ }\href
  {\doibase 10.1038/nphoton.2011.15} {\bibfield  {journal} {\bibinfo  {journal}
  {Nature Photonics}\ }\textbf {\bibinfo {volume} {5}},\ \bibinfo {pages} {273}
  (\bibinfo {year} {2011})}\BibitemShut {NoStop}%
\bibitem [{\citenamefont {Lodahl}\ \emph {et~al.}(2015)\citenamefont {Lodahl},
  \citenamefont {Mahmoodian},\ and\ \citenamefont
  {Stobbe}}]{lodahl_interfacing_2015}%
  \BibitemOpen
  \bibfield  {author} {\bibinfo {author} {\bibfnamefont {P.}~\bibnamefont
  {Lodahl}}, \bibinfo {author} {\bibfnamefont {S.}~\bibnamefont {Mahmoodian}},
  \ and\ \bibinfo {author} {\bibfnamefont {S.}~\bibnamefont {Stobbe}},\ }\href
  {\doibase 10.1103/RevModPhys.87.347} {\bibfield  {journal} {\bibinfo
  {journal} {Reviews of Modern Physics}\ }\textbf {\bibinfo {volume} {87}},\
  \bibinfo {pages} {347} (\bibinfo {year} {2015})}\BibitemShut {NoStop}%
\bibitem [{\citenamefont {T{\"o}rm{\"a}}\ and\ \citenamefont
  {Barnes}(2015)}]{torma_strong_2015}%
  \BibitemOpen
  \bibfield  {author} {\bibinfo {author} {\bibfnamefont {P.}~\bibnamefont
  {T{\"o}rm{\"a}}}\ and\ \bibinfo {author} {\bibfnamefont {W.~L.}\ \bibnamefont
  {Barnes}},\ }\href {\doibase 10.1088/0034-4885/78/1/013901} {\bibfield
  {journal} {\bibinfo  {journal} {Rep. Prog. Phys.}\ }\textbf {\bibinfo
  {volume} {78}},\ \bibinfo {pages} {013901} (\bibinfo {year}
  {2015})}\BibitemShut {NoStop}%
\bibitem [{\citenamefont {Puthumpally-Joseph}\ \emph
  {et~al.}(2014)\citenamefont {Puthumpally-Joseph}, \citenamefont {Sukharev},
  \citenamefont {Atabek},\ and\ \citenamefont
  {Charron}}]{puthumpally-joseph_dipole-induced_2014}%
  \BibitemOpen
  \bibfield  {author} {\bibinfo {author} {\bibfnamefont {R.}~\bibnamefont
  {Puthumpally-Joseph}}, \bibinfo {author} {\bibfnamefont {M.}~\bibnamefont
  {Sukharev}}, \bibinfo {author} {\bibfnamefont {O.}~\bibnamefont {Atabek}}, \
  and\ \bibinfo {author} {\bibfnamefont {E.}~\bibnamefont {Charron}},\ }\href
  {\doibase 10.1103/PhysRevLett.113.163603} {\bibfield  {journal} {\bibinfo
  {journal} {Phys. Rev. Lett.}\ }\textbf {\bibinfo {volume} {113}},\ \bibinfo
  {pages} {163603} (\bibinfo {year} {2014})}\BibitemShut {NoStop}%
\bibitem [{\citenamefont {Puthumpally-Joseph}\ \emph
  {et~al.}(2015)\citenamefont {Puthumpally-Joseph}, \citenamefont {Atabek},
  \citenamefont {Sukharev},\ and\ \citenamefont
  {Charron}}]{puthumpally-joseph_theoretical_2015}%
  \BibitemOpen
  \bibfield  {author} {\bibinfo {author} {\bibfnamefont {R.}~\bibnamefont
  {Puthumpally-Joseph}}, \bibinfo {author} {\bibfnamefont {O.}~\bibnamefont
  {Atabek}}, \bibinfo {author} {\bibfnamefont {M.}~\bibnamefont {Sukharev}}, \
  and\ \bibinfo {author} {\bibfnamefont {E.}~\bibnamefont {Charron}},\ }\href
  {\doibase 10.1103/PhysRevA.91.043835} {\bibfield  {journal} {\bibinfo
  {journal} {Phys. Rev. A}\ }\textbf {\bibinfo {volume} {91}},\ \bibinfo
  {pages} {043835} (\bibinfo {year} {2015})}\BibitemShut {NoStop}%
\bibitem [{\citenamefont {Sukharev}\ and\ \citenamefont
  {Nitzan}(2017)}]{sukharev_optics_2017}%
  \BibitemOpen
  \bibfield  {author} {\bibinfo {author} {\bibfnamefont {M.}~\bibnamefont
  {Sukharev}}\ and\ \bibinfo {author} {\bibfnamefont {A.}~\bibnamefont
  {Nitzan}},\ }\href {\doibase 10.1088/1361-648X/aa85ef} {\bibfield  {journal}
  {\bibinfo  {journal} {J. Phys.: Condens. Matter}\ }\textbf {\bibinfo {volume}
  {29}},\ \bibinfo {pages} {443003} (\bibinfo {year} {2017})}\BibitemShut
  {NoStop}%
\bibitem [{\citenamefont {Vasa}\ and\ \citenamefont
  {Lienau}(2018)}]{vasa_strong_2018}%
  \BibitemOpen
  \bibfield  {author} {\bibinfo {author} {\bibfnamefont {P.}~\bibnamefont
  {Vasa}}\ and\ \bibinfo {author} {\bibfnamefont {C.}~\bibnamefont {Lienau}},\
  }\href {\doibase 10.1021/acsphotonics.7b00650} {\bibfield  {journal}
  {\bibinfo  {journal} {ACS Photonics}\ }\textbf {\bibinfo {volume} {5}},\
  \bibinfo {pages} {2} (\bibinfo {year} {2018})}\BibitemShut {NoStop}%
\bibitem [{\citenamefont {Dicke}(1954)}]{dicke_coherence_1954}%
  \BibitemOpen
  \bibfield  {author} {\bibinfo {author} {\bibfnamefont {R.~H.}\ \bibnamefont
  {Dicke}},\ }\href {\doibase 10.1103/PhysRev.93.99} {\bibfield  {journal}
  {\bibinfo  {journal} {Phys. Rev.}\ }\textbf {\bibinfo {volume} {93}},\
  \bibinfo {pages} {99} (\bibinfo {year} {1954})}\BibitemShut {NoStop}%
\bibitem [{\citenamefont {Andreev}\ \emph {et~al.}(1980)\citenamefont
  {Andreev}, \citenamefont {Emel'yanov},\ and\ \citenamefont {Il'inski{\u
  i}}}]{andreev_collective_1980}%
  \BibitemOpen
  \bibfield  {author} {\bibinfo {author} {\bibfnamefont {A.~V.}\ \bibnamefont
  {Andreev}}, \bibinfo {author} {\bibfnamefont {V.~I.}\ \bibnamefont
  {Emel'yanov}}, \ and\ \bibinfo {author} {\bibfnamefont {Y.~A.}\ \bibnamefont
  {Il'inski{\u i}}},\ }\href {\doibase 10.1070/PU1980v023n08ABEH005024}
  {\bibfield  {journal} {\bibinfo  {journal} {Sov. Phys. Usp.}\ }\textbf
  {\bibinfo {volume} {23}},\ \bibinfo {pages} {493} (\bibinfo {year}
  {1980})}\BibitemShut {NoStop}%
\bibitem [{\citenamefont {Oppel}\ \emph {et~al.}(2014)\citenamefont {Oppel},
  \citenamefont {Wiegner}, \citenamefont {Agarwal},\ and\ \citenamefont {von
  Zanthier}}]{oppel_directional_2014}%
  \BibitemOpen
  \bibfield  {author} {\bibinfo {author} {\bibfnamefont {S.}~\bibnamefont
  {Oppel}}, \bibinfo {author} {\bibfnamefont {R.}~\bibnamefont {Wiegner}},
  \bibinfo {author} {\bibfnamefont {G.~S.}\ \bibnamefont {Agarwal}}, \ and\
  \bibinfo {author} {\bibfnamefont {J.}~\bibnamefont {von Zanthier}},\ }\href
  {\doibase 10.1103/PhysRevLett.113.263606} {\bibfield  {journal} {\bibinfo
  {journal} {Phys. Rev. Lett.}\ }\textbf {\bibinfo {volume} {113}},\ \bibinfo
  {pages} {263606} (\bibinfo {year} {2014})}\BibitemShut {NoStop}%
\bibitem [{\citenamefont {Dirac}(1927{\natexlab{a}})}]{dirac_quantum_1927}%
  \BibitemOpen
  \bibfield  {author} {\bibinfo {author} {\bibfnamefont {P.~A.~M.}\
  \bibnamefont {Dirac}},\ }\href {\doibase 10.1098/rspa.1927.0071} {\bibfield
  {journal} {\bibinfo  {journal} {Proc. R. Soc. Lond. A}\ }\textbf {\bibinfo
  {volume} {114}},\ \bibinfo {pages} {710} (\bibinfo {year}
  {1927}{\natexlab{a}})}\BibitemShut {NoStop}%
\bibitem [{\citenamefont {Dirac}(1927{\natexlab{b}})}]{dirac_quantum_1927-1}%
  \BibitemOpen
  \bibfield  {author} {\bibinfo {author} {\bibfnamefont {P.~A.~M.}\
  \bibnamefont {Dirac}},\ }\href {\doibase 10.1098/rspa.1927.0039} {\bibfield
  {journal} {\bibinfo  {journal} {Proc. R. Soc. Lond. A}\ }\textbf {\bibinfo
  {volume} {114}},\ \bibinfo {pages} {243} (\bibinfo {year}
  {1927}{\natexlab{b}})}\BibitemShut {NoStop}%
\bibitem [{\citenamefont {Milonni}(1976)}]{milonni_semiclassical_1976}%
  \BibitemOpen
  \bibfield  {author} {\bibinfo {author} {\bibfnamefont {P.~W.}\ \bibnamefont
  {Milonni}},\ }\href {\doibase 10.1016/0370-1573(76)90037-5} {\bibfield
  {journal} {\bibinfo  {journal} {Physics Reports}\ }\textbf {\bibinfo {volume}
  {25}},\ \bibinfo {pages} {1} (\bibinfo {year} {1976})}\BibitemShut {NoStop}%
\bibitem [{\citenamefont {Kapral}\ and\ \citenamefont
  {Ciccotti}(1999)}]{kapral_mixed_1999}%
  \BibitemOpen
  \bibfield  {author} {\bibinfo {author} {\bibfnamefont {R.}~\bibnamefont
  {Kapral}}\ and\ \bibinfo {author} {\bibfnamefont {G.}~\bibnamefont
  {Ciccotti}},\ }\href {\doibase 10.1063/1.478811} {\bibfield  {journal}
  {\bibinfo  {journal} {J. Chem. Phys.}\ }\textbf {\bibinfo {volume} {110}},\
  \bibinfo {pages} {8919} (\bibinfo {year} {1999})}\BibitemShut {NoStop}%
\bibitem [{\citenamefont {Tully}(1998)}]{tully_mixed_1998}%
  \BibitemOpen
  \bibfield  {author} {\bibinfo {author} {\bibfnamefont {J.~C.}\ \bibnamefont
  {Tully}},\ }\href {http://link.aip.org/link/?JCPSA6/110/8919/1} {\bibfield
  {journal} {\bibinfo  {journal} {Faraday Discuss.}\ }\textbf {\bibinfo
  {volume} {110}},\ \bibinfo {pages} {407} (\bibinfo {year}
  {1998})}\BibitemShut {NoStop}%
\bibitem [{\citenamefont {Tully}(1990)}]{tully_molecular_1990}%
  \BibitemOpen
  \bibfield  {author} {\bibinfo {author} {\bibfnamefont {J.~C.}\ \bibnamefont
  {Tully}},\ }\href {\doibase 10.1063/1.459170} {\bibfield  {journal} {\bibinfo
   {journal} {The Journal of Chemical Physics}\ }\textbf {\bibinfo {volume}
  {93}},\ \bibinfo {pages} {1061} (\bibinfo {year} {1990})}\BibitemShut
  {NoStop}%
\bibitem [{\citenamefont {Wang}\ \emph {et~al.}(1998)\citenamefont {Wang},
  \citenamefont {Sun},\ and\ \citenamefont {Miller}}]{wang_semiclassical_1998}%
  \BibitemOpen
  \bibfield  {author} {\bibinfo {author} {\bibfnamefont {H.}~\bibnamefont
  {Wang}}, \bibinfo {author} {\bibfnamefont {X.}~\bibnamefont {Sun}}, \ and\
  \bibinfo {author} {\bibfnamefont {W.~H.}\ \bibnamefont {Miller}},\ }\href
  {\doibase 10.1063/1.476447} {\bibfield  {journal} {\bibinfo  {journal} {J.
  Chem. Phys.}\ }\textbf {\bibinfo {volume} {108}},\ \bibinfo {pages} {9726}
  (\bibinfo {year} {1998})}\BibitemShut {NoStop}%
\bibitem [{\citenamefont {Wang}\ \emph {et~al.}(1999)\citenamefont {Wang},
  \citenamefont {Song}, \citenamefont {Chandler},\ and\ \citenamefont
  {Miller}}]{wang_semiclassical_1999}%
  \BibitemOpen
  \bibfield  {author} {\bibinfo {author} {\bibfnamefont {H.}~\bibnamefont
  {Wang}}, \bibinfo {author} {\bibfnamefont {X.}~\bibnamefont {Song}}, \bibinfo
  {author} {\bibfnamefont {D.}~\bibnamefont {Chandler}}, \ and\ \bibinfo
  {author} {\bibfnamefont {W.~H.}\ \bibnamefont {Miller}},\ }\href {\doibase
  10.1063/1.478388} {\bibfield  {journal} {\bibinfo  {journal} {J. Chem.
  Phys.}\ }\textbf {\bibinfo {volume} {110}},\ \bibinfo {pages} {4828}
  (\bibinfo {year} {1999})}\BibitemShut {NoStop}%
\bibitem [{\citenamefont {Ziolkowski}\ \emph {et~al.}(1995)\citenamefont
  {Ziolkowski}, \citenamefont {Arnold},\ and\ \citenamefont
  {Gogny}}]{ziolkowski_ultrafast_1995}%
  \BibitemOpen
  \bibfield  {author} {\bibinfo {author} {\bibfnamefont {R.~W.}\ \bibnamefont
  {Ziolkowski}}, \bibinfo {author} {\bibfnamefont {J.~M.}\ \bibnamefont
  {Arnold}}, \ and\ \bibinfo {author} {\bibfnamefont {D.~M.}\ \bibnamefont
  {Gogny}},\ }\href {\doibase 10.1103/PhysRevA.52.3082} {\bibfield  {journal}
  {\bibinfo  {journal} {Phys. Rev. A}\ }\textbf {\bibinfo {volume} {52}},\
  \bibinfo {pages} {3082} (\bibinfo {year} {1995})}\BibitemShut {NoStop}%
\bibitem [{\citenamefont {Slavcheva}\ \emph {et~al.}(2002)\citenamefont
  {Slavcheva}, \citenamefont {Arnold}, \citenamefont {Wallace},\ and\
  \citenamefont {Ziolkowski}}]{slavcheva_coupled_2002}%
  \BibitemOpen
  \bibfield  {author} {\bibinfo {author} {\bibfnamefont {G.}~\bibnamefont
  {Slavcheva}}, \bibinfo {author} {\bibfnamefont {J.~M.}\ \bibnamefont
  {Arnold}}, \bibinfo {author} {\bibfnamefont {I.}~\bibnamefont {Wallace}}, \
  and\ \bibinfo {author} {\bibfnamefont {R.~W.}\ \bibnamefont {Ziolkowski}},\
  }\href {\doibase 10.1103/PhysRevA.66.063418} {\bibfield  {journal} {\bibinfo
  {journal} {Phys. Rev. A}\ }\textbf {\bibinfo {volume} {66}},\ \bibinfo
  {pages} {063418} (\bibinfo {year} {2002})}\BibitemShut {NoStop}%
\bibitem [{\citenamefont {Fratalocchi}\ \emph {et~al.}(2008)\citenamefont
  {Fratalocchi}, \citenamefont {Conti},\ and\ \citenamefont
  {Ruocco}}]{fratalocchi_three-dimensional_2008}%
  \BibitemOpen
  \bibfield  {author} {\bibinfo {author} {\bibfnamefont {A.}~\bibnamefont
  {Fratalocchi}}, \bibinfo {author} {\bibfnamefont {C.}~\bibnamefont {Conti}},
  \ and\ \bibinfo {author} {\bibfnamefont {G.}~\bibnamefont {Ruocco}},\ }\href
  {\doibase 10.1103/PhysRevA.78.013806} {\bibfield  {journal} {\bibinfo
  {journal} {Phys. Rev. A}\ }\textbf {\bibinfo {volume} {78}},\ \bibinfo
  {pages} {013806} (\bibinfo {year} {2008})}\BibitemShut {NoStop}%
\bibitem [{\citenamefont {Sukharev}\ and\ \citenamefont
  {Nitzan}(2011)}]{sukharev_numerical_2011}%
  \BibitemOpen
  \bibfield  {author} {\bibinfo {author} {\bibfnamefont {M.}~\bibnamefont
  {Sukharev}}\ and\ \bibinfo {author} {\bibfnamefont {A.}~\bibnamefont
  {Nitzan}},\ }\href {\doibase 10.1103/PhysRevA.84.043802} {\bibfield
  {journal} {\bibinfo  {journal} {Phys. Rev. A}\ }\textbf {\bibinfo {volume}
  {84}},\ \bibinfo {pages} {043802} (\bibinfo {year} {2011})}\BibitemShut
  {NoStop}%
\bibitem [{\citenamefont {Miller}(1978)}]{miller_classical/semiclassical_1978}%
  \BibitemOpen
  \bibfield  {author} {\bibinfo {author} {\bibfnamefont {W.~H.}\ \bibnamefont
  {Miller}},\ }\href {\doibase 10.1063/1.436793} {\bibfield  {journal}
  {\bibinfo  {journal} {The Journal of Chemical Physics}\ }\textbf {\bibinfo
  {volume} {69}},\ \bibinfo {pages} {2188} (\bibinfo {year}
  {1978})}\BibitemShut {NoStop}%
\bibitem [{\citenamefont {Li}\ \emph {et~al.}(2018{\natexlab{a}})\citenamefont
  {Li}, \citenamefont {Nitzan}, \citenamefont {Sukharev}, \citenamefont
  {Martinez}, \citenamefont {Chen},\ and\ \citenamefont
  {Subotnik}}]{li_mixed_2018}%
  \BibitemOpen
  \bibfield  {author} {\bibinfo {author} {\bibfnamefont {T.~E.}\ \bibnamefont
  {Li}}, \bibinfo {author} {\bibfnamefont {A.}~\bibnamefont {Nitzan}}, \bibinfo
  {author} {\bibfnamefont {M.}~\bibnamefont {Sukharev}}, \bibinfo {author}
  {\bibfnamefont {T.}~\bibnamefont {Martinez}}, \bibinfo {author}
  {\bibfnamefont {H.-T.}\ \bibnamefont {Chen}}, \ and\ \bibinfo {author}
  {\bibfnamefont {J.~E.}\ \bibnamefont {Subotnik}},\ }\href {\doibase
  10.1103/PhysRevA.97.032105} {\bibfield  {journal} {\bibinfo  {journal} {Phys.
  Rev. A}\ }\textbf {\bibinfo {volume} {97}},\ \bibinfo {pages} {032105}
  (\bibinfo {year} {2018}{\natexlab{a}})}\BibitemShut {NoStop}%
\bibitem [{\citenamefont {Provazza}\ and\ \citenamefont
  {Coker}(2018)}]{provazza_communication:_2018}%
  \BibitemOpen
  \bibfield  {author} {\bibinfo {author} {\bibfnamefont {J.}~\bibnamefont
  {Provazza}}\ and\ \bibinfo {author} {\bibfnamefont {D.~F.}\ \bibnamefont
  {Coker}},\ }\href {\doibase 10.1063/1.5031788} {\bibfield  {journal}
  {\bibinfo  {journal} {The Journal of Chemical Physics}\ }\textbf {\bibinfo
  {volume} {148}},\ \bibinfo {pages} {181102} (\bibinfo {year}
  {2018})}\BibitemShut {NoStop}%
\bibitem [{\citenamefont {Parandekar}\ and\ \citenamefont
  {Tully}(2006)}]{parandekar_detailed_2006}%
  \BibitemOpen
  \bibfield  {author} {\bibinfo {author} {\bibfnamefont {P.~V.}\ \bibnamefont
  {Parandekar}}\ and\ \bibinfo {author} {\bibfnamefont {J.~C.}\ \bibnamefont
  {Tully}},\ }\href {\doibase 10.1021/ct050213k} {\bibfield  {journal}
  {\bibinfo  {journal} {J. Chem. Theory Comput.}\ }\textbf {\bibinfo {volume}
  {2}},\ \bibinfo {pages} {229} (\bibinfo {year} {2006})}\BibitemShut {NoStop}%
\bibitem [{\citenamefont {Jain}\ and\ \citenamefont
  {Subotnik}(2018)}]{jain_vibrational_2018}%
  \BibitemOpen
  \bibfield  {author} {\bibinfo {author} {\bibfnamefont {A.}~\bibnamefont
  {Jain}}\ and\ \bibinfo {author} {\bibfnamefont {J.~E.}\ \bibnamefont
  {Subotnik}},\ }\href {\doibase 10.1021/acs.jpca.7b09018} {\bibfield
  {journal} {\bibinfo  {journal} {J. Phys. Chem. A}\ }\textbf {\bibinfo
  {volume} {122}},\ \bibinfo {pages} {16} (\bibinfo {year} {2018})}\BibitemShut
  {NoStop}%
\bibitem [{\citenamefont {Weisskopf}\ and\ \citenamefont
  {Wigner}(1930)}]{weisskopf_berechnung_1930}%
  \BibitemOpen
  \bibfield  {author} {\bibinfo {author} {\bibfnamefont {V.}~\bibnamefont
  {Weisskopf}}\ and\ \bibinfo {author} {\bibfnamefont {E.}~\bibnamefont
  {Wigner}},\ }\href {\doibase 10.1007/BF01336768} {\bibfield  {journal}
  {\bibinfo  {journal} {Z. Physik}\ }\textbf {\bibinfo {volume} {63}},\
  \bibinfo {pages} {54} (\bibinfo {year} {1930})}\BibitemShut {NoStop}%
\bibitem [{\citenamefont {Scully}\ and\ \citenamefont
  {Zubairy}(1997)}]{scully_quantum_1997}%
  \BibitemOpen
  \bibfield  {author} {\bibinfo {author} {\bibfnamefont {M.~O.}\ \bibnamefont
  {Scully}}\ and\ \bibinfo {author} {\bibfnamefont {M.~S.}\ \bibnamefont
  {Zubairy}},\ }\href@noop {} {\emph {\bibinfo {title} {Quantum {Optics}}}}\
  (\bibinfo  {publisher} {Cambridge University Press},\ \bibinfo {year}
  {1997})\BibitemShut {NoStop}%
\bibitem [{\citenamefont {Power}\ and\ \citenamefont
  {Zienau}(1959)}]{power_coulomb_1959}%
  \BibitemOpen
  \bibfield  {author} {\bibinfo {author} {\bibfnamefont {E.~A.}\ \bibnamefont
  {Power}}\ and\ \bibinfo {author} {\bibfnamefont {S.}~\bibnamefont {Zienau}},\
  }\href {\doibase 10.1098/rsta.1959.0008} {\bibfield  {journal} {\bibinfo
  {journal} {Phil. Trans. R. Soc. Lond. A}\ }\textbf {\bibinfo {volume}
  {251}},\ \bibinfo {pages} {427} (\bibinfo {year} {1959})}\BibitemShut
  {NoStop}%
\bibitem [{\citenamefont {Atkins}\ and\ \citenamefont
  {Woolley}(1970)}]{atkins_interaction_1970}%
  \BibitemOpen
  \bibfield  {author} {\bibinfo {author} {\bibfnamefont {P.~W.}\ \bibnamefont
  {Atkins}}\ and\ \bibinfo {author} {\bibfnamefont {R.~G.}\ \bibnamefont
  {Woolley}},\ }\href {\doibase 10.1098/rspa.1970.0192} {\bibfield  {journal}
  {\bibinfo  {journal} {Proc. R. Soc. Lond. A}\ }\textbf {\bibinfo {volume}
  {319}},\ \bibinfo {pages} {549} (\bibinfo {year} {1970})}\BibitemShut
  {NoStop}%
\bibitem [{\citenamefont {Cohen-Tannoudji}\ \emph {et~al.}(1997)\citenamefont
  {Cohen-Tannoudji}, \citenamefont {Dupont-Roc},\ and\ \citenamefont
  {Grynberg}}]{cohen-tannoudji_photons_1997}%
  \BibitemOpen
  \bibfield  {author} {\bibinfo {author} {\bibfnamefont {C.}~\bibnamefont
  {Cohen-Tannoudji}}, \bibinfo {author} {\bibfnamefont {J.}~\bibnamefont
  {Dupont-Roc}}, \ and\ \bibinfo {author} {\bibfnamefont {G.}~\bibnamefont
  {Grynberg}},\ }\href
  {https://www.wiley.com/en-us/Photons+and+Atoms%3A+Introduction+to+Quantum+Electrodynamics-p-9780471184331}
  {\emph {\bibinfo {title} {Photons and {Atoms}: {Introduction} to {Quantum}
  {Electrodynamics}}}}\ (\bibinfo  {publisher} {Wiley},\ \bibinfo {year}
  {1997})\BibitemShut {NoStop}%
\bibitem [{\citenamefont {Nitzan}(2006)}]{nitzan_chemical_2006}%
  \BibitemOpen
  \bibfield  {author} {\bibinfo {author} {\bibfnamefont {A.}~\bibnamefont
  {Nitzan}},\ }\href {\doibase 10.1002/cphc.200700074} {\emph {\bibinfo {title}
  {Chemical {Dynamics} in {Condensed} {Phases}: {Relaxation}, {Transfer}, and
  {Reactions} in {Condensed} {Molecular} {Systems}}}}\ (\bibinfo  {publisher}
  {Oxford University Press, New York},\ \bibinfo {year} {2006})\BibitemShut
  {NoStop}%
\bibitem [{\citenamefont {Mukamel}(1999)}]{mukamel_principles_1999}%
  \BibitemOpen
  \bibfield  {author} {\bibinfo {author} {\bibfnamefont {S.}~\bibnamefont
  {Mukamel}},\ }\href@noop {} {\emph {\bibinfo {title} {Principles of
  {Nonlinear} {Optics} and {Spectroscopy}}}}\ (\bibinfo  {publisher} {Oxford
  University Press},\ \bibinfo {year} {1999})\BibitemShut {NoStop}%
\bibitem [{Note1()}]{Note1}%
  \BibitemOpen
  \bibinfo {note} {In QED, the Coulomb interaction between particles $\alpha $
  and $\beta $ can be expressed as\protect \citep
  {cohen-tannoudji_photons_1997} \protect \[ \protect \cc@accent
  {"705E}{V}_{\protect \mathrm {Coul}}=\protect \frac {1}{\epsilon _{0}}\DOTSI
  \intop \ilimits@ \protect \mathrm {d}v\protect \cc@accent {"705E}{\protect
  \mathbf {P}}_{\parallel }^{\left (\alpha \right )}\left (\protect \mathbf
  {r}\right )\cdot \protect \cc@accent {"705E}{\protect \mathbf {P}}_{\parallel
  }^{\left (\beta \right )}\left (\protect \mathbf {r}\right ). \protect \]
  Consider a quantum subsystem composed of a single electron within a
  semiclassical approximation. The Coulomb self-interaction energy in Eq.~(\ref
  {eq:Power-Zienau-Woolley}) is \protect \[ \protect \cc@accent
  {"705E}{V}_{\protect \mathrm {self}}=\protect \frac {1}{\epsilon _{0}}\DOTSI
  \intop \ilimits@ \protect \mathrm {d}v\protect \mathbf {P}_{\parallel }\left
  (\protect \mathbf {r},t\right )\cdot \protect \mathbf {\protect \cc@accent
  {"705E}{P}}_{\parallel }\left (\protect \mathbf {r}\right ). \protect \] If
  we add this term to the Hamiltonian in Eq.~(\ref {eq:electronic_Hamiltonian})
  and substitute $\protect \mathbf {E}_{\perp }=\protect \mathbf {E}+\protect
  \frac {1}{\epsilon _{0}}\protect \mathbf {P}_{\parallel }$, we find that the
  Coulomb self energy is canceled, yielding Eq.~(\ref
  {eq:electric_Hamiltonian_withself}) \protect \[ \protect \cc@accent
  {"705E}{H}^{\protect \mathrm {el}}+\protect \cc@accent {"705E}{V}_{\protect
  \mathrm {self}}=\protect \cc@accent {"705E}{H}_{s}-\DOTSI \intop \ilimits@
  \protect \mathrm {d}v\protect \mathbf {E}\left (\protect \mathbf {r},t\right
  )\cdot \protect \mathbf {\protect \cc@accent {"705E}{P}}\left (\protect
  \mathbf {r}\right ). \protect \] For dynamics propagated with the
  semiclassical electronic Hamiltonian in Eq.~(\ref
  {eq:electric_Hamiltonian_withself}), the conserved energy becomes \protect \[
  \begin {split}U_{\protect \mathrm {tot}} & =\protect \mathrm {Tr}_{s}\left
  (\protect \cc@accent {"705E}{\rho }\left (t\right )H_{s}\right )+\\ & \DOTSI
  \intop \ilimits@ \protect \mathrm {d}v\left (\protect \frac {\epsilon
  _{0}}{2}\protect \mathbf {E}\left (\protect \mathbf {r},t\right
  )^{2}+\protect \frac {1}{2\mu _{0}}\protect \mathbf {B}\left (\protect
  \mathbf {r},t\right )^{2}\right ) \end {split} \protect \]}\BibitemShut
  {NoStop}%
\bibitem [{\citenamefont {Li}\ \emph {et~al.}(2018{\natexlab{b}})\citenamefont
  {Li}, \citenamefont {Chen}, \citenamefont {Sukharev}, \citenamefont
  {Nitzan},\ and\ \citenamefont {Subotnik}}]{li_ehrenfest+r_2018}%
  \BibitemOpen
  \bibfield  {author} {\bibinfo {author} {\bibfnamefont {T.~E.}\ \bibnamefont
  {Li}}, \bibinfo {author} {\bibfnamefont {H.-T.}\ \bibnamefont {Chen}},
  \bibinfo {author} {\bibfnamefont {M.}~\bibnamefont {Sukharev}}, \bibinfo
  {author} {\bibfnamefont {A.}~\bibnamefont {Nitzan}}, \ and\ \bibinfo {author}
  {\bibfnamefont {J.~E.}\ \bibnamefont {Subotnik}},\ }\href@noop {} {\
  (\bibinfo {year} {2018}{\natexlab{b}})},\ \bibinfo {note} {(in
  preparation)}\BibitemShut {NoStop}%
\bibitem [{\citenamefont {Neuhauser}\ and\ \citenamefont
  {Lopata}(2007)}]{neuhauser_molecular_2007}%
  \BibitemOpen
  \bibfield  {author} {\bibinfo {author} {\bibfnamefont {D.}~\bibnamefont
  {Neuhauser}}\ and\ \bibinfo {author} {\bibfnamefont {K.}~\bibnamefont
  {Lopata}},\ }\href {\doibase 10.1063/1.2790436} {\bibfield  {journal}
  {\bibinfo  {journal} {The Journal of Chemical Physics}\ }\textbf {\bibinfo
  {volume} {127}},\ \bibinfo {pages} {154715} (\bibinfo {year}
  {2007})}\BibitemShut {NoStop}%
\bibitem [{\citenamefont {Lopata}\ and\ \citenamefont
  {Neuhauser}(2009{\natexlab{a}})}]{lopata_nonlinear_2009}%
  \BibitemOpen
  \bibfield  {author} {\bibinfo {author} {\bibfnamefont {K.}~\bibnamefont
  {Lopata}}\ and\ \bibinfo {author} {\bibfnamefont {D.}~\bibnamefont
  {Neuhauser}},\ }\href {\doibase 10.1063/1.3167407} {\bibfield  {journal}
  {\bibinfo  {journal} {The Journal of Chemical Physics}\ }\textbf {\bibinfo
  {volume} {131}},\ \bibinfo {pages} {014701} (\bibinfo {year}
  {2009}{\natexlab{a}})}\BibitemShut {NoStop}%
\bibitem [{\citenamefont {Lopata}\ and\ \citenamefont
  {Neuhauser}(2009{\natexlab{b}})}]{lopata_multiscale_2009}%
  \BibitemOpen
  \bibfield  {author} {\bibinfo {author} {\bibfnamefont {K.}~\bibnamefont
  {Lopata}}\ and\ \bibinfo {author} {\bibfnamefont {D.}~\bibnamefont
  {Neuhauser}},\ }\href {\doibase 10.1063/1.3082245} {\bibfield  {journal}
  {\bibinfo  {journal} {The Journal of Chemical Physics}\ }\textbf {\bibinfo
  {volume} {130}},\ \bibinfo {pages} {104707} (\bibinfo {year}
  {2009}{\natexlab{b}})}\BibitemShut {NoStop}%
\bibitem [{\citenamefont {Deinega}\ and\ \citenamefont
  {Seideman}(2014)}]{deinega_self-interaction-free_2014}%
  \BibitemOpen
  \bibfield  {author} {\bibinfo {author} {\bibfnamefont {A.}~\bibnamefont
  {Deinega}}\ and\ \bibinfo {author} {\bibfnamefont {T.}~\bibnamefont
  {Seideman}},\ }\href {\doibase 10.1103/PhysRevA.89.022501} {\bibfield
  {journal} {\bibinfo  {journal} {Phys. Rev. A}\ }\textbf {\bibinfo {volume}
  {89}},\ \bibinfo {pages} {022501} (\bibinfo {year} {2014})}\BibitemShut
  {NoStop}%
\bibitem [{\citenamefont {Chen}\ \emph {et~al.}(2018)\citenamefont {Chen},
  \citenamefont {Li}, \citenamefont {Sukharev}, \citenamefont {Nitzan},\ and\
  \citenamefont {Subotnik}}]{chen_ehrenfest+r_2018-2}%
  \BibitemOpen
  \bibfield  {author} {\bibinfo {author} {\bibfnamefont {H.-T.}\ \bibnamefont
  {Chen}}, \bibinfo {author} {\bibfnamefont {T.~E.}\ \bibnamefont {Li}},
  \bibinfo {author} {\bibfnamefont {M.}~\bibnamefont {Sukharev}}, \bibinfo
  {author} {\bibfnamefont {A.}~\bibnamefont {Nitzan}}, \ and\ \bibinfo {author}
  {\bibfnamefont {J.~E.}\ \bibnamefont {Subotnik}},\ }\href@noop {} {\
  (\bibinfo {year} {2018})},\ \bibinfo {note} {(to be published)}\BibitemShut
  {NoStop}%
\bibitem [{Note2()}]{Note2}%
  \BibitemOpen
  \bibinfo {note} {Formally, the rescaling direction in Eqs.~(\ref
  {eq:rescaling_E}) and (\ref {eq:rescaling_B}) are motivated by a comparison
  of the electrodynamical quantum\textendash classical Liouville equation
  (QCLE) and Ehrenfest dynamics in the framework of mixed quantum-classical
  theory (to be published).}\BibitemShut {Stop}%
\bibitem [{Note3()}]{Note3}%
  \BibitemOpen
  \bibinfo {note} {Note that the self-interference length strongly depends on
  dimensionality and is much smaller in 3D than in 1D.}\BibitemShut {Stop}%
\bibitem [{\citenamefont {Peslherbe}\ and\ \citenamefont
  {Hase}(1994)}]{peslherbe_analysis_1994}%
  \BibitemOpen
  \bibfield  {author} {\bibinfo {author} {\bibfnamefont {G.~H.}\ \bibnamefont
  {Peslherbe}}\ and\ \bibinfo {author} {\bibfnamefont {W.~L.}\ \bibnamefont
  {Hase}},\ }\href {\doibase 10.1063/1.466648} {\bibfield  {journal} {\bibinfo
  {journal} {The Journal of Chemical Physics}\ }\textbf {\bibinfo {volume}
  {100}},\ \bibinfo {pages} {1179} (\bibinfo {year} {1994})}\BibitemShut
  {NoStop}%
\bibitem [{\citenamefont {Brieuc}\ \emph {et~al.}(2016)\citenamefont {Brieuc},
  \citenamefont {Bronstein}, \citenamefont {Dammak}, \citenamefont {Depondt},
  \citenamefont {Finocchi},\ and\ \citenamefont
  {Hayoun}}]{brieuc_zero-point_2016}%
  \BibitemOpen
  \bibfield  {author} {\bibinfo {author} {\bibfnamefont {F.}~\bibnamefont
  {Brieuc}}, \bibinfo {author} {\bibfnamefont {Y.}~\bibnamefont {Bronstein}},
  \bibinfo {author} {\bibfnamefont {H.}~\bibnamefont {Dammak}}, \bibinfo
  {author} {\bibfnamefont {P.}~\bibnamefont {Depondt}}, \bibinfo {author}
  {\bibfnamefont {F.}~\bibnamefont {Finocchi}}, \ and\ \bibinfo {author}
  {\bibfnamefont {M.}~\bibnamefont {Hayoun}},\ }\href {\doibase
  10.1021/acs.jctc.6b00684} {\bibfield  {journal} {\bibinfo  {journal} {J.
  Chem. Theory Comput.}\ }\textbf {\bibinfo {volume} {12}},\ \bibinfo {pages}
  {5688} (\bibinfo {year} {2016})}\BibitemShut {NoStop}%
\bibitem [{\citenamefont {Taflove}\ and\ \citenamefont
  {Taflove}(1998)}]{taflove_advances_1998}%
  \BibitemOpen
  \bibfield  {author} {\bibinfo {author} {\bibfnamefont {A.}~\bibnamefont
  {Taflove}}\ and\ \bibinfo {author} {\bibfnamefont {A.}~\bibnamefont
  {Taflove}},\ }\href
  {https://www.scholars.northwestern.edu/en/publications/advances-in-computational-electrodynamics-the-finite-difference-t}
  {{\selectlanguage {English}\emph {\bibinfo {title} {Advances in
  {Computational} {Electrodynamics}: {The} {Finite}-{Difference}
  {Time}-{Domain} {Method}}}}}\ (\bibinfo  {publisher} {Artech House},\
  \bibinfo {year} {1998})\BibitemShut {NoStop}%
\bibitem [{\citenamefont {Mollow}(1969)}]{mollow_power_1969}%
  \BibitemOpen
  \bibfield  {author} {\bibinfo {author} {\bibfnamefont {B.~R.}\ \bibnamefont
  {Mollow}},\ }\href {\doibase 10.1103/PhysRev.188.1969} {\bibfield  {journal}
  {\bibinfo  {journal} {Phys. Rev.}\ }\textbf {\bibinfo {volume} {188}},\
  \bibinfo {pages} {1969} (\bibinfo {year} {1969})}\BibitemShut {NoStop}%
\bibitem [{\citenamefont {Huo}\ and\ \citenamefont
  {Coker}(2012)}]{huo_consistent_2012}%
  \BibitemOpen
  \bibfield  {author} {\bibinfo {author} {\bibfnamefont {P.}~\bibnamefont
  {Huo}}\ and\ \bibinfo {author} {\bibfnamefont {D.~F.}\ \bibnamefont
  {Coker}},\ }\href {\doibase 10.1063/1.4748316} {\bibfield  {journal}
  {\bibinfo  {journal} {The Journal of Chemical Physics}\ }\textbf {\bibinfo
  {volume} {137}},\ \bibinfo {pages} {22A535} (\bibinfo {year}
  {2012})}\BibitemShut {NoStop}%
\bibitem [{\citenamefont {Kim}\ \emph {et~al.}(2008)\citenamefont {Kim},
  \citenamefont {Nassimi},\ and\ \citenamefont
  {Kapral}}]{kim_quantum-classical_2008}%
  \BibitemOpen
  \bibfield  {author} {\bibinfo {author} {\bibfnamefont {H.}~\bibnamefont
  {Kim}}, \bibinfo {author} {\bibfnamefont {A.}~\bibnamefont {Nassimi}}, \ and\
  \bibinfo {author} {\bibfnamefont {R.}~\bibnamefont {Kapral}},\ }\href
  {\doibase 10.1063/1.2971041} {\bibfield  {journal} {\bibinfo  {journal} {J.
  Chem. Phys.}\ }\textbf {\bibinfo {volume} {129}},\ \bibinfo {pages} {084102}
  (\bibinfo {year} {2008})}\BibitemShut {NoStop}%
\bibitem [{\citenamefont {McDonald}(1997)}]{mcdonald_relation_1997}%
  \BibitemOpen
  \bibfield  {author} {\bibinfo {author} {\bibfnamefont {K.~T.}\ \bibnamefont
  {McDonald}},\ }\href {\doibase 10.1119/1.18723} {\bibfield  {journal}
  {\bibinfo  {journal} {American Journal of Physics}\ }\textbf {\bibinfo
  {volume} {65}},\ \bibinfo {pages} {1074} (\bibinfo {year}
  {1997})}\BibitemShut {NoStop}%
\bibitem [{\citenamefont {Panofsky}\ and\ \citenamefont
  {Phillips}(2005)}]{panofsky_classical_2005}%
  \BibitemOpen
  \bibfield  {author} {\bibinfo {author} {\bibfnamefont {W.~K.~H.}\
  \bibnamefont {Panofsky}}\ and\ \bibinfo {author} {\bibfnamefont
  {M.}~\bibnamefont {Phillips}},\ }\href
  {https://trove.nla.gov.au/work/16412826} {\emph {\bibinfo {title} {Classical
  electricity and magnetism}}},\ \bibinfo {edition} {2nd}\ ed.\ (\bibinfo
  {publisher} {Dover Publications},\ \bibinfo {year} {2005})\BibitemShut
  {NoStop}%
\bibitem [{\citenamefont {Griffiths}(2014)}]{griffiths_introduction_2014}%
  \BibitemOpen
  \bibfield  {author} {\bibinfo {author} {\bibfnamefont {D.~J.}\ \bibnamefont
  {Griffiths}},\ }\href@noop {} {\emph {\bibinfo {title} {Introduction to
  {Electrodynamics}}}}\ (\bibinfo  {publisher} {Pearson Higher Ed.},\ \bibinfo
  {year} {2014})\BibitemShut {NoStop}%
\bibitem [{Note4()}]{Note4}%
  \BibitemOpen
  \bibinfo {note} {Note that, in 1D, $\protect \mathbf {P}^{\protect \text
  {1D}}$ is always transverse.}\BibitemShut {Stop}%
\bibitem [{\citenamefont {Tully}(2012)}]{tully_perspective:_2012}%
  \BibitemOpen
  \bibfield  {author} {\bibinfo {author} {\bibfnamefont {J.~C.}\ \bibnamefont
  {Tully}},\ }\href {\doibase 10.1063/1.4757762} {\bibfield  {journal}
  {\bibinfo  {journal} {J. Chem. Phys.}\ }\textbf {\bibinfo {volume} {137}},\
  \bibinfo {pages} {22A301} (\bibinfo {year} {2012})}\BibitemShut {NoStop}%
\bibitem [{\citenamefont {Donoso}\ and\ \citenamefont
  {Martens}(2001)}]{donoso_quantum_2001}%
  \BibitemOpen
  \bibfield  {author} {\bibinfo {author} {\bibfnamefont {A.}~\bibnamefont
  {Donoso}}\ and\ \bibinfo {author} {\bibfnamefont {C.~C.}\ \bibnamefont
  {Martens}},\ }\href {\doibase 10.1103/PhysRevLett.87.223202} {\bibfield
  {journal} {\bibinfo  {journal} {Phys. Rev. Lett.}\ }\textbf {\bibinfo
  {volume} {87}},\ \bibinfo {pages} {223202} (\bibinfo {year}
  {2001})}\BibitemShut {NoStop}%
\bibitem [{\citenamefont {Donoso}\ \emph {et~al.}(2003)\citenamefont {Donoso},
  \citenamefont {Zheng},\ and\ \citenamefont
  {Martens}}]{donoso_simulation_2003}%
  \BibitemOpen
  \bibfield  {author} {\bibinfo {author} {\bibfnamefont {A.}~\bibnamefont
  {Donoso}}, \bibinfo {author} {\bibfnamefont {Y.}~\bibnamefont {Zheng}}, \
  and\ \bibinfo {author} {\bibfnamefont {C.~C.}\ \bibnamefont {Martens}},\
  }\href {\doibase 10.1063/1.1597496} {\bibfield  {journal} {\bibinfo
  {journal} {The Journal of Chemical Physics}\ }\textbf {\bibinfo {volume}
  {119}},\ \bibinfo {pages} {5010} (\bibinfo {year} {2003})}\BibitemShut
  {NoStop}%
\end{thebibliography}%

\end{document}